%% file: elp_edm.tex
\documentclass[12pt,dvips]{article}
\usepackage{rotating}
\usepackage{axodraw}
\usepackage{epsfig}
\usepackage{color}
\usepackage{cite}
\input def_colors

\def\theequation{\arabic{section}.\arabic{equation}}

\newcommand{\imag}{\Im {\rm m}}

\newcommand{\lsim}{\raisebox{-0.13cm}{~\shortstack{$<$ \\[-0.07cm] $\sim$}}~}
\newcommand{\gsim}{\raisebox{-0.13cm}{~\shortstack{$>$ \\[-0.07cm] $\sim$}}~}

\begin{document}

\def\thefootnote{\fnsymbol{footnote}}

\begin{flushright}
{\tt CERN-PH-TH/2008-175}, {\tt MAN/HEP/2008/22 }\\
{\tt arXiv:0808.1819} \\
August 2008
\end{flushright}

\begin{center}
{\bf {\LARGE Electric Dipole Moments in the MSSM\\[3mm]
Reloaded } }
%\\[3.mm]
\end{center}

\medskip

\begin{center}{\large
John~Ellis$^a$,
Jae~Sik~Lee$^{b,c,d}$ and
Apostolos~Pilaftsis$^e$}
\end{center}

\begin{center}
{\em $^a$Theory Division, CERN, CH-1211 Geneva 23, Switzerland}\\[0.2cm]
{\em $^b$Physics Division, National Center for Theoretical Sciences, 
Hsinchu, Taiwan}\\[0.2cm]
{\em $^c$Department of Physics and Center for Mathematics and
  Theoretical Physics,}\\ 
{\em National Central University, Chung-Li, Taiwan}\\[0.2cm]
{\em $^d$Institute of Physics, Academia Sinica, Taipei, Taiwan}\\[0.2cm]
{\em $^e$School of Physics and Astronomy, University of Manchester,}\\
{\em Manchester M13 9PL, United Kingdom}
\end{center}

\bigskip\bigskip

\centerline{\bf ABSTRACT}

\noindent  
We  present a  detailed study  of the  Thallium, neutron,  Mercury and
deuteron electric  dipole moments  (EDMs) in the  CP-violating Minimal
Supersymmetric extension  of the Standard Model~(MSSM).   We take into
account   the  complete   set   of  one-loop   graphs,  the   dominant
Higgs-mediated  two-loop diagrams,  the complete  CP-odd dimension-six
Weinberg  operator and  the Higgs-mediated  four-fermion  operators.  We
improve upon earlier calculations by including the resummation effects
due to CP-violating Higgs-boson mixing and to threshold corrections to
the  Yukawa couplings  of all  up-  and down-type  quarks and  charged
leptons.   As  an  application  of  our  study,  we  analyse  the  EDM
constraints  on the  CPX, trimixing  and Maximally  CP-  and Minimally
Flavour-Violating (MCPMFV)  scenarios.  Cancellations may  occur among
the CP-violating contributions to the three measured EDMs arising from
the 6  CP-violating phases  in the MCPMFV  scenario, leaving  open the
possibility  of relatively large  contributions to  other CP-violating
observables.  The analytic expressions for the EDMs are implemented in
an updated version of the code {\tt CPsuperH2.0}.

\newpage

\section{Introduction}

With the imminent  advent of the LHC, we are  entering an exciting
era for probing new physics at the TeV scale. If new physics is indeed
observed at this scale, the  questions of its flavour and CP structure
will  immediately become  very critical. The non-observation  of  the Thallium  
($^{205}{\rm Tl}$)~\cite{Regan:2002ta},    neutron~($n$)~\cite{Baker:2006ts},   and
Mercury ($^{199}{\rm Hg}$)~\cite{Romalis:2000mg} electric-dipole moments (EDMs) already provide
remarkably  tight bounds on  possible new  CP-violating phases  beyond the
Cabibbo--Kobayashi--Maskawa  (CKM)  one of  the  Standard Model  (SM). 
Complementary to  the direct
explorations  at the  LHC, a  new generation  of  precision low-energy
experiments  is also  expected to  play  an important  role.  The  new
precision experiments will place much stronger indirect constraints on
the possible CP and flavour  structure of models of TeV-scale physics.
In particular,  if  the proposed experiment  searching  for  a deuteron  ($^2{\rm
H}^+$)          EDM           achieves          the          projected
sensitivity~\cite{Semertzidis:2003iq,OMS}, this  will improve the existing
bounds on  possible CP-violating chromoelectric operators  by 
several orders of magnitude~\cite{Lebedev:2004va}.

One of  the theoretically best-motivated  scenarios of new  physics is
Supersymmetry   (SUSY)~\cite{SUSY}.    Its   minimal  realization,   the   Minimal
Supersymmetric  extension  of the  Standard  Model  (MSSM), with  SUSY
broken softly at the TeV scale, addresses the naturalness of the gauge
hierarchy,  predicts  gauge-coupling  unification, provides  a  viable
candidate for Cold  Dark Matter (CDM) and may  help explain the baryon
asymmetry in  the Universe (BAU)  via a first-order  electroweak phase
transition~\cite{BAU}.  An essential role in the generation of the BAU could
be played by the new CP-odd  phases that appear in the MSSM~\cite{BAU}.  However,
the  non-observation of  EDMs severely  constrains  these CP-violating
phases~\cite{Ibrahim:2007fb}.

The  aim  of  this  paper  is  to present  a  detailed  study  of  the
$^{205}{\rm Tl}$, $n$, $^{199}{\rm Hg}$  and $^2{\rm H}^+$ EDMs in the
CP-violating  MSSM. We  include  in  our study  the  complete set  of
one-loop  graphs~\cite{EFN,Abel:2001vy},  the dominant  Higgs-mediated
two-loop    diagrams~\cite{Chang:1998uc,Pilaftsis:2002fe,Ellis:2005ika}
of  the  Barr--Zee type~\cite{BZ}  and  the Higgs-mediated  four-fermion
operators~\cite{Pilaftsis:2002fe,LP,Demir:2003js},  originally studied
by  Barr~\cite{SBarr} within  a two-Higgs  doublet model~\footnote{However, 
we do not include here the contributions of
a possible non-zero CP-violating QCD vacuum parameter $\theta$.}.  We improve
upon     earlier     calculations~\cite{Dai:1990xh,Dicus:1989va}    by
calculating    the     complete    CP-odd    dimension-six    Weinberg
operator~\cite{Weinberg:1989dx}  and by including  resummation effects
due  to CP-violating Higgs-boson  mixing~\cite{APLB} and  to threshold
corrections to  the Yukawa couplings  of the up- and  down-type quarks
and  charged leptons~\cite{Banks}.   We  then use  this compendium  to
derive  representative constraints on  the CP-violating  parameters of
phenomenologically relevant  benchmarks in the  MSSM, such as  the CPX
scenario~\cite{Carena:2000ks},              the              trimixing
scenario~\cite{Ellis:2004fs,Ellis:2005ika}  and the  general Maximally
CP-        and       Minimally        Flavour-Violating       (MCPMFV)
framework~\cite{Ellis:2007kb},  for  selected sets  of  values of  the
CP-conserving parameters. Clearly, sufficiently small values of the
CP-violating parameters must be compatible with the experimental
upper limits on the EDMs but, as we shall illustrate with explicit examples, 
larger values may also be allowed by non-trivial cancellations.

For  the   presentation  of  our  analytic  results,   we  follow  the
conventions   and   notations   of   {\tt   CPsuperH}~\cite{cpsuperh},
especially for  the masses  and mixing matrices  of the  neutral Higgs
bosons  and SUSY  particles.   We  note parenthetically that the  new
version  of {\tt  CPsuperH}, {\tt  CPsuperH2.0}, includes  an improved
treatment of  Higgs-boson propagators and Higgs  couplings, and enables
numerical predictions for a number of flavour-changing-neutral-current
(FCNC)  $B$-meson observables, including CP-violating effects.
On the basis of the  results in  this  work, we
further  improve  the  code  {\tt  CPsuperH2.0}  by  implementing  the
computation of the Thallium, neutron, Mercury and deuteron EDMs in the
CP-violating MSSM.

The layout of the paper is as follows. Section~2 presents all formulae
relevant   to  the   one-loop  contributions   to  the   electric  and
chromoelectric dipole  moments of the charged leptons  and quarks that
result    from    chargino-,    neutralino-,    and    gluino-mediated
diagrams. Non-holomorphic threshold  effects on the light-quark Yukawa
couplings have been appropriately  resummed, as these are the dominant
source of  higher-order corrections.   In Section~3, we  calculate the
CP-odd  dimension-six  three-gluon   Weinberg  operator,  taking  into
account loop diagrams  involving $t$ and $b$ quarks  and Higgs bosons,
in addition  to the previously known  loop effects due to  $t$ and $b$
squarks and gluinos.  In addition, we present analytic results for the
Higgs-mediated  four-fermion operators  and the  dominant Higgs-mediated
two-loop diagrams.  In Section~4 we compute the $^{205}{\rm Tl}$, $n$,
$^{199}{\rm  Hg}$ and  $^2{\rm H}^+$  EDMs in  the  CP-violating MSSM.
Section~5 presents illustrative  constraints on key soft SUSY-breaking
parameters and CP-odd phases in  the CPX, the trimixing and the MCPMFV
scenarios.  We summarize our conclusions in Section~6.

\setcounter{equation}{0}
\section{One-Loop EDMs of Leptons and Quarks}

At the  one-loop level,  the charged leptons,  $e$, $\mu$ and $\tau$,  and the
light quarks,  $u$, $d$ and $s$,  can have EDMs  induced by charginos,
neutralinos and gluinos. The $u$,  $d$ and $s$ quarks may also develop
chromoelectric dipole moments (CEDMs) via the corresponding squark and
gluino loop  diagrams. In this Section we  exhibit analytical formulae
for the one-loop EDMs of charged leptons and light quarks, and the CEDMs
of the quarks.

We denote the  EDM of a fermion  by $d^E_f$ and the CEDM  of a quark
by $d^C_q$. The relevant (C)EDM interaction Lagrangian is given by
\begin{equation}
 \label{CEDM}
{\cal L}_{\rm
  (C)EDM}\ =\
  -\; \frac{i}{2}\,d^E_f\,F^{\mu\nu}\,\bar{f}\,\sigma_{\mu\nu}\gamma_5\,f\  
-\ \frac{i}{2}\,d_q^C\,G^{a\,\mu\nu}\,\bar{q}\,\sigma_{\mu\nu}\gamma_5 T^a\,q\,,
\end{equation}
where  $F^{\mu\nu}$ and  $G^{a\,\mu\nu}$ are  the  electromagnetic and
strong  field strengths, respectively,  and the $T^a=\lambda^a/2$  are the
generators    of    the     SU(3)$_C$    group.     The    interaction
Lagrangian~(\ref{CEDM}) leads to a matrix element of the form:
\begin{equation}
i{\cal    M}\ =\ -d^E_f\,\epsilon(q)\cdot(p+p^\prime)\,\bar{u}(p^\prime
)\,\gamma_5\,u(p)\; ,
\end{equation}
where  $p=q+p^\prime$  and   $\epsilon(q)\cdot  p  =  \epsilon(q)\cdot
p^\prime$,  since  $\epsilon(q)\cdot  q=0$.   We  use  the  convention
$\sigma^{\mu\nu}    =    \frac{i}{2}[\gamma^\mu\,,\gamma^\nu]   =    i
(\gamma^\mu\gamma^\nu -g^{\mu\nu})$.

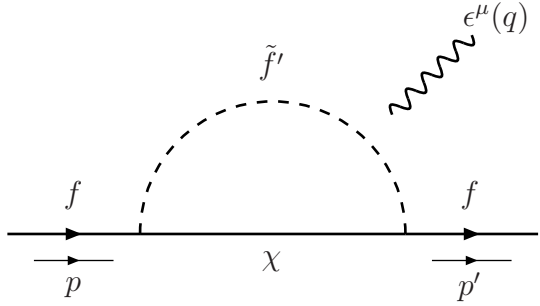
\begin{figure}[t]
\begin{center}
\begin{picture}(250,120)(0,0)
%-
\SetWidth{1.0}
\ArrowLine(25,25)(75,25)
\Line(75,25)(175,25)
\ArrowLine(175,25)(225,25)
\DashCArc(125,25)(50,0,180){4}
\Photon(170,70)(200,100){3}{5}

\SetWidth{0.5}
\ArrowLine(35,15)(65,15)\Text(50,5)[]{$p$}\Text(50,40)[]{$f$}
\ArrowLine(185,15)(215,15)\Text(200,5)[]{$p^\prime$}\Text(200,40)[]{$f$}
\Text(125,15)[]{$\chi$}
\Text(125,90)[]{$\tilde{f'}$}
\Text(210,107)[]{$\epsilon^\mu (q)$}
%-
\end{picture} 
\end{center}
\smallskip
\noindent
\caption{\it Generic Feynman diagram for the EDM $(d^E_f)^\chi$ of the fermion $f$
induced by $\chi$ exchange effects. The photon line can be attached
to the sfermion $\tilde{f'}$ line,
or to the $\chi$ line if $\chi=\tilde\chi^\pm_i$.}
\label{fig:EDM}
\end{figure}

To set  our coupling notations  and normalisations, we write  down the
generic interaction of a chargino $\tilde{\chi}^\pm_{1,2}$, neutralino
$\tilde{\chi}^0_{1,2,3,4}$   or  gluino   $\tilde{g}^a$,  collectively
denoted by $\chi$, with a fermion $f$ and sfermion $\tilde{f'}_{1,2}$,
as follows:\footnote{Here  the convention for the  couplings $g_L$ and
$g_R$   is  different   from  that   used   in  \cite{Ibrahim:2007fb}:
$g_L=R_{ik}^*$ and $g_R=L_{ik}^*$.}
\begin{equation}
  \label{chiff}
{\cal L}_{\chi f \tilde{f'}}\ =\ g^{\chi f\tilde{f}'}_{L\, ij}\, 
(\bar{\chi}_i P_L f)\, \tilde{f}^{\prime *}_j\ +\
g^{\chi f\tilde{f}'}_{R\, ij}\, (\bar{\chi}_i P_R f)\,
\tilde{f}^{\prime *}_j\ 
+\ {\rm h.c.}
\end{equation}
Likewise, the interaction Lagrangians for the couplings of a photon
$A^\mu$ with $\chi$ and $\tilde{f}'$ read:
\begin{equation}
 \label{chifA}
{\cal L}_{\chi\chi A}\ =\ -e\,Q_\chi\,(\bar{\chi}\gamma_\mu \chi)
A^\mu\, ,\qquad
{\cal L}_{\tilde{f'}\tilde{f'}A}\ =\
-ie\,Q_{\tilde{f'}}\,\tilde{f}^{\prime *}
\stackrel {\leftrightarrow} {\partial}_\mu \tilde{f'} A^\mu\; .
\end{equation}
Employing~(\ref{chiff})    and   (\ref{chifA})    and    taking   into
consideration  the diagrams  of Fig.~\ref{fig:EDM},  we  calculate the
one-loop fermion EDM,
\begin{equation}
\left(\frac{d^E_f}{e}\right)^\chi\ =\ \frac{m_{\chi_i}}{16\pi^2
  m_{\tilde{f}'_j}^2} 
\imag\Big[ \Big(g^{\chi f\tilde{f}'}_{R\, ij}\Big)^* 
  g^{\chi f\tilde{f}'}_{L\, ij} \Big]
\left[Q_{\chi}\, A(m_{\chi_i}^2/m_{\tilde{f}'_j}^2) + Q_{\tilde{f'}}\, 
B(m_{\chi_i}^2/m_{\tilde{f}'_j}^2)\right]\,,
\end{equation}
where
\begin{eqnarray}
A(r)  \!&=&\! \frac{1}{2(1-r)^2}\left(3-r+\frac{2 \ln{r}}{1-r}\right)\,, \quad
B(r) \ =\ \frac{1}{2(1-r)^2}\left(1+r+\frac{2 r \ln{r}}{1-r}\right)\; ,
\end{eqnarray}
with $A(1)=-1/3$  and $B(1)=1/6$.  We  have checked that  our analytic
expressions for the one-loop EDMs are in agreement with~\cite{Ibrahim:2007fb}
and~\cite{Abel:2001vy}.

%---------------------  EDM Diagrams (START)  ---------------------------%
%---------------------  EDM Diagrams (END)  ---------------------------%

We  now  present  the  individual one-loop  contributions  of  charginos
$\tilde{\chi}^\pm$     to    the     EDMs    of     charged    leptons
$(d_l^E/e)^{\tilde{\chi}^\pm}$,             up-type             quarks
$(d_u^E/e)^{\tilde{\chi}^\pm}$        and       down-type       quarks
$(d_d^E/e)^{\tilde{\chi}^\pm}$. In detail, these are given by
\begin{eqnarray}
  \label{eEDMchi}
\left(\frac{d^E_l}{e}\right)^{\tilde\chi^\pm} \!\!\!&=&\!
\frac{1}{16\pi^2}\sum_i \frac{m_{\tilde{\chi}^\pm_i}}{m_{\tilde{\nu}_l}^2}\,
\imag[(g_{R\,i}^{\tilde{\chi}^\pm l \tilde{\nu}})^*\,
g_{L\,i}^{\tilde{\chi}^\pm l \tilde{\nu}}]\;
Q_{\tilde{\chi}^-}\,
A({m_{\tilde{\chi}^\pm_i}^2}/{m_{\tilde{\nu}_l}^2})\,,\\
  \label{uEDMchi}
\left(\frac{d^E_u}{e}\right)^{\tilde\chi^\pm} \!\!\!&=&\!
\frac{1}{16\pi^2}\sum_{i,j} \frac{m_{\tilde{\chi}^\pm_i}}{m_{\tilde{d}_j}^2}\,
\imag[(g_{R\,ij}^{\tilde{\chi}^\pm u \tilde{d}})^*\,
g_{L\,ij}^{\tilde{\chi}^\pm u \tilde{d}}]\, 
\Big[Q_{\tilde{\chi}^+}\, A({m_{\tilde{\chi}^\pm_i}^2}/{m_{\tilde{d}_j}^2})\:
+\: Q_{\tilde{d}}B({m_{\tilde{\chi}^\pm_i}^2}/{m_{\tilde{d}_j}^2})\Big]\,,\\
  \label{dEDMchi}
\left(\frac{d^E_d}{e}\right)^{\tilde\chi^\pm} \!\!\!&=&\!
\frac{1}{16\pi^2}\sum_{i,j} \frac{m_{\tilde{\chi}^\pm_i}}{m_{\tilde{u}_j}^2}\,
\imag[(g_{R\,ij}^{\tilde{\chi}^\pm d \tilde{u}})^*\,
g_{L\,ij}^{\tilde{\chi}^\pm d \tilde{u}}]\, 
\Big[Q_{\tilde{\chi}^-}\, A({m_{\tilde{\chi}^\pm_i}^2}/{m_{\tilde{u}_j}^2})\:
+\:
Q_{\tilde{u}}B({m_{\tilde{\chi}^\pm_i}^2}/{m_{\tilde{u}_j}^2})\Big]\,,\qquad 
\end{eqnarray}
where  the electric-charge assignments  for the  loop  particles are:
$Q_{\tilde{\chi}^\pm} = \pm1$, $Q_{\tilde{u}} = 2/3$, $Q_{\tilde{d}} =
-1/3$, and
\begin{eqnarray}
  \label{echi}
g_{L\,i}^{\tilde{\chi}^\pm l \tilde{\nu}} \!&=&\! -g
(C_R)_{i1}\,,\hspace{5.2cm}
g_{R\,i}^{\tilde{\chi}^\pm l \tilde{\nu}}\ =\  h_l^* (C_L)_{i2}\,,\\
  \label{uchi}
g_{L\,ij}^{\tilde{\chi}^\pm u \tilde{d}} \!&=&\! 
-g (C_L)_{i1}^* (U^{\tilde{d}})_{1j}^*
+h_d (C_L)_{i2}^* (U^{\tilde{d}})_{2j}^*\,, \qquad\,
g_{R\,ij}^{\tilde{\chi}^\pm u \tilde{d}}\ =\ 
h_u^* (C_R)_{i2}^* (U^{\tilde{d}})_{1j}^*\,,\\
  \label{dchi}
g_{L\,ij}^{\tilde{\chi}^\pm d \tilde{u}} \!&=&\! - g (C_R)_{i1} 
(U^{\tilde{u}})_{1j}^* + h_u (C_R)_{i2} (U^{\tilde{u}})_{2j}^*\,,\qquad 
g_{R\,ij}^{\tilde{\chi}^\pm d \tilde{u}}\ =\
h_d^* (C_L)_{i2} (U^{\tilde{u}})_{1j}^*\; .
\end{eqnarray} 
We note       that      the      coupling       coefficients      defined
in~(\ref{echi})--(\ref{dchi})                                     appear
in~(\ref{eEDMchi})--(\ref{dEDMchi}), respectively.

Correspondingly,  the  contributions of  neutralinos  to  the EDMs  of
charged    leptons   $(d_l^E/e)^{\tilde{\chi}^0}$,    up-type   quarks
$(d_u^E/e)^{\tilde{\chi}^0}$        and        down-type        quarks
$(d_d^E/e)^{\tilde{\chi}^0}$ may conveniently be expressed as
\begin{equation}
\left(\frac{d^E_f}{e}\right)^{\tilde\chi^0}\ =\
\frac{1}{16\pi^2}\sum_{i,j} \frac{m_{\tilde{\chi}^0_i}}{m_{\tilde{f}_j}^2}\,
\imag[(g_{R\,ij}^{\tilde{\chi}^0 f \tilde{f}})^*\,
g_{L\,ij}^{\tilde{\chi}^0 f \tilde{f}}]\,
Q_{\tilde{f}}\, B({m_{\tilde{\chi}^0_i}^2}/{m_{\tilde{f}_j}^2})\; ,
\end{equation}
with $f=l,u,d$. The neutralino-fermion-sfermion couplings are
\begin{eqnarray}
g_{L\,ij}^{\tilde{\chi}^0 f \tilde{f}} \! &=&\! 
-\sqrt{2}\, g\, T_3^f\, N_{i2}^* (U^{\tilde{f}})_{1j}^*
-\sqrt{2}\, g\, t_W\, (Q_f-T_3^f) N_{i1}^* (U^{\tilde{f}})_{1j}^*
-h_f N_{i\alpha}^* (U^{\tilde{f}})_{2j}^*\,,\nonumber\\
g_{R\,ij}^{\tilde{\chi}^0 f \tilde{f}} \! &=&\!
\sqrt{2}\, g\, t_W\, Q_f\, N_{i1} (U^{\tilde{f}})_{2j}^*
-h_f^* N_{i\alpha} (U^{\tilde{f}})_{1j}^*\,, 
\end{eqnarray}
where the Higgsino index $\alpha=3\,(f=l,d)$ or $ 4\,(f=u)$, $T_3^{l,d}=-1/2$
and $T_3^{u}=+1/2$.

In addition to charginos and neutralinos, gluinos also contribute to
the quark EDMs $(d_q^E/e)^{\,\tilde{g}}$.
In the gluino mass basis, $(d_q^E/e)^{\,\tilde{g}}$ is given by
\begin{equation}
\left(\frac{d_q^E}{e}\right)^{\,\tilde{g}} =
\frac{1}{3\pi^2} \sum_j \frac{|M_3|}{m_{\tilde{q}_j}^2}\,
\imag[(g_{R\,j}^{\tilde{g} q \tilde{q}})^* g_{L\,j}^{\tilde{g} q \tilde{q}}]\,
Q_{\tilde{q}}B({|M_3|^2}/{m_{\tilde{q}_j}^2})\,,
\end{equation}
where the gluino-quark-squark couplings are given by
\begin{eqnarray}
g_{L\,j}^{\tilde{g} q \tilde{q}} \!&=&\!
-\frac{g_s}{\sqrt{2}}\,{\rm e}^{-i\Phi_3/2} (U^{\tilde{q}})^*_{1j}\,, \qquad
g_{R\,j}^{\tilde{g} q \tilde{q}} \ =\
+\frac{g_s}{\sqrt{2}}\,{\rm e}^{+i\Phi_3/2} (U^{\tilde{q}})^*_{2j}\, .
\end{eqnarray}
 
As well as the  EDMs, the  chargino, neutralino  and gluino  loops can
produce       non-vanishing      CEDMs      for       the      quarks,
$(d_q^C)^{\chi^\pm,\chi^0,\tilde{g}}$.  Their individual contributions
are as follows:
\begin{eqnarray}
\left(d^C_u\right)^{\tilde\chi^\pm}&=&
\frac{g_s}{16\pi^2}\sum_{i,j} \frac{m_{\tilde{\chi}^\pm_i}}{m_{\tilde{d}_j}^2}\,
\imag[(g_{R\,ij}^{\tilde{\chi}^\pm u \tilde{d}})^*\,
g_{L\,ij}^{\tilde{\chi}^\pm u \tilde{d}}]\,
B({m_{\tilde{\chi}^\pm_i}^2}/{m_{\tilde{d}_j}^2})\,,
\nonumber \\
\left(d^C_d\right)^{\tilde\chi^\pm}&=&
\frac{g_s}{16\pi^2}\sum_{i,j}
\frac{m_{\tilde{\chi}^\pm_i}}{m_{\tilde{u}_j}^2}\, 
\imag[(g_{R\,ij}^{\tilde{\chi}^\pm d \tilde{u}})^*\,
g_{L\,ij}^{\tilde{\chi}^\pm d \tilde{u}}]\,
B({m_{\tilde{\chi}^\pm_i}^2}/{m_{\tilde{u}_j}^2})\,, \nonumber \\
\left(d^C_{q=u,d}\right)^{\tilde\chi^0}&=&
\frac{g_s}{16\pi^2}\sum_{i,j} \frac{m_{\tilde{\chi}^0_i}}{m_{\tilde{q}_j}^2}\,
\imag[(g_{R\,ij}^{\tilde{\chi}^0 q \tilde{q}})^*\,
g_{L\,ij}^{\tilde{\chi}^0 q \tilde{q}}]\,
B({m_{\tilde{\chi}^0_i}^2}/{m_{\tilde{q}_j}^2})\,, \nonumber \\
(d^C_{q=u,d})^{\,\tilde{g}} &=&
-\frac{g_s}{8\pi^2} \sum_j \frac{|M_3|}{m_{\tilde{q}_j}^2}\,
\imag[(g_{R\,j}^{\tilde{g} q \tilde{q}})^* g_{L\,j}^{\tilde{g} q \tilde{q}}]\,
C({|M_3|^2}/{m_{\tilde{q}_j}^2})\,,
\end{eqnarray}
where 
\begin{equation}
C(r)\  \equiv \ 
\frac{1}{6(1-r)^2}\left(10r-26+\frac{2r\ln{r}}{1-r} - 
     \frac{18\ln{r}}{1-r}\right)\,,  
\end{equation}
so that $C(1)=19/18$.

Finally,  it is  important to  stress that  there  are non-holomorphic
threshold corrections to the light-quark Yukawa couplings that occur in
the  chargino   and  neutralino  couplings.    These  corrections  are
proportional to the strong  coupling $\alpha_s$, and become significant
at large  $\tan\beta$. We therefore resum these  effects by redefining
the light-quark Yukawa couplings as follows:
\begin{eqnarray}
h_u \!&=&\! \frac{\sqrt{2} m_u}{vs_\beta}\,\frac{1}{1+\Delta_u/t_\beta}\,, \ \ \
h_c\ =\ \frac{\sqrt{2} m_c}{vs_\beta}\,\frac{1}{1+\Delta_c/t_\beta}\,, \nonumber \\
h_d \!&=&\! \frac{\sqrt{2} m_d}{vc_\beta}\,\frac{1}{1+\Delta_d t_\beta}\,, \ \ \ \ \
h_s\ =\ \frac{\sqrt{2} m_s}{vc_\beta}\,\frac{1}{1+\Delta_s t_\beta}\,,
\label{eq:Yukawa_light}
\end{eqnarray}
where
\begin{eqnarray}
\Delta_u \!&=&\! \frac{2\alpha_s}{3\pi}\,\mu^*M_3^*\,
I(M_{\tilde{U}_1}^2,M_{\tilde{Q}_1}^2,|M_3|^2)\,, \ \ \
\Delta_c\ =\ \frac{2\alpha_s}{3\pi}\,\mu^*M_3^*\,
I(M_{\tilde{U}_2}^2,M_{\tilde{Q}_2}^2,|M_3|^2)\,, \nonumber \\
\Delta_d \!&=&\! \frac{2\alpha_s}{3\pi}\,\mu^*M_3^*\,
I(M_{\tilde{D}_1}^2,M_{\tilde{Q}_1}^2,|M_3|^2)\,, \ \ \
\Delta_s\ =\ \frac{2\alpha_s}{3\pi}\,\mu^*M_3^*\,
I(M_{\tilde{D}_2}^2,M_{\tilde{Q}_2}^2,|M_3|^2)\,,\qquad
\end{eqnarray}
and
\begin{equation}
  \label{Ixyz}
I(x,y,z)\  \equiv \ \frac{xy\,\ln (x/y)\: +\: yz\,\ln (y/z)\: +\:
             xz\, \ln (z/x)}{(x-y)\,(y-z)\,(x-z)}\ 
\end{equation}
is  the one-loop function  that takes  account of  the non-holomorphic
threshold corrections.

\setcounter{equation}{0}
\section{Higher-Order Contributions to EDMs}

Beyond  the  one-loop  single-particle   level,  there  are  many  CP-violating
operators that can have significant effects on the EDMs. Of particular
importance  are the gluino--  and Higgs-mediated  dimension-6 Weinberg
operator,  the Higgs-exchange four-fermion  operators, and  the two-loop
Higgs-mediated Barr--Zee-type  diagrams. We present  detailed analytic
expressions for all those contributions in this Section.

\subsection{Weinberg Operator}

%---------------------  Weinberg operator Diagrams (START)  -----------------%
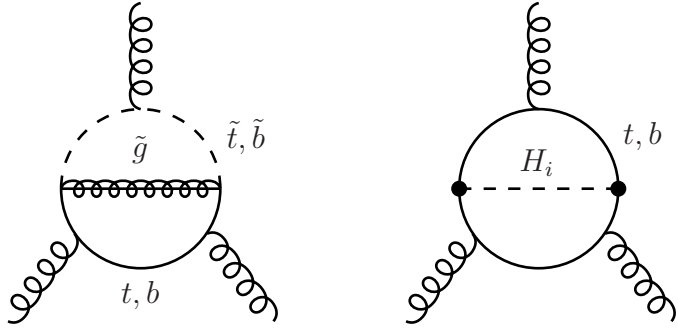
\begin{figure}[t]
\begin{center}
\begin{picture}(250,120)(0,0)
%-
\SetWidth{1.0}
\CArc(50,50)(30,180,360) \Text(50,10)[]{$t,b$}
\DashCArc(50,50)(30,0,180){5} \Text(90,70)[]{$\tilde{t},\tilde{b}$}
\Line(20,50)(80,50)
\Gluon(20,50)(80,50){3}{8} \Text(50,65)[]{$\tilde{g}$}
\Gluon(50,80)(50,120){4}{4}
\Gluon(25,33)(0,0){4}{4}
\Gluon(75,33)(100,0){4}{4}

\CArc(200,50)(30,0,360) \Text(240,70)[]{$t,b$}
\DashLine(170,50)(230,50){5} \Text(200,60)[]{$H_i$}
\Gluon(200,80)(200,120){4}{4}
\Gluon(175,33)(150,0){4}{4}
\Gluon(225,33)(250,0){4}{4}
\Vertex(170,50){3}\Vertex(230,50){3}
%-
\end{picture}
\end{center}
\smallskip
\noindent
\caption{\it Typical Feynman diagrams for $(d^{\,G})^{\tilde{g}}$ and
$(d^{\,G})^{H}$.  The $H_i$ lines denote all three neutral Higgs
bosons including CP-violating Higgs-boson mixing. Heavy dots indicate
resummation of threshold corrections to the corresponding Yukawa
couplings.}
\label{fig:Weinberg}
\end{figure}
%---------------------  Weinberg operator Diagrams (END)  -----------------------%

\begin{figure}[t]
\hspace{ 0.0cm}
\vspace{-0.5cm}
\centerline{\epsfig{figure=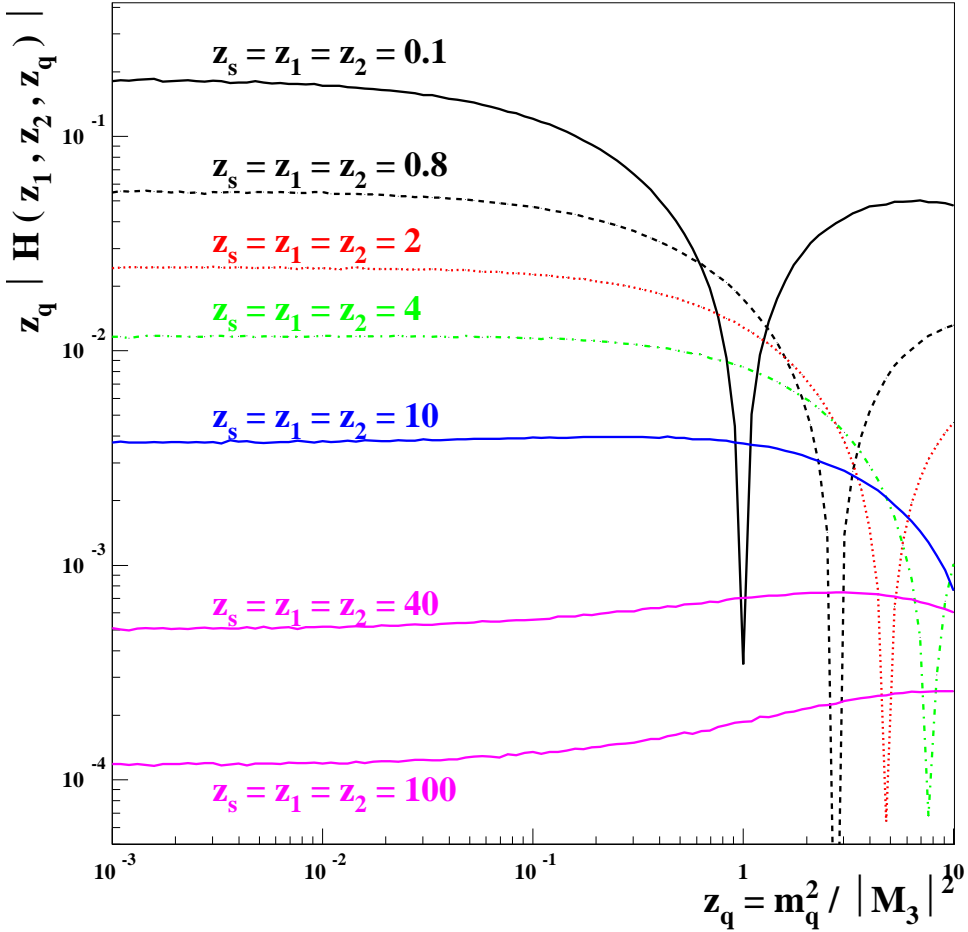,height=14.0cm,width=14.0cm}}
\vspace{-0.5cm}
\caption{\it The functional dependence of $z_q | H(z_1,z_2,z_q) |$ on $z_q$ for
several values of $z_s\equiv z_1=z_2$. When $z_s \leq 4$, $H(z_1,z_2,z_q)$
becomes negative when $z_q$ is larger than a specific value.}
\label{fig:zqh}
\end{figure}

We  start  by  considering  first  the  gluonic  dimension-6  Weinberg
operator.  This is described by the interaction Lagrangian~\footnote{Note  the 
coefficient  $d^{\,G}$ has  a different  sign from
that     in~\cite{Weinberg:1989dx}:    $d^{\,G}     =    -C$.}~\cite{Weinberg:1989dx}:
\begin{equation}
{\cal L}_{\rm Weinberg}\ =\
%-\frac{1}{6}\,d^{\,G}\,f_{abc}\, G^a_{\mu\rho}\, G^{b\,\rho}_{~~\nu}\,
%G^c_{\lambda\sigma}\, 
%\epsilon^{\mu\nu\lambda\sigma}
%=
\frac{1}{3}\,d^{\,G}\,f_{abc}\,G^a_{\rho\mu}\,\tilde{G}^{b\,\mu\nu}\,
{G^c}_{\nu}^{~~\rho}\; , 
\end{equation}
where $\tilde{G}^{\mu\nu} = \frac{1}{2} \epsilon^{\mu\nu\lambda\sigma}
G_{\lambda\sigma}$ is the dual  of the SU(3)$_c$ field-strength tensor
$G_{\lambda\sigma}$.  In  the MSSM,  $d^{\,G}$ is induced  at two-loop
order  by  $\tilde{g}$ and  $\tilde{t},\tilde{b}$,  and at  three-loop
order by CP-violating Higgs-boson mixing and non-holomorphic threshold
corrections to the top- and bottom-quark Yukawa couplings.  Hence, the
Weinberg operator $d^{\,G}$ is the sum of two terms:
\begin{equation}
  \label{dG}
d^{\,G}\ =\ (d^{\,G})^{\tilde{g}}\: +\: (d^{\,G})^{H}\; ,
\end{equation}
as illustrated in Fig.~\ref{fig:Weinberg}.
The  first term,  $(d^{\,G})^{\tilde{g}}$, is  the quark-squark-gluino
exchange contribution and is given by \cite{Dai:1990xh}
\begin{eqnarray} 
(d^{\,G})^{\tilde{g}} \!&=&\! -\frac{3}{2\pi}\left(\frac{g_s}{4\pi
    |M_3|}\right)^3 \sum_{q=t,b} m_q 
\left\{\sum_j \frac{m_{\tilde{q}_j}^2}{|M_3|^2}
\imag[(g_{R\,j}^{\tilde{g} q \tilde{q}})^* g_{L\,j}^{\tilde{g} q \tilde{q}}]
\right\} \nonumber \\
&& \times
H({m_{\tilde{q}_1}^2}/|M_3|^2,{m_{\tilde{q}_2}^2}/|M_3|^2,m_q^2/|M_3|^2)\,,
\end{eqnarray}
where
\begin{equation}
H(z_1,z_2,z_q)\ =\ \frac{1}{2}\int_0^1{\rm d}x\int_0^1{\rm d}u\int_0^1{\rm d}y\,
x(1-x)u\,\frac{N_1 N_2}{D^4}
\end{equation}
and
\begin{eqnarray}
N_1 \!&=&\! u(1-x)+z_qx(1-x)(1-u)-2ux[z_1y+z_2(1-y)]\,, \nonumber \\
N_2 \!&=&\! (1-x)^2(1-u)^2+u^2-\frac{1}{9}\,x^2(1-u)^2\,, \nonumber \\
D   \!&=&\! u(1-x)+z_qx(1-x)(1-u)+ux[z_1y+z_2(1-y)]\; .
\end{eqnarray}
Figure~\ref{fig:zqh}  shows  the   functional  dependence  of  $z_q  |
H(z_1,z_2,z_q)|$ on $z_q$ for several values of $z_s\equiv z_1 = z_2$.
Our   results    are   at    variance   with   those    presented   
in~\cite{Dai:1990xh}.  For instance, we find that when $z_s \leq 4$,
$H(z_1,z_2,z_q)$   becomes   negative   beyond   certain   values   of
$z_q$~\footnote{We thank Oleg Lebedev and Pran Nath for
useful comparisons and comments regarding the loop function $H(z_1,z_2,z_q)$.}.

The loop function $H(z_1,z_2,z_q)$ simplifies considerably in specific
regions of  the parameter space.  Specifically, if $z_1=z_2=  z_s$ and
the limit $z_q\to 0$ is taken, the following function may be defined:
\begin{eqnarray}
\widehat{H}_0(z_s) \!&\equiv&\! \lim_{z_q\to 0} z_q H(z_s,z_s,z_q)
\nonumber \\
&=&\frac{1}{18(1-z_s)^4}\left[
2(1-z_s)(1+11z_s)-(1-16z_s-9z_s^2)\,\ln z_s\right]\;,
\end{eqnarray}
with  $\widehat{H}_0(1)=5/108$.   Hence,   for  $z_q  \lsim  0.1$  and
$(z_2-z_1)^2/(z_2+z_1)^2\ll  1$,  the  loop function  $H(z_1,z_2,z_q)$
may be approximated as follows:
\begin{equation}
H(z_1,z_2,z_q)\ \approx\ 
\frac{1}{z_q}\left[\widehat{H}_0\left(\frac{z_1+z_2}{2}\right)\: 
+\: \frac{(z_2-z_1)^2}{(z_1+z_2)^2}\,\widehat{H}_1
\left(\frac{z_1+z_2}{2}\right) \right]\; , 
\label{eq:zqh}
\end{equation}
where
\begin{eqnarray}
\widehat{H}_1(z_s) \! &\equiv &\!
\frac{1}{108(1-z_s)^6}
\left[ (1-z_s)(1+7z_s+295z_s^2+177z_s^3) 
\right. \nonumber \\ && \hspace{2.6cm} \left. 
+6z_s^2(21+50z_s+9z_s^2)\ln(z_s)\right]\,,
\end{eqnarray}
with $\widehat{H}_1(1)=11/1080$.

The second  term in~(\ref{dG}), $(d^{\,G})^{H}$, is  the neutral Higgs
contribution  \cite{Weinberg:1989dx,Dicus:1989va}, which  may  be cast
into the form:
\begin{equation}
(d^{\,G})^{H}\ =\ \frac{4\sqrt{2}\, G_F\, g_s^3}{(4\pi)^4}
\sum_{q=t,b} \left[\sum_i g^S_{H_i\bar{q}q}\,
  g^P_{H_i\bar{q}q}\,h(z_{iq})\right]\,, 
\end{equation}
where $z_{iq} \equiv M_{H_i}^2/m_q^2$ and
\begin{equation}
h(z)\ =\ \frac{1}{4}\int_0^1{\rm d}x\int_0^1{\rm d}u\,
\frac{u^3 x^3 (1-x)}{[x(1-ux)+z(1-u)(1-x)]^2}\; ,
\end{equation}
with $h(0)=1/16$. We note that for the loop function $h(z)$ we 
follow~\cite{Dicus:1989va}, whose result is smaller by  a factor 2 than the one
given in~\cite{Weinberg:1989dx}.

\subsection{CP-Odd Four-Fermion Interactions}

%---------------  Four-fermion operator Diagrams (START)  -----------------%
\begin{figure}[t]
\begin{center}
\begin{picture}(250,120)(0,0)
%-
\SetWidth{1.0}
\Line(50,0)(250,0) \Text(100,10)[]{$f$}\Text(200,10)[]{$f$}
\Line(50,120)(250,120) \Text(100,110)[]{$f^\prime$}\Text(200,110)[]{$f^\prime$}
\DashLine(150,0)(150,120){5} \Text(170,60)[]{$H_i$}
\Vertex(150,0){4}\Vertex(150,120){4}
%-
\end{picture}
\end{center}
\smallskip
\noindent
\caption{\it  Feynman diagrams for CP-odd four-fermion operators.  The $H_i$
line  denotes all  three neutral  Higgs bosons  including CP-violating
Higgs-boson  mixing.  Heavy  dots  indicate resummation  of  threshold
corrections to the corresponding Yukawa couplings.}
\label{fig:four-fermion}
\end{figure}
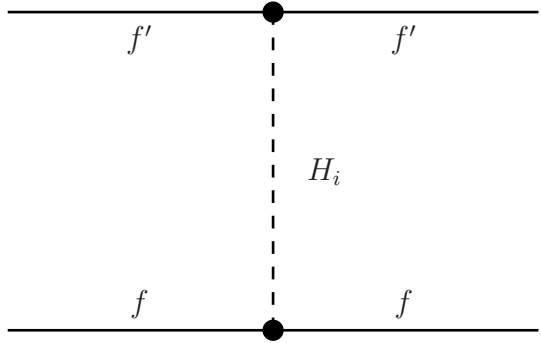
%---------------------  Four-fermion operator Diagrams (END)

CP-odd four-fermion interactions play  a significant role in the EDMs.
These interactions may be generically described by the Lagrangian
\begin{equation}
  \label{L4f}
{\cal     L}_{\rm      4f}\ =\  \sum_{f,f'}     C_{ff'}     (\bar{f}f)
(\bar{f'}i\gamma_5f')\; .
\end{equation}
The  CP-odd  four-fermion operators  in~(\ref{L4f})  are generated  by
CP-violating  neutral Higgs-boson  mixing  in the  $t$-channel and  by
CP-violating       Yukawa       threshold       corrections,       see
Fig.~\ref{fig:four-fermion}.   The   combined  effect  of   these  two
contributions gives rise to the CP-odd coefficients
\begin{equation}
  \label{eq:cff} 
(C_{ff'})^H\ =\ g_f\, g_{f'}\,\sum_i
\frac{g^S_{H_i\bar{f}f}\,g^P_{H_i\bar{f'}f'}}{M_{H_i}^2}\; ,
\end{equation}
where   $g_f=m_f/v$   with   $v=2M_W/g$   for   $f=l,d,u$.    Possible
sub-dominant contributions  from box diagrams~\cite{Demir:2003js} have
been neglected.

\subsection{Barr--Zee Graphs}

%---------------------  Barr-Zee Diagrams (START)  -------------------%
\begin{figure}[t]
\begin{center}
\begin{picture}(400,300)(0,70)
\SetWidth{1.0}
%-
\ArrowLine(10,230)(45,260)\Text(30,235)[lb]{$f$}
\DashLine(45,260)(68,285){5}\Vertex(67,284.5){3}
%\Vertex(56,272){3} \Text(53,275)[rb]{$a, \phi_{1,2}$}
\Text(53,275)[rb]{$H_i$}
\Photon(92,285)(115,260){3}{3}\Text(105,280)[l]{$\gamma,(g)$}
\ArrowLine(45,260)(115,260)\Text(80,251)[l]{$f$}
\ArrowLine(115,260)(150,230)\Text(118,235)[lb]{$f$}
\Photon(80,320)(80,350){3}{3}\Text(86,340)[l]{$\gamma,(g)$}
\DashArrowArc(80,300)(20,0,360){3}
%\Text(55,300)[r]{$\tilde{t},\tilde{t}^*,\tilde{b},\tilde{b}^*$}
\Text(55,310)[r]{$\tilde{\tau},\tilde{t},\tilde{b}$}

%\Text(75,220)[]{\bf (a)}

\ArrowLine(210,230)(245,260)\Text(230,235)[lb]{$f$}
\DashLine(245,260)(268,285){5}\Vertex(267,284.5){3}
%\Vertex(256,272){3} \Text(253,275)[rb]{$a, \phi_{1,2}$}
\Text(253,275)[rb]{$H_i$}
\Photon(301,300)(315,260){3}{4}\Text(315,280)[l]{$\gamma,(g)$}
\Photon(301,300)(315,330){3}{3}\Text(315,335)[r]{$\gamma,(g)$}
\ArrowLine(245,260)(315,260)\Text(280,251)[l]{$f$}
\ArrowLine(315,260)(350,230)\Text(318,235)[lb]{$f$}
\DashArrowArc(280,300)(20,0,360){3}
%\Text(255,300)[r]{$\tilde{t},\tilde{t}^*,\tilde{b},\tilde{b}^*$}
\Text(255,310)[r]{$\tilde{\tau},\tilde{t},\tilde{b}$}

%\Text(275,220)[]{\bf (b)}

\ArrowLine(10,70)(45,100)\Text(30,75)[lb]{$f$}
\DashLine(45,100)(68,125){5}\Vertex(67,124.5){3}
%\Vertex(56,112){3} \Text(53,115)[rb]{$a, \phi_{1,2}$}
\Text(53,115)[rb]{$H_i$}
\Photon(92,125)(115,100){3}{3}\Text(105,120)[l]{$\gamma,(g)$}
\ArrowLine(45,100)(115,100)\Text(80,91)[l]{$f$}
\ArrowLine(115,100)(150,70)\Text(118,75)[lb]{$f$}
\Photon(80,160)(80,190){3}{3}\Text(86,180)[l]{$\gamma,(g)$}
\ArrowArc(80,140)(20,0,360)
\Text(55,150)[r]{$\tau,t,b$}

%\Text(75,60)[]{\bf (c)}

\ArrowLine(210,70)(245,100)\Text(230,75)[lb]{$f$}
\DashLine(245,100)(268,125){5}
%\Vertex(256,112){3} \Text(253,115)[rb]{$a, \phi_{1,2}$}
\Text(253,115)[rb]{$H_i$}
\Photon(292,125)(315,100){3}{3}\Text(305,120)[l]{$\gamma$}
\ArrowLine(245,100)(315,100)\Text(280,91)[l]{$f$}
\ArrowLine(315,100)(350,70)\Text(318,75)[lb]{$f$}
\Photon(280,160)(280,190){3}{3}\Text(286,180)[l]{$\gamma$}
\ArrowArc(280,140)(20,0,360)
\Text(260,150)[r]{$\chi^\pm$}

%\Text(275,60)[]{\bf (d)}
%-
\end{picture}
\end{center}
\smallskip
\noindent
\caption{\it  Barr-Zee  diagrams: the  $H_i$  lines  denote all  three
neutral Higgs bosons,  including CP-violating Higgs-boson mixing, and heavy
dots   indicate   resummation   of   threshold  corrections   to   the
corresponding Yukawa couplings.}
\label{fig:Barr-Zee}
\end{figure}
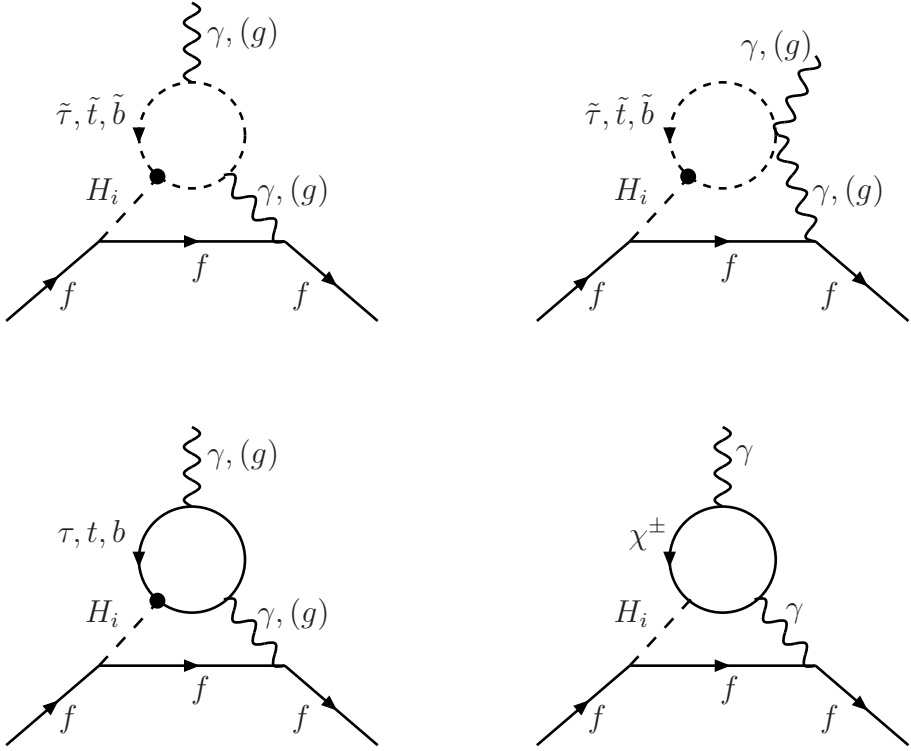
%---------------------  Bar-Zee Diagrams (END) -----------------------%

\begin{figure}

\begin{center}
\begin{picture}(450,150)(0,0)
\SetWidth{0.8}

\Gluon(10,50)(50,50){3}{5}\ArrowArc(70,50)(20,0,90)\ArrowArc(70,50)(20,90,180)
\ArrowArc(70,50)(20,180,360)\Gluon(90,50)(130,50){3}{5}
\DashLine(70,70)(70,120){4}
\ArrowLine(10,120)(70,120)\ArrowLine(70,120)(130,120)
%\GCirc(70,95){5}{0.7}
\Vertex(70,70){3}
\Text(75,95)[l]{$H_i$}
\Text(20,125)[b]{$e^-$}\Text(120,125)[b]{$e^-$}
\Text(20,40)[t]{$g$}\Text(120,40)[t]{$g$}
\Text(70,25)[t]{$t,b$}

%\Text(70,-10)[]{\bf (a)}

\Gluon(200,50)(240,50){3}{5}\DashArrowArc(260,50)(20,0,90){3}
\DashArrowArc(260,50)(20,90,180){3}
\DashArrowArc(260,50)(20,180,360){3}\Gluon(280,50)(320,50){3}{5}
\DashLine(260,70)(260,120){4}
\ArrowLine(200,120)(260,120)\ArrowLine(260,120)(320,120)
%\GCirc(260,95){5}{0.7}
\Vertex(260,70){3}
\Text(265,95)[l]{$H_i$}
\Text(210,125)[b]{$e^-$}\Text(310,125)[b]{$e^-$}
\Text(210,40)[t]{$g$}\Text(310,40)[t]{$g$}
\Text(260,25)[t]{$\tilde{t},\tilde{b}$}

%\Text(340,70)[]{$+$}

\Gluon(360,30)(390,30){3}{3}\DashArrowArc(390,50)(20,-90,90){3}
\DashArrowArc(390,50)(20,90,270){3}\Gluon(390,30)(420,30){3}{3}
\DashLine(390,70)(390,120){4}
\ArrowLine(360,120)(390,120)\ArrowLine(390,120)(420,120)
%\GCirc(390,95){5}{0.7}
\Vertex(390,70){3}
\Text(395,95)[l]{$H_i$}
\Text(365,125)[b]{$e^-$}\Text(415,125)[b]{$e^-$}
\Text(365,25)[t]{$g$}\Text(415,25)[t]{$g$}
\Text(420,50)[l]{$\tilde{t},\tilde{b}$}

%\Text(345,-10)[]{\bf (b)}

\end{picture}
\end{center}
\vspace{0.cm}
\caption{\em Feynman graphs contributing to $(C_S)^g$: the $H_i$
lines denote all three neutral Higgs bosons, including CP-violating
Higgs-boson mixing, and heavy dots indicate resummation of threshold
corrections to the corresponding Yukawa couplings.}
\label{fig:CSG}
\end{figure}
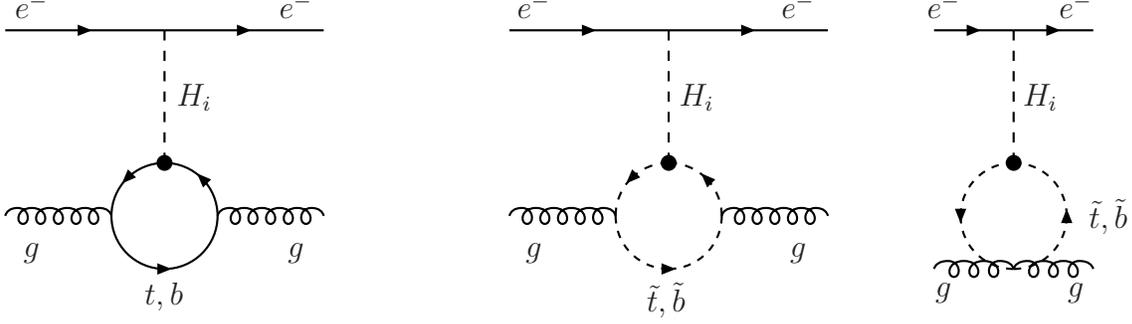

Finally,  there  are   additional  Higgs-boson  quantum  effects  that
contribute significantly  to the EDMs beyond the  one-loop level.  For
the Thallium EDM, these are  the two-loop Barr--Zee graphs, denoted as
$(d^E_e)^H$,  and  the   CP-odd  electron-nucleon  interaction  ${\cal
L}_{C_S}=C_S\,\bar{e}i\gamma_5                                      e\,
\bar{N}N$~\cite{Chang:1998uc,Pilaftsis:2002fe},  which  is induced  by
CP-violating  gluon-gluon-Higgs  couplings,  $(C_S)^g$.  As  shown  in
Fig.~\ref{fig:Barr-Zee},  the electron EDM  $(d^E_e)^H$ is  induced by
CP-violating phases of third-generation  fermions and sfermions and of
charginos. More explicitly, $(d^E_e)^H$ is given by
\begin{eqnarray}
\left(\frac{d_e^E}{e}\right)^H \!&=&\! \sum_{q=t,b}\Bigg\{
\frac{3\alpha_{\rm
em}\,Q_q^2\,m_e}{32\pi^3}\sum_{i=1}^3\frac{g^P_{H_ie^+e^-}}{M_{H_i}^2}
\sum_{j=1,2} g_{H_i\tilde{q}_j^*\tilde{q}_j}\,F(\tau_{\tilde{q}_ji})
\nonumber \\
&&\! +\frac{3\alpha_{\rm em}^2\,Q_q^2\,m_e}{8\pi^2s_W^2M_W^2}
\sum_{i=1}^3\left[
g^P_{H_ie^+e^-} g^S_{H_i\bar{q}q}\,f(\tau_{qi})
+g^S_{H_ie^+e^-} g^P_{H_i\bar{q}q}\,g(\tau_{qi})
\right]\Bigg\}  \nonumber \\
&&\! +\
\frac{\alpha_{\rm
    em}\,m_e}{32\pi^3}\sum_{i=1}^3\frac{g^P_{H_ie^+e^-}}{M_{H_i}^2} 
\sum_{j=1,2} g_{H_i\tilde{\tau}_j^*\tilde{\tau}_j}\,F(\tau_{\tilde{\tau}_ji})
\nonumber \\
&&\! +\
\frac{\alpha_{\rm em}^2\,m_e}{8\pi^2s_W^2M_W^2}
\sum_{i=1}^3\left[
g^P_{H_ie^+e^-} g^S_{H_i\tau^+\tau^-}\,f(\tau_{\tau i})
+g^S_{H_ie^+e^-} g^P_{H_i\tau^+\tau^-}\,g(\tau_{\tau i})
\right] \nonumber \\
&&\! +\ \frac{\alpha_{\rm em}^2\,m_e}{4\sqrt{2}\pi^2s_W^2M_W} \nonumber \\
&&\hspace{0.5cm}\times
\sum_{i=1}^3\sum_{j=1,2}\frac{1}{m_{\chi^\pm_j}}\left[
g^P_{H_ie^+e^-} g^S_{H_i\chi^+_j\chi^-_j}\,f(\tau_{\chi_j^\pm i})
+g^S_{H_ie^+e^-} g^P_{H_i\chi^+_j\chi^-_j}\,g(\tau_{\chi_j^\pm i})
\right]\,,\nonumber \\
\end{eqnarray}
with $\tau_{xi} =  m_x^2/M_{H_i}^2$.   The
Higgs-mediated  two-loop  quark   EDMs  $(d^E_{q=u,d,s})^H$  are  also
calculated similarly.   In the  above, the two-loop  functions $F(\tau
)$, $f(\tau )$, and $g(\tau )$ are given by
\begin{eqnarray}
F(\tau) &=& \int_0^{1} dx\ \frac{x(1-x)}{\tau\: -\: x(1-x)}\
\ln \bigg[\,\frac{x(1-x)}{\tau}\,\bigg]\, , \nonumber \\
f(\tau) &=& \frac{\tau}{2}\, \int_0^{1} dx\ \frac{1\: -\: 2x(1-x)}{x(1-x)\: -\: \tau}\
\ln \bigg[\,\frac{x(1-x)}{\tau}\,\bigg]\, , \nonumber \\
g(\tau) &=& \frac{\tau}{2}\, \int_0^{1} dx\ \frac{1}{x(1-x)\: -\: \tau}\
\ln \bigg[\,\frac{x(1-x)}{\tau}\,\bigg]\,.
\end{eqnarray}
There    are    subleading    two-loop    contributions    which    we
neglect~\cite{TwoLoopEdms,THW}.     Instead,    we     consider    the
gluon-gluon-Higgs contribution to  $C_S$, see Fig.~\ref{fig:CSG}. This
is given by
\begin{equation}
(C_S)^g = (0.1\,{\rm GeV})\,
%\frac{m_e\pi\alpha_{\rm em}}{s_W^2M_W^2}
\frac{m_e}{v^2}
\sum_{i=1}^3 \frac{g^S_{H_igg}g^P_{H_i\bar{e}e}}{M_{H_i}^2}\,,
\label{eq:csg}
\end{equation}
where                                                          $\langle
N|\frac{\alpha_s}{8\pi}G^{a,\mu\nu}G^a_{\mu\nu}|N\rangle    =-(0.1~{\rm
GeV})\bar{N}N$   is   used,   and   we  use   the   tree-level   value
$g^P_{H_i\bar{e}e}=-\tan\beta\, O_{ai}$. In addition, $g^S_{H_igg}$ is
the scalar form factor $S_i^g$ in the heavy (s)quark limit:
\begin{equation}
g^S_{H_igg}=\sum_{q=t,b}\left\{\frac{2\,x_q}{3}g_{H_i\bar{q}q}^S-\frac{v^2}{12}
\sum_{j=1,2} \frac{g_{H_i\tilde{q}_j^*\tilde{q}_j}}{m_{\tilde{q}_j}^2}
\right\}\,,
\label{eq:gsHgg}
\end{equation}
where $x_t=1$ and $x_b=(1-0.25\kappa)$ with $\kappa\equiv\langle N |
m_s \bar{s} s | N \rangle/220~{\rm MeV} \simeq 0.50\pm0.25$
[cf.~(\ref{eq:cs4f})].

%---------------------  C_S^g Diagrams (START)  -------------------------%
%---------------------  C_S^g Diagrams (END)  ---------------------------%

Apart from EDMs, the two-loop  Barr-Zee graphs also generate CEDMs
for the $u$  and $d$ quarks. These additional  contributions are given
by
\begin{eqnarray}
\left(d_{q_l}^C\right)^H \!&=&\! - \sum_{q=t,b}\Bigg\{
\frac{g_s\,\alpha_s\,m_{q_l}}{64\pi^3}
\sum_{i=1}^3\frac{g^P_{H_i\bar{q}_lq_l}}{M_{H_i}^2}
\sum_{j=1,2} g_{H_i\tilde{q}_j^*\tilde{q}_j}\,F(\tau_{\tilde{q}_ji})
\nonumber \\
&&~~ +\frac{g_s\,\alpha_s\,\alpha_{\rm em}\,m_{q_l}}{16\pi^2s_W^2M_W^2}
\sum_{i=1}^3\left[
g^P_{H_i\bar{q}_lq_l} g^S_{H_i\bar{q}q}\,f(\tau_{qi})
+g^S_{H_i\bar{q}_lq_l} g^P_{H_i\bar{q}q}\,g(\tau_{qi})
\right]\Bigg\}\,,  %\nonumber \\
\end{eqnarray}
with $q_l=u,d$.

\setcounter{equation}{0}
\section{Thallium, Neutron, Mercury and Deuteron EDMs}

In  this Section we  present analytic  expressions for  the Thallium,
neutron,  Mercury  and  deuteron  EDMs  in terms  of  the  constituent
particle (C)EDMs  and the coefficients of  the dimension-six
Weinberg operator and the four-fermion operators.

\subsection{Thallium}

We first consider the  atomic EDM $d_{\rm Tl}$ of $^{205}$Tl.  This
receives   contributions   mainly   from   two
terms~\cite{KL,Pospelov:2005pr}:
\begin{equation}
d_{\rm Tl}\,[e\,{\rm cm}]\ =\ -585\cdot d_e^E\,[e\,{\rm cm}]\:
-\: 8.5\times 10^{-19}\,[e\,{\rm cm}]\cdot (C_S\,{\rm TeV}^2)\: +\ \cdots\,,
\end{equation}
where $d_e^E$ is the electron EDM, which is given by the sum
\begin{equation}
d_e^E = (d_e^E)^{\tilde{\chi}^\pm}
    + (d_e^E)^{\tilde{\chi}^0} 
    + (d_e^E)^{H} \; .
\end{equation}
The coefficient $C_S$ is calculated as
\begin{equation}
C_S\ =\ (C_S)^{4f}\: +\: (C_S)^g\; ,
\end{equation}
where the (down-type) quark contribution is given by \cite{Demir:2003js}
\begin{equation}
(C_S)^{4f}\ =\ C_{de}\frac{29\,{\rm MeV}}{m_d}\:
+\: C_{se}\frac{\kappa\times 220\,{\rm MeV}}{m_s}\; .
%+C_{be}\frac{66\,{\rm MeV} (1-0.25\kappa)}{m_b}\,.
\label{eq:cs4f}
\end{equation}
We  neglect  the  $u$-  and  $c$-quark contributions  and  absorb  the
$b$-quark  contribution  into  $(C_S)^g$  together with  the  top  and
heavy-squark contributions [cf.~(\ref{eq:gsHgg})].  We recall that the
ratio      $C_{qe}/m_q       \propto      m_e/v^2$      is      scale-
invariant~\cite{Demir:2003js}.     More    details   concerning    the
calculations of these coefficients are given in Appendix~\ref{sec:cs}.

\subsection{Neutron}

We  now  turn to  the  calculation of  the  neutron  EDM, $d_n$.   Its
magnitude is somewhat  uncertain, because of non-perturbative dynamics
at the  hadron level. We  consider three different hadronic  
approaches  for  computing $d_n$:  (i)~the  Chiral Quark  Model~(CQM),
(ii)~the Parton Quark Model~(PQM) and (iii)~the QCD sum-rule
technique.

The  CQM  is  a  non-relativistic  model, where  the  quark  EDMs  are
estimated via naive dimensional analysis (NDA)~\cite{NDA}, and
the neutron EDM is given by
\begin{eqnarray}
  \label{dnCQM}
d_n \!&=&\! \frac{4}{3}\,d^{\rm NDA}_d\: -\: 
                           \frac{1}{3}\,d^{\rm NDA}_u\,, \nonumber \\
d^{\rm NDA}_{q=u,d} \!&=&\! \eta^E\,d_q^E\: +\: \eta^C\,\frac{e}{4\pi}\,d_q^C
\: +\: \eta^G\,\frac{e\Lambda}{4\pi}\,d^{\,G}\,, \nonumber \\
d_{q=u,d}^{E\,(C)} \!&=&\! (d_q^{E\,(C)})^{\tilde{\chi}^\pm}\:
+\: (d_q^{E\,(C)})^{\tilde{\chi}^0}\:
+\: (d_q^{E\,(C)})^{\,\tilde{g}}\; 
+\: (d_q^{E\,(C)})^{\,H}\; ,
\end{eqnarray}
where  the chiral  symmetry breaking  scale $\Lambda  \simeq 1.19~{\rm
GeV}$ and the $\eta^{E,C,G}$  are QCD correction factors that describe
the renormalization-group (RG) evolution of $d^{E,C}_q$ and $d^G$ from
the electroweak  (EW) scale, e.g.,  the $Z$-boson mass $M_Z$,  down to
the  hadronic   scale.   These   QCD  correction  factors   are  given
in~\cite{Ibrahim:1997gj} as
\begin{equation}
\eta^E \simeq 1.53\,, \ \ 
\eta^C \simeq \eta^G \simeq 3.4\; .
\end{equation}
We note that the  EDM operators $d^{E,C}_q$ and $d^G$ in~(\ref{dnCQM})
are computed at the EW scale.

Another approach  to computing  the neutron EDM  is based on  the PQM,
which   uses  low-energy   data  related   to   the  constituent-quark
contributions  to  the  proton  spin combined  with  isospin  symmetry
\cite{Ellis:1996dg}.  In the PQM, the neutron EDM is given by
\begin{eqnarray}
  \label{dnPQM}
d_n  \!&=&\! \eta^E\, (\Delta^{\rm PQM}_d\,d_d^E\: +\:
\Delta^{\rm PQM}_u\,d_u^E+\Delta^{\rm PQM}_s\,d_s^E)\;,
\end{eqnarray}
where we  consider the following  particular values, to  contrast with
the  PQM predictions~\footnote{We note  that isospin  symmetry between
the    neutron    $n$    and    the   proton    $p$    implies    that
$\Delta_d=(\Delta_u)_p=4/3$,              $\Delta_u=(\Delta_d)_p=-1/3$.
Furthermore,  in the  relativistic Naive  Quark Model  (NQM),  one has
$\Delta_s=(\Delta_s)_p=0$.  }
\begin{equation}
\Delta^{\rm PQM}_d =0.746\,,\qquad
\Delta^{\rm PQM}_u =-0.508\,,\qquad
\Delta^{\rm PQM}_s =-0.226\;.
\end{equation}
As  in the  CQM, $d^{E}_q\,$'s in~(\ref{dnPQM})
are evaluated at the EW scale.

The third approach  to computing the neutron EDM  employs QCD sum rule
techniques~\cite{qcdsumrule1,qcdsumrule2,Demir:2002gg,Demir:2003js,
Olive:2005ru}.  With the  aid of these techniques, the  neutron EDM is
determined by
\begin{eqnarray}
  \label{dnQCD}
d_n \!&=&\! d_n(d_q^E\,,d_q^C)\: +\: d_n(d^{\,G})\: +\: d_n(C_{bd})\:
+\: \cdots\,, \nonumber \\
d_n(d_q^E\,,d_q^C) \!&=&\! (1.4\pm 0.6)\,(d_d^E-0.25\,d_u^E)\: 
+\: (1.1\pm 0.5)\,e\,(d_d^C+0.5\,d_u^C)/g_s\,,
\nonumber \\ 
d_n(d^{\,G}) \!&\sim&\! \pm\, e\, (20\pm 10)~{\rm MeV} \,d^{\,G}\,,
\nonumber \\ 
d_n (C_{bd}) \!&\sim &\! \pm\, e\, 2.6\times 10^{-3}~{\rm GeV}^2\,
\left[\frac{C_{bd}}{m_b}\: +\: 0.75\frac{C_{db}}{m_b}\right]\; ,
\end{eqnarray}
where  $d_q^E$ and $d_q^C$  should be  evaluated at  the EW  scale and
$d^{\,G}$  at  the 1  GeV  scale~\footnote{Here  we  make use  of  the
relation: $d^G\big|_{1~{\rm  GeV}} \simeq (\eta^G/0.4)\, d^G\big|_{\rm
EW}  \simeq 8.5\,  d^G\big|_{\rm EW}$~\cite{Demir:2002gg}.}.   We note
that the contribution  of $d^G$ to $d_n$ is a  factor $\sim 2$ smaller
than in the  CQM. We calculate the coefficients  $C_{bd}$ and $C_{db}$
by  means of~(\ref{eq:cff})  and evaluate  the coefficients  $g_b$ and
$g_d$,  at the  energy scales  $m_b$  and $1$  GeV, respectively.  For
definiteness, in our numerical estimates we assume that both $d^{\,G}$
and $C_{bd}$ contribute positively to $d_n$.

The comparisons between the results obtained in these three approaches
indicates  the significance of  the non-perturbative  uncertainties in
calculating   $d_n$.   Related  uncertainties   appear  also   in  the
calculations  of EDMs  of  nuclei. In  Section~\ref{EDMs}, we  present
numerical estimates of  the neutron EDM based on the  CQM, the PQM and
QCD sum-rules.

\subsection{Mercury}

Using QCD sum rules~\cite{Demir:2003js,Olive:2005ru} to calculate
the Mercury EDM, one finds that 
\begin{eqnarray}
  \label{eq:mercuryedm}
d_{\rm Hg} \!&=&\! 7\times 10^{-3}\,e\,(d_u^C-d_d^C)/g_s\ +\ 
10^{-2}\,d_e^E \nonumber \\
\!&&\! -\ 1.4\times 10^{-5}\,e\,{\rm GeV}^2\,
\left[\frac{0.5C_{dd}}{m_d}+3.3\kappa\frac{C_{sd}}{m_s}
+(1-0.25\kappa)\frac{C_{bd}}{m_b} \right] \nonumber \\
\!&&\! +\ (3.5\times 10^{-3}~{\rm GeV})\,e\,C_S \nonumber \\
\!&&\! +\ (4\times 10^{-4}~{\rm GeV})\,e\,\left[C_P+
\left(\frac{Z-N}{A}\right)_{\rm Hg}\,C^\prime_P\right]\,.
\end{eqnarray}
The parameters $C_P$ and $C^\prime_P$  are the couplings of the CP-odd
singlet and  triplet electron-nucleon interactions,  respectively, and
are described by the interaction Lagrangian
\begin{equation} 
{\cal       L}_{C_P}\ =\ C_P\,\bar{e}e\,\bar{N}i\gamma_5 N\: +\: 
C^\prime_P\,\bar{e}e\,\bar{N}i\gamma_5 \tau_3  N\; .
\end{equation}  
Making use of the SU(2) isospin  symmetry of the nucleon $N = (p, n)$,
one may evaluate the  triplet contribution $C'_P$, which is suppressed
by a  factor $\left[(Z-N)/A\right]_{\rm Hg}=-0.2$ with  respect to the
singlet  one  $C_P$.   In  detail, the  contribution  of  four-fermion
interactions to $C_P$ and $C^\prime_P$ is given by
\begin{eqnarray}
C_P \!&=&\! (C_P)^{4f}\ \simeq\ 
-\,375~{\rm MeV}\,\sum_{q=c,s,t,b} \frac{C_{eq}}{m_q}\,,\nonumber \\
C^\prime_P \!&=&\! (C^\prime_P)^{4f} \
\simeq\ -\,806~{\rm MeV}\,\frac{C_{ed}}{m_d}\,
-\,181~{\rm MeV}\,\sum_{q=c,s,t,b} \frac{C_{eq}}{m_q}\,,
\end{eqnarray}
where the  $u$-quark contribution has  been neglected. For  a detailed
discussion, see Appendix~\ref{sec:cp}.

On  the  basis  of above  results,  we  improve upon  an  earlier
calculation of the Mercury EDM~\cite{Pospelov:2005pr}\footnote{We also
correct the errors in the contributions of $\bar{g}^{(1)}_{\pi NN}$ to
$d_{\rm Hg}$  and of four-fermion  interactions to $\bar{g}^{(1)}_{\pi
NN}$~\cite{private:PospelovRitz}.}.  More  explicitly,  including  the
CP-odd triplet electron-nucleon interaction,  the Mercury EDM is given
by
\begin{eqnarray}
d_{\rm Hg} \!&=&\!
(1.8\times 10^{-3}~{\rm GeV}^{-1})\,e\,\bar{g}^{(1)}_{\pi NN}
+10^{-2} d_e^E %\nonumber \\ &&
+(3.5\times 10^{-3} {\rm GeV})\,e\,C_S \nonumber \\
\!&&\!+\ (4\times 10^{-4}~{\rm GeV})\,e\,\left[C_P+
\left(\frac{Z-N}{A}\right)_{\rm Hg}\,C^\prime_P\right]\,,\nonumber \\
\bar{g}^{(1)}_{\pi NN} \!&=&\!
2^{+4}_{-1}\times 10^{-12}\,\frac{(d_u^C-d_d^C)/g_s}{10^{-26}{\rm cm}}\,
\frac{|\langle \bar{q} q\rangle |}{(225\,{\rm MeV})^3}\,, \nonumber \\
\bar{g}^{(1)}_{\pi NN} \!&\sim&\!
-8\times 10^{-3} {\rm GeV}^3\,
\left[\frac{0.5C_{dd}}{m_d}+3.3\kappa\frac{C_{sd}}{m_s}
+(1-0.25\kappa)\frac{C_{bd}}{m_b} \right]\,,
\end{eqnarray}
where    ${\cal   L}_{\pi    NN}   \supset    \bar{g}^{(1)}_{\pi   NN}
\bar{N}N\pi^0$.  We note   that  the  factors   $1.8\times  10^{-3}~{\rm
GeV}^{-1}$ and  $-8\times 10^{-3}~{\rm GeV}^3$ are  known only up to  50 \%
accuracy~\cite{private:PospelovRitz}.

\subsection{Deuteron}

Finally, the  deuteron EDM may be  calculated by using  QCD sum rules.
We  include   the   contributions   from   the   pion-nucleon-nucleon
isospin-triplet              coupling              $\bar{g}^{(1)}_{\pi
NN}$~\cite{Khriplovich:1999qr},  the  constituent  proton and  neutron
EDMs~\cite{qcdsumrule2},  and   the  dimension-six  Weinberg  operator
\cite{Weinberg:1989dx,Demir:2002gg}.  The   deuteron  EDM  $d_D$ is
then found to be~\cite{Lebedev:2004va}
\begin{eqnarray}
d_{D} \!&=&\! d_D^{\,\pi NN}+d_D(d_n,d_p)+d_D(d^G) \,,\nonumber \\
d_D^{\,\pi NN} \!&= &\!
-\ \frac{e\, g_{\pi NN}\, 
\bar{g}^{(1)}_{\pi NN}}{12\pi m_\pi}\, \frac{1+\xi}{(1+2\xi)^2}\
\simeq\ -\, (1.3\pm 0.3)\,e\, \bar{g}^{(1)}_{\pi NN}\,{\rm
  GeV}^{-1}\,, \nonumber \\ 
d_D(d_n,d_p) \!&\simeq&\! (0.5\pm 0.3)(d^E_u+d^E_d)
-(0.6\pm 0.3)\,e\,[(d^C_u-d^C_d)/g_s+0.3(d^C_u+d^C_d)/g_s] \,,\nonumber \\
d_D(d^G)\, &\simeq & d_n(d^G)+d_p(d^G)\ \sim\
\pm\, e\, (20\pm 10)~{\rm MeV} \,d^{\,G}\; ,
\end{eqnarray}
where $g_{\pi NN}\simeq 13.45$ and $\xi=\sqrt{m_p\epsilon}/m_\pi$, with
$\epsilon =2.23$  MeV being  the deuteron binding  energy.  Collecting
all the above  intermediate results, we may write  $d_D$ in the more
compact form:
\begin{eqnarray}
d_{D} \!&\simeq &\! -\left[5^{+11}_{-3} + (0.6\pm 0.3)
  \right]\,e\,(d^C_u-d^C_d)/g_s 
\nonumber \\ 
\!&&\! -(0.2\pm 0.1)\,e\,(d^C_u+d^C_d)/g_s\:
+\: (0.5 \pm 0.3) (d^E_u+d^E_d) \nonumber \\
\!&&\! +(1\pm 0.2)\times 10^{-2}\,e\,{\rm GeV}^2\,
\left[\frac{0.5C_{dd}}{m_d}+3.3\kappa\frac{C_{sd}}{m_s}
+(1-0.25\kappa)\frac{C_{bd}}{m_b} \right] \nonumber \\
&& \pm\ e\, (20\pm 10)~{\rm MeV} \,d^{\,G}\; .
\end{eqnarray}
In  the above,  $d^{\,G}$  is evaluated  at the  $1$ GeV scale,
and the coupling coefficients $g_{d,s,b}$ that occur in $C_{dd,sd,bd}$
are computed at energies 1~GeV, 1~GeV and $m_b$, respectively.
All other  EDM  operators  are calculated  at  the EW  scale.
We observe that  the leading  dependence of $d_D$  on $d^C_{u,d}$  is the
same as in $d_{\rm Hg}$.  In the numerical estimates given in the next
Section, we assume that $d^{\,G}$ contributes positively to $d_D$.

\setcounter{equation}{0}
\section{EDM Constraints}\label{EDMs}

In this Section, we present illustrative constraints on key soft
SUSY-breaking parameters and CP phases in the trimixing, the CPX and
the MCPMFV scenarios.
We use the following current experimental limits on the Thallium~\cite{Regan:2002ta},
neutron~\cite{Baker:2006ts}, and Mercury~\cite{Romalis:2000mg} EDMs:
\begin{eqnarray}
|d_{\rm Tl}| &<& 9\times 10^{-25}~e\,{\rm cm}\,, \nonumber \\
|d_{\rm Hg}| &<& 2\times 10^{-28}~e\,{\rm cm}\,, \nonumber \\
|d_{\rm n}|  &<& 3\times 10^{-26}~e\,{\rm cm}\,.
\end{eqnarray}
On the other hand, the projected sensitivity to the deuteron EDM
is~\cite{Semertzidis:2003iq}
\begin{equation}
\left|d_D\right| < (1-3)\times 10^{-27}~e\,{\rm cm}\,. 
\end{equation}
For our numerical study, we take $3 \times 10^{-27}~e\,{\rm cm}$ as a
representative expected value. However, we note that much better
statistical precision at the level of $10^{-29}~e\,{\rm cm}$
may be possible in principle~\cite{OMS}.

\subsection{Trimixing Scenario}

The trimixing scenario is characterized by large $\tan\beta$ and a
light charged Higgs boson, resulting in a strongly-mixed system of
three neutral Higgs bosons with mass differences smaller than the
decay widths~\cite{Ellis:2004fs}:
\begin{eqnarray}
  \label{eq:Trimixing}
&&\tan\beta=50, \ \ M_{H^\pm}=155~~{\rm GeV},
\nonumber \\
&&M_{\tilde{Q}_3} = M_{\tilde{U}_3} = M_{\tilde{D}_3} =
M_{\tilde{L}_3} = M_{\tilde{E}_3} = M_{\rm SUSY} = 0.5 ~~{\rm TeV},
\nonumber \\
&& |\mu|=0.5 ~~{\rm TeV}, \ \
|A_{t,b,\tau}|=1 ~~{\rm TeV},   \ \
|M_2|=|M_1|=0.3~~{\rm TeV}, \ \ |M_3|=1 ~~{\rm TeV},
\nonumber \\
&&
\Phi_\mu = 0^\circ, \ \
\Phi_1 = \Phi_2 = 0^\circ\,, \ \
\Phi_{A_\tau}=\Phi_{A_e}=\Phi_{A_u}=\Phi_{A_c}=\Phi_{A_d}=\Phi_{A_s}=0^\circ\,.
\end{eqnarray}
Note that,  in this  scenario, only two independent CP-violating  phases generate
EDMs:  $\Phi_A=\Phi_{A_t}=\Phi_{A_b}$ and  $\Phi_3$.  In addition,  we
introduce a common hierarchy factor $\rho$ between the masses of the first two and
third generations:
\begin{equation}
M_{\tilde{X}_{1,2}}=\rho\, M_{\tilde{X}_3}\,,
\end{equation}
with $X=Q,U,D,L,E$.  For the size of first two generation $A$ terms,
we take $|A_e|=|A_\tau|$, $|A_{u,c}|=|A_t|$, and
$|A_{d,s}|=|A_b|$. Note that in this scenario, only $d^E_{u,d,s}$ and
$d^C_{u,d}$ depend on the hierarchy factor $\rho$ through the resummed
threshold corrections~(\ref{eq:Yukawa_light}).

In  Fig.~\ref{fig:tledm.tri},  we  show  the  absolute  value  of  the
Thallium  EDM  divided  by  its  current experimental  limit,  in  the
$\Phi_3$-$\Phi_A$ plane  (left) and as  a function of  $\Phi_3$ taking
$\Phi_A=60^\circ$ (right).   In the left  frame, the plane  is divided
into 4  regions: $|d_{\rm Tl}/d^{\rm EXP}|<1$  (black), $1\leq |d_{\rm
Tl}/d^{\rm  EXP}|<10$  (red),  $10\leq |d_{\rm  Tl}/d^{\rm  EXP}|<100$
(green),  and  $100\leq   |d_{\rm  Tl}/d^{\rm  EXP}|$  (magenta).  The
unshaded region is not allowed  theoretically. In the right frame, the
constituent contributions from the electron EDM $d^E_e$ and the CP-odd
electron-nucleon  interaction $C_S$ are  shown in  the thin  solid and
dashed lines,  respectively.  The thick  solid line is for  the total.
Note   the  non-trivial   cancellation  between   $d^E_e$   and  $C_S$
contributions around $\Phi_3=275^\circ$ in the right frame.

In Fig.~\ref{fig:n1edm.tri}, we show the neutron EDM calculated in the
CQM for  two values of  the hierarchy factor $\rho$:  $\rho=1$ (upper)
and 3 (lower).   In the left frames, the regions are  shaded as in the
Thallium  case,  Fig.~\ref{fig:tledm.tri}. In  the  right frames,  the
constituent contributions  from the quark EDMs  $d^E_{u,d}$, the quark
CEDMs $d^C_{u,d}$, the  Weinberg operator $d^G$ are shown  in the thin
solid,  dashed, and dotted  lines, respectively.  The thick  lines are
again for the total. We note first that $d^E_{u,d}$ and $d^C_{u,d}$ do
not    vanish,    even    though    we   are    taking    $\Phi_{1,2}=
\Phi_{A_{u,d}}=0^\circ$  in  this   scenario.   This  is  because  the
non-vanishing  $\Phi_3$  enters   the  Yukawa  couplings  through  the
threshold  corrections and  the two-loop  Higgs-mediated  graphs.  The
$d^{E,C}_{u,d}$ contributions decreases  for larger $\rho$, whilst the
$d^G$ contribution is independent of $\rho$.  For larger $\rho=3$, the
$d^G$   contribution   is   almost  dominating.    The   non-vanishing
$\rho$-independent   $d^{E,C}_{u,d}$  at  $\Phi_3=0^\circ\,,180^\circ$
comes  from  the  two-loop  Higgs-mediated  diagrams.   A  non-trivial
cancellation occurs,  for example,  between the $d^C_{u,d}$  and $d^G$
contributions around $\Phi_3=10^\circ$ in the lower-right frame.

In Fig.~\ref{fig:n2edm.tri}, we display  the neutron EDM calculated in
the  PQM.   The  shaded  regions  and   lines  are  the   same  as  in
Fig.~\ref{fig:n1edm.tri}, except for  the constituent contributions in
the right frames: the thin solid, dashed, and dotted lines are for the
contributions from the EDMs of $u$, $d$, and $s$ quarks, respectively.
In  this  model,   the  neutron  EDM  is  dominated   by  the  $d^E_s$
contribution  and  decreases as  $\rho$  increases.  The  dips in  the
lower-right   frames  are   due  to   the  cancellation   between  the
contributions to $d^E_s$ from the Higgs-mediated and other diagrams.

Figure~\ref{fig:n3edm.tri} shows  the predictions for  the neutron EDM
calculated using the  QCD sum rule approach. In  the right frames, the
thin  solid,  dashed,  dotted,  and  dash-dotted  lines  are  for  the
constituent   contributions  $d^E_{u,d}$,   $d^C_{u,d}$,   $d^G$,  and
$C_{bd,db}$,   respectively.   Compared   to   the  CQM,   $d^C_{u,d}$
contribution  is about  3 times  larger and  $d^G$ one  about  2 times
smaller.  The  $C_{bd,db}$ contribution is significant  because of the
large  value of  $\tan\beta$  and  the light  Higgs  spectrum in  this
scenario.

Comparing     Figs.~\ref{fig:n1edm.tri},    \ref{fig:n2edm.tri}    and
\ref{fig:n3edm.tri},  we  see that  they  yield qualitatively  similar
overall results, but with important detail differences. In particular,
the  appearances  and   locations  of  non-trivial  cancellations  are
model-dependent.

Finally,  Fig.~\ref{fig:hgedm.tri} gives  our numerical  estimates for
the Mercury EDM. In the  right frames, the thin solid, dashed, dotted,
and  dash-dotted  lines are  for  the  constituent contributions  from
$d^E_e$,     $d^C_{u,d}$,     $C_{4f}\equiv     C_{dd,sd,bd}$,     and
$C_{S,P}^{(\prime)}$, respectively.  When $\rho=1$, the Mercury EDM is
dominated   by  $d^C_{u,d}$.   However,  as   $\rho$   increases,  the
$d^C_{u,d}$  contribution  decreases  while other  four  contributions
remain the same.  Cancellations  occurs more easily for larger $\rho$,
in which  case all the  contributions become more or  less comparable,
and we see a non-trivial example in the lower-right frame.

\subsection{CPX Scenario}

In the CPX scenario~\cite{Carena:2000ks}, the product of $\mu$ and the
third-generation $A$ terms are larger than the common SUSY scale of
the third-generation squarks by a factor of 8:
\begin{eqnarray}
&& \hspace{-3.0cm}
M_{\tilde{Q}_3} = M_{\tilde{U}_3} = M_{\tilde{D}_3} =
M_{\tilde{L}_3} = M_{\tilde{E}_3} = M_{\rm SUSY}\,,
\nonumber \\
&& \hspace{-3.0cm}
|\mu|=4\,M_{\rm SUSY}\,, \ \
|A_{t,b,\tau}|=2\,M_{\rm SUSY} \,, \ \
|M_3|=1 ~~{\rm TeV}\,.
\label{eq:CPX}
\end{eqnarray}
As an example, we have fixed $|M_2|=2|M_1|=100$ GeV, and taken
the charged Higgs-boson pole mass
$M_{H^\pm}=300$ GeV and the common SUSY scale $M_{\rm SUSY}=0.5$ TeV,
but the parameter $\tan\beta$ is varied.  The $A$-term phases of
first two generations are set to vanish, as in the trimixing scenario:
$\Phi_{A_\tau}=\Phi_{A_e}=\Phi_{A_u}=\Phi_{A_c}=\Phi_{A_d}=\Phi_{A_s}=0^\circ$.
For the size of the $A$ terms of the first two generations, we also take
$|A_e|=|A_\tau|$, $|A_{u,c}|=|A_t|$, and $|A_{d,s}|=|A_b|$.  As for the CP-violating
phases, initially we vary two phases generating EDMs, $\Phi_A$
and $\Phi_3$, taking $\Phi_1=\Phi_2=0$.  The effects of non-trivial
$\Phi_1$ and $\Phi_2$ are described later.

Taking $\Phi_1=\Phi_2=0^\circ$, in Figs.~\ref{fig:tledm.cpx},
\ref{fig:n1edm.cpx}, \ref{fig:n2edm.cpx}, \ref{fig:n3edm.cpx}, and
\ref{fig:hgedm.cpx}, we show the Thallium, neutron, and Mercury EDMs
on the $\Phi_3-\Phi_A$ plane (left) and as functions of $\Phi_3$
taking $\Phi_A=90^\circ$ (right).  We take two values of $\tan\beta=5$
(upper) and 50 (lower) but with fixed $\rho=1$.  The shaded regions
and the lines are the same as in Figs.  \ref{fig:tledm.tri},
\ref{fig:n1edm.tri}, \ref{fig:n2edm.tri}, \ref{fig:n3edm.tri}, and
\ref{fig:hgedm.tri}, respectively.

We  observe the  Thallium EDM,  shown in  Fig.~\ref{fig:tledm.cpx}, is
dominated by the two-loop  Higgs-mediated electron EDM for both cases,
with  only mild  dependence on  $\Phi_3$.  For  small  $\tan\beta$, we
always  have  $|d_{\rm  Tl}/d^{\rm   EXP}|  <  10$,  independently  of
$\Phi_{3,A}$, as  can be  seen from the  upper-left frame.   For large
$\tan\beta$, the sub-leading $C_S$  contribution becomes larger by two
orders of magnitude, whilst the electron EDM contribution is larger by
one order of magnitude as seen by comparing the two right frames.

Turning   now   to   the   neutron   EDM  in   the   CQM,   shown   in
Fig.~\ref{fig:n1edm.cpx},  we see  that the  three  contributions from
$d^E_{u,d}$, $d^C_{u,d}$, and $d^G$ are comparable. The most important
contributions to the down-quark EDM $d^E_d$ and CEDM $d^C_d$ come from
the   one-loop   gluino  diagrams,   explaining   the  mild   $\Phi_A$
dependence. The different dependence on $\Phi_3$ for large $\tan\beta$
is  due  to  the  enhanced  two-loop  Higgs-mediated  contribution  to
$d^{E,C}_d$, and we note a non-trivial cancellation in the lower-right
panel  when $\Phi_3  \sim  325^\circ$.   The neutron  EDM  in the  PQM,
Fig.~\ref{fig:n2edm.cpx},  and that  found  using the  QCD sum  rules,
Fig.~\ref{fig:n3edm.cpx}, are  somewhat larger  in general and  show a
similar behaviour, due to the dominance by the $d^E_s$ and the $d^C_d$
contributions, respectively.
The Mercury EDM, Fig.~\ref{fig:hgedm.cpx}, is also dominated by the
$d^C_d$ contribution. However, the sub-dominant
contributions from $d^E_e$, $C_{4f}\equiv C_{dd,sd,bd}$, and
$C_{S,P}^{(\prime)}$ become larger for large $\tan\beta$.

We now study the effects of non-zero $\Phi_1$ and $\Phi_2$, and
some cancellation properties, varying the common hierarchy factor $\rho$
and assuming maximal CP violation in $\Phi_A$ and $\Phi_3$, i.e.,
$\Phi_A=\Phi_3=90^\circ$. The one-loop contributions decrease as
$\rho$ increases but the contributions from the Weinberg operator and
the two-loop Higgs mediated diagrams remain constant.

In the upper frames of Fig.~\ref{fig:tledm.cpx.rho}, we show the
Thallium EDM as a function of $\rho$ for two values of $\tan\beta$: 5
(upper-left) and 50 (upper-right). The four lines are for
$(\Phi_1,\Phi_2)=(0^\circ,0^\circ)$ (solid), $(90^\circ,0^\circ)$
(dashed), $(0^\circ,90^\circ)$ (dotted), and $(0^\circ,270^\circ)$
(dash-dotted).  When $(\Phi_1,\Phi_2)=(0^\circ,0^\circ)$, the Thallium
EDM is independent of $\rho$ because the main contribution from the
electron EDM is dominated by the two-loop Higgs-mediated diagrams, see
Fig.~\ref{fig:tledm.cpx}.  The two dips around $\rho=1.5$ and 4 when
$(\Phi_1,\Phi_2)=(90^\circ,0^\circ)$ and $(0^\circ,90^\circ)$ are due
to the cancellations between the $\rho$-independent two-loop Higgs and
neutralino and chargino contributions to the electron EDM,
respectively. In the lower frames we show explicitly the
chargino-Higgs cancellation in the $(0^\circ,90^\circ)$ case.  In the
lower frames, the thick solid line is for the total electron EDM and
the thin solid, dashed, horizontal dash-dotted lines are for the
chargino, neutralino, and Higgs contributions to it, respectively.

In Fig.~\ref{fig:n123edm.cpx.rho},  we compare the  three calculations
of the neutron EDM: the CQM (upper row), the PQM (middle row), and the
QCD  sum-rule technique (lower  row).  The  lines are  the same  as in
Fig.~\ref{fig:tledm.cpx.rho}.           The         cases         with
$(\Phi_1,\Phi_2)=(0^\circ,0^\circ)$  (solid)  and $(90^\circ,0^\circ)$
(dashed)  are  hardly  distinguishable   from  each  other.   The  PQM
calculation  (middle) is most  sensitive to  $\Phi_2$, whilst  the QCD
sum-rule approach  shows the least  sensitivity.  We observe  that the
value of $\rho$ where the cancellation occurs varies on the models and
approaches.  In all cases,  the neutron  EDM calculations  saturate to
certain values determined  by the $\rho$-independent contribution from
the  Weinberg operator  or the  two-loop Higgs  mediated  diagrams, as
shown below.

In Figs.~\ref{fig:n13edm.cpx.rho} and \ref{fig:n2edm.cpx.rho}, we show
details     of     the     neutron     EDM    taking     the     cases
$(\Phi_1,\Phi_2)=(0^\circ,90^\circ)$     and    $(0^\circ,270^\circ)$,
respectively,  as  examples.   In  Fig.~\ref{fig:n13edm.cpx.rho},  the
thick line is for the total neutron EDM and the thin solid, dashed and
horizontal dotted  lines are  for the contributions  from the  EDMs of
quarks $d^E_{u,d}$,  the CEDMs $d^C_{u,d}$, and  the Weinberg operator
$d^G$, respectively.  The EDMs and CEDMs are dominated by those of the
down quark. In the lower-right frame, the lower horizontal dash-dotted
line  is  for  the  contribution  from  the  bottom-down  four-fermion
operator,  $C_{bd,db}$. We  observe the  dips of  the thick  lines are
determined  by the  interplay of  the three  main  contributions.  For
example,  in the  upper-left frames  with  $\tan\beta=5$, cancellation
occurs at $\rho=2$ in the CQM.  On the other hand, in the QCD sum-rule
approach where the contribution from $d^G$ ($d^C_{u,d}$) is suppressed
(enhanced) compared  to the CQM, the cancellation  occurs at $\rho\sim 3.6$.
Taking into account of the  uncertainty involved in the calculation of
the $d^G$ contribution, the $\rho$ value where the cancellation occurs
may  change  by  $\sim  \pm   1$,  at  least.   The  more  significant
contribution from  $d^E_{u,d}$ in the CQM explains why
this calculation is  more sensitive to $\Phi_2$ than  the QCD sum-rule
approach.  The dips in the thin solid lines for $d^E_{u,d}$ are due to
three-way cancellations  among the one-loop  gluino, one-loop chargino
and  two-loop Higgs-mediated  diagrams.
For large $\tan\beta$, no cancellation occurs due to the dominance of $d^C_{u,d}$.

Figure~\ref{fig:n2edm.cpx.rho}  displays our  numerical  estimates for
the neutron EDM in the PQM when $(\Phi_1,\Phi_2)=(0^\circ,270^\circ)$.
In the  upper frames, the thick lines  are for the total  EDM, and the
thin solid,  dashed, and dotted  lines are for the  contributions from
the  EDMs of  the  up $d^E_u$,  down  $d^E_d$, and  the strange  quark
$d^E_s$,  respectively. In  the lower  frame, we  show  the dominating
contribution from the strange-quark EDM $d^E_s$ (thick solid) together
with  its   constituent  contributions  from   the  one-loop  chargino
$\chi^\pm$ (thin  solid), one-loop neutralino  $\chi^0$ (thin dashed),
one-loop gluino $\tilde{g}$ (thin dotted), and two-loop Higgs-mediated
$H^0$ (horizontal thin dashed-dotted) diagrams.  We observe that there
is   a  cancellation   between   the  chargino,   gluino,  and   Higgs
contributions at $\rho \sim 4$.

The above examples demonstrate that the interpretation of neutron
EDM measurements is subject to uncertainties in non-perturbative QCD,
which are reflected in the specific models discussed.

In Fig.~\ref{fig:hgedm.cpx.rho}, we show the Mercury EDM. It shows
barely any sensitivity to $\Phi_1$ and $\Phi_2$, due to
the dominance by the down-quark CEDM $d^C_d$,
which is dominated by the one-loop gluino and two-loop Higgs-mediated
diagrams, as shown in the lower frames. 
In the lower frames,
the lines are the same as in the lower frames of
Fig.~\ref{fig:n2edm.cpx.rho}.

Taking into account the uncertainty of the $d^G$ contribution to
the neutron EDM, and that involved in the Mercury EDM calculation, there
is a possibility of evading all the three EDM constraints by taking
$(\Phi_1,\Phi_2) \sim (0^\circ,90^\circ)$ and $\rho\sim 4$ in the CPX
scenario with $\Phi_A=\Phi_3=90^\circ$ when $\tan\beta=5$, if one
calculates the neutron EDM using the QCD sum-rule approach.

Finally, in Fig.~\ref{fig:ddedm.cpx.rho} we show the deuteron EDM. It
is not sensitive to $\Phi_1$ and $\Phi_2$, as it is dominated by the $d^G$
and $d^C_d$ contributions. At high $\rho$, the EDM is given by the sum
of the $d^G$ contribution and the contribution to $d^C_d$ from the
two-loop Higgs-mediated diagrams. In the lower frames, the thick line
is for the total EDM and the thin solid, dashed, horizontal dashed,
horizontal dash-dotted lines are for the contributions from
$d^E_{u,d}$, $d^C_{u,d}$, $C_{4f}\equiv C_{dd,sd,bd}$, and $d^G$.

\subsection{MCPMFV Scenario}

In the MCPMFV scenario, in contrast to the trimixing and CPX
scenarios, the soft SUSY-breaking parameters are specified at the the gauge
coupling unification (GUT) scale where the MFV condition is
imposed~\cite{Ellis:2007kb,Argyrou:2008rn}.  This scenario has a total of 19
parameters, including 6 CP-violating phases and 13 real mass
parameters. As a numerical example, in order to study the effects of
the CP-violating phases on EDMs in this framework, we consider a
CP-violating variant of a SPS1a-like~\cite{SPS} scenario:
\begin{eqnarray}
&&\left|M_{1,2,3}\right|=250~~{\rm GeV}\,, \nonumber \\
&&M^2_{H_u}=M^2_{H_d}=\widetilde{M}^2_Q=\widetilde{M}^2_U=\widetilde{M}^2_D
=(100~~{\rm GeV})^2\,, \nonumber \\
&&\widetilde{M}^2_L=\widetilde{M}^2_E=(200~~{\rm GeV})^2\,, \nonumber \\
&&\left|A_u\right|=\left|A_d\right|=\left|A_e\right|=100~~{\rm GeV}\,,
\label{eq:cpsps1a}
\end{eqnarray}
with $\tan\beta\,(M_{\rm SUSY})=40$.  As for the CP-violating phases,
we adopt the convention that $\Phi_\mu=0^\circ$, and we vary separately the three
phases of the gaugino mass parameters, $\Phi_{1,2,3}$, taking however
vanishing $A$-term phases at the GUT scale: $\Phi^{\rm
GUT}_{A_u}=\Phi^{\rm GUT}_{A_d}=\Phi^{\rm GUT}_{A_e}=0^\circ$.  At the
low-energy $M_{\rm SUSY}$ scale, $A$-term phases may be
generated by the CP-violating phases of the gaugino mass parameters
$\Phi_{1,2,3}$, even though we are taking real $A$ terms at the GUT
scale~\cite{Goto:1998iy,Ellis:2007kb}.

In the upper-left frame of Fig.~\ref{fig:tledm.sps1a.p2}, we show the
Thallium EDM as a function of $\Phi_2$ for several values of
$(\Phi_1,\Phi_3)$: $(0^\circ,0^\circ)$ (solid), $(0^\circ,90^\circ)$
(dashed), $(90^\circ,0^\circ)$ (dotted), and $(270^\circ,0^\circ)$
(dash-dotted). We see the Thallium EDM is nearly independent of
$\Phi_3$, cf., the two overlapping lines for
$(\Phi_1,\Phi_3)=(0^\circ,0^\circ)$ and $(0^\circ,90^\circ)$. This is
because the Thallium EDM is dominated by the electron EDM. In the
upper-right, lower-left, and lower-right frames, we show the electron
EDM as a function of $\Phi_2$ when $(\Phi_1,\Phi_3) =
(0^\circ,90^\circ)$, $(90^\circ,0^\circ)$, and $(270^\circ,90^\circ)$,
respectively.  In the lower frames, we observe the neutralino
contribution becomes less dependent on $\Phi_2$ when $\Phi_1=90^\circ$
and $270^\circ$ and cancels the chargino one around $\Phi_2=\pm
4^\circ$, resulting in the dips in the upper-left frame.

Figure~\ref{fig:n1edm.sps1a.p2}  presents numerical  estimates  of the
neutron EDM in the CQM as  a function of $\Phi_2$. We observe that the
EDM is nearly independent of $\Phi_1$ and $\Phi_3$. This is because of
the dominance of the down-quark EDM $d^E_d$, see the upper-right frame
with $(\Phi_1,\Phi_3)=(0^\circ,0^\circ)$ in  which the thin solid line
for $d^E_{u,d}$ is  overlapped by the total thick  line.  The electric
EDM of the  down quark still plays a dominant  role even when $\Phi_3$
has a non-trivial value, because of an accidental cancellation between
the contributions  from the Weinberg operator $d^G$  (dotted line) and
the down-quark CEDM $d^C_d$ (dashed line) as shown in the lower frames
for     $(\Phi_1,\Phi_3)=(180^\circ,90^\circ)$     (lower-left)    and
$(180^\circ,270^\circ)$ (lower-right). This accidental cancellation is
lifted when the EDM is  calculated using the QCD sum-rule approach, in
which $d^G$ and  $d^C_d$ contribute differently to the  total EDM, see
the lower-frames of Fig.~\ref{fig:n3edm.sps1a.p2}.

In the upper-left frame  of Fig.~\ref{fig:n2edm.sps1a.p2}, we show the
neutron EDM  calculated in the  PQM. We find  that the neutron  EDM is
nearly independent of $\Phi_3$.   The dominant contribution comes from
the  strange-quark EDM  $d^E_s$,  and  we show  it  together with  the
constituent    contributions    in    the    other    frames    taking
$(\Phi_1,\Phi_3)=(0^\circ,0^\circ)$                      (upper-right),
$(0^\circ,10^\circ)$     (lower-left),     and    $(0^\circ,90^\circ)$
(lower-right).  The thick lines  are for  the total  EDM and  the thin
solid, dashed, dotted, and dash-dotted lines are for the contributions
from  the one-loop  chargino $\chi^\pm$,  neutralino  $\chi^0$, gluino
$\tilde{g}$,  and the  two-loop Higgs-mediated  $H^0$  diagrams.  When
$\Phi_3=0^\circ$, $d^E_s$  is dominated by  the chargino contribution,
see the upper-right frame.  As shown in the lower frames, the $\Phi_3$
dependence of  the strange-quark EDM  is largely cancelled in  the two
main  chargino  and  gluino  contributions,  explaining  the  $\Phi_3$
independence of  the neutron  EDM shown in  the upper-left  frame.  We
note that  this cancellation  resulting in $\Phi_3$  independence also
occurs in the  down-quark EDM $d^E_d$, see the  thin solid $d^E_{u,d}$
lines  in  the  lower  frames of  Figs.  \ref{fig:n1edm.sps1a.p2}  and
\ref{fig:n3edm.sps1a.p2}.

Figure~\ref{fig:n3edm.sps1a.p2}  gives the  predicted  values for  the
neutron  EDM  calculated  using  the  QCD sum-rule  approach.  In  the
upper-right  and  lower   frames,  the  lines  are  the   same  as  in
Fig.~\ref{fig:n1edm.sps1a.p2},  but   for  different  combinations  of
$\Phi_1$ and  $\Phi_3$.  The non-trivial  three-way cancellation among
the  contributions from $d^E_{u,d}$,  $d^C_{u,d}$, and  $d^G$ explains
the  dips   in  the   upper-left  frame  when   $\Phi_3=20^\circ$  and
$340^\circ$.

In Fig.~\ref{fig:hgedm.sps1a.p2},  we present numerical  estimates for
the Mercury EDM. In the upper-right  and lower frames, we see that the
dominant  contributions come  from the  electron EDM  $d^E_e$  and the
down-quark CEDM $d^C_d$. The  cancellation between them results in the
dips  in  the upper-left  frame.   Note  that $\Phi_3=90^\circ$  could
generate an EDM larger than the current experimental limit by a factor
$\sim$ 400, nearly independent  of $\Phi_2$, see the nearly horizontal
thin solid line in the upper-left frame.

Finally,  Fig.~\ref{fig:ddedm.sps1a.p2} shows  theoretical predictions
for the deuteron EDM. The  deuteron EDM is dominated by the down-quark
CEDM  $d^C_d$. The  sub-leading  contributions are  from the  Weinberg
operator $d^G$ and the down-quark EDM $d^E_d$.  In the upper-right and
lower frames,  we show the  constituent contributions to  $d^C_d$. The
chargino  and neutralino contributions  are dominant  when $\Phi_3=0$,
see the upper-right and lower-left frames. However, as $\Phi_3$ grows,
the  gluino  contribution   rapidly  increases,  see  the  lower-right
frame.  We  note  that   the  dip  around  $\Phi_2=-20^\circ$  in  the
upper-left frame when $(\Phi_1,\Phi_3)=(0^\circ,10^\circ)$ is due to a
cancellation between the $d^C_d$ and $d^E_d$ contributions.

\section{Conclusions}

We have  performed a fully-fledged analysis of  the Thallium, neutron,
Mercury and deuteron EDMs within the general CP-violating framework of
the  MSSM.   In our  analysis,  we have  taken  into  account the  the
complete set of one-loop graphs, the dominant Higgs-mediated Barr--Zee
diagrams, the complete CP-odd  dimension-six Weinberg operator and the
Higgs-mediated  four-fermion  operators.  Our  study  has also  improved
earlier calculations  in two important aspects.   First, it includes
CP-violating  Higgs-boson mixing  effects and,  secondly,  it properly
implements  resummation effects  due to  threshold corrections  to the
Yukawa couplings of all up-  and down-type quarks and charged leptons.
Not only do these two effects turn out to be significant for the existing
one-  and  two-loop  EDMs,  but  also they give  rise  to  additional
higher-order  contributions  within the  MSSM,  such  as the  original
Weinberg operator  induced by $t$ and  $b$ quarks with  Higgs bosons in
the loop. In addition, we have improved the Mercury EDM calculation by
including the contribution due  to the CP-odd triplet electron-nucleon
interaction.

Having  established the latest  state-of-art theoretical  framework as
described above, we have then explored the EDM constraints on the CPX,
trimixing and  MCPMFV scenarios. Clearly, sufficiently small values of the 
CP-violating parameters survive the experimental upper limits on the EDMs.
However,  we have also found
that larger values of the CP-violating parameters may be allowed exceptionally by
accidental cancellations among the CP-violating
contributions to the three measured EDMs, i.e., $d_{\rm Tl}$, $d_n$ and
$d_{\rm Hg}$, in  all the above scenarios.  In  detail, the results of
our analysis may be summarized as follows.

First, we studied the trimixing  scenario. We have explored the impact
on the  three measured  EDMs resulting from  an hierarchy of  the soft
SUSY-breaking  masses between  the  first two  and third  generations,
$\rho$, and the  CP phases $\Phi_A=\Phi_{A_t}=\Phi_{A_b}$ and $\Phi_3$
related  to  the  third  generation  $A$-terms  and  the  gluino  mass
parameter,  respectively.  In the  case  of  the  Thallium EDM,  large
CP-violating phases  are allowed due  to the cancellation  between the
contributions from  the CP-odd electron-nucleon  interaction $C_S$ and
the two-loop Higgs-mediated electron EDM $(d^E_e)^H$.
%Such a cancellation is found to be independent of $\rho$.
The two contributions are found to be independent of $\rho$ in the scenario under
consideration.

The neutron  EDM has been  computed within three different  models and
approaches:  (i)  the CQM,  (ii)~the  PQM  and  (iii) a  QCD  sum-rule
approach. In the  CQM and QCD sum-rule approaches,  the neutron EDM is
dominated by  the EDM and CEDM  of the down quark  $d^{E,C}_d$ and the
contribution from  the dimension-six Weinberg operator  $d^G$.  In the
QCD sum  rule approach, $d^G$ ($d^C_d$) becomes  less (more) important
than  in  the  CQM.   The  down-quark EDM  and  CEDM  $d^{E,C}_d$  are
generated through  the resummed threshold  corrections at one  loop by
$\Phi_3$ and the two-loop Higgs-mediated diagrams.  The latter remains
the same  for large $\rho$.   The $d^{E,C}_d$ contributions  depend on
$\rho$,  whilst  the one  due  to  $d^G$  is $\rho$-independent.   The
bottom-down four-quark interaction  $C_{bd,db}$ is also independent of
$\rho$  and  becomes  significant   in  the  QCD  sum  rule  approach.
Interestingly,  cancellations may occur  among $d^{E,C}_d$,  $d^G$ and
$C_{bd,db}$  more easily when  $\rho$ is  large.  In  the PQM,  on the
other  hand, the  neutron EDM  is dominated  by the  strange-quark EDM
$d^E_s$, which  is generated by one-loop threshold  corrections at the
one-loop  level   and  by   the  two-loop  Barr--Zee   graphs.  Again,
cancellations  occur  between the  two  contributions  when $\rho$  is
large.  All the  three  different hadronic  models  or approaches  for
computing the  neutron EDM  give comparable estimates.  However, there
are   significant   differences   of   detail,   implying   that   the
interpretation of neutron EDM measurements is model-dependent.

As far as the Mercury EDM  in the trimixing scenario is concerned, the
$d$-quark CEDM  $d^C_d$ was found to  be dominant for  small values of
$\rho$,  whilst it  gets suppressed  for large  values of  $\rho$.  In
addition,  the   $\rho$-independent  contributions  from   $C_S$,  the
four-quark  operators $C_{dd,sd,bd}$, the  CP-odd singlet  and triplet
electron-nucleon  interactions,  $C_P$  and  $C^{\prime}_P$,  and  the
electron  EDM   become  all   relevant,  leading  to   an  interesting
cancellation pattern.

In the  CPX scenario, we have  analyzed the dependence of  EDMs on the
hierarchy factor $\rho$,  the CP phases $\Phi_A=\Phi_{A_t}=\Phi_{A_b}$
and  $\Phi_{1,2,3}$,   for  a  relatively  low  and   large  value  of
$\tan\beta$:  $\tan\beta=5$ and  $50$.   In the  absence  of any  mass
hierarchy between  the first two  and third generation of  squarks and
sleptons,  i.e., $\rho=1$,  and for  $\Phi_1=\Phi_2=0^\circ$,  we have
found that the  Thallium EDM is dominated by  the electron EDM, whilst
$C_S$    starts    becoming     important    only    for    $\tan\beta
\stackrel{>}{{}_\sim} 10$.  Correspondingly,  the neutron EDM receives
the largest contribution from $d^{E,C}_d$ and $d^G$ in the CQM and the
QCD sum-rule approach. The main contributions to $d^{E,C}_d$ come from
the   one-loop  gluino  diagrams   and  the   two-loop  Higgs-mediated
diagrams. The  latter become important  when $\tan\beta$ is  large. In
the PQM,  the neutron  EDM mainly results  from $d^E_s$.  Finally, the
Mercury EDM receives its biggest contribution from $d^C_d$, whilst the
contribution due to
%$d^E_e$ becomes significant only for large $\tan\beta$.
$d^E_e$ becomes important only for large $\tan\beta$.

Varying  $\rho$ and the CP  phases $\Phi_1$ and  $\Phi_2$, we have
found that  several cancellations can  occur within the  CPX scenario.
Specifically,  one-loop  neutralino   and  chargino  effects  and  the
two-loop Barr--Zee  graphs may add  up destructively and  suppress the
electron  EDM  $d^E_e$.  Likewise,  cancellations  among the  one-loop
chargino and  gluino graphs and two-loop  Higgs-mediated diagrams lead
to  suppressed strange-  and  down-quark EDMs  $d^E_{d,s}$  and to  an
equally  small  down-quark  CEDM  $d^C_d$.  As  a  consequence  of  the
%suppressed $d^E_{d,s}$ and $d^C_d$, 
suppressed $d^E_e$, $d^E_s$ and $d^C_d$, 
the Thallium, neutron (in the PQM)
and  Mercury  EDMs  were  all  found  to come  out  well  below  their
experimental limits.  A similar result has been obtained in the CQM or
QCD  sum-rule approach,  where the  Weinberg operator  $d^G$  plays an
important role.
In particular,  we have demonstrated explicitly  the possibility
of evading all the  three EDM constraints from  $d_{\rm Tl}$,
$d_n$  and $d_{\rm  Hg}$  
%in the  CPX  scenario, if  one assumes  that
%$(\Phi_1,\Phi_2)   \sim   (0^\circ,90^\circ)$,  $\Phi_A \sim \Phi_3 \sim 90^\circ$,
%$\rho\sim 4$,  $\tan\beta \sim 5$, 
in the  CPX  scenario with $\Phi_A \sim \Phi_3 \sim 90^\circ$, if  one assumes  that
$(\Phi_1,\Phi_2)   \sim   (0^\circ,90^\circ)$ and
$\rho\sim 4$ when  $\tan\beta = 5$, 
provided  $d_n$ is calculated  using QCD
sum-rule techniques.
Finally, the deuteron  EDM~\cite{Semertzidis:2003iq,OMS}, whose size crucially depends  on $d^G$ and
%$d^C_{u,d}$, can constrain dramatically, by a factor of about 100, 
%this region where the contributions
%from the different CP-violating phases cancel.
$d^C_{d}$, can constrain dramatically the CP-violating phases
by a factor of about 100~\cite{Lebedev:2004va}.
 
A third benchmark scenario of the  MSSM that has been analyzed was
the  MCPMFV scenario. For  definiteness, we  have used SPS1a-like
input  parameters  given  at  the  GUT  scale,  for  which  $\Phi^{\rm
GUT}_{A_u}=\Phi^{\rm   GUT}_{A_u}=\Phi^{\rm   GUT}_{A_e}=0^\circ$  and
$\tan\beta=40$. This choice results in a relatively light SUSY spectrum,
where phases for the $A$-terms  are generated by RG  running down
to the electroweak scale.  Within this particular MCPMFV  scenario, we have
studied  the implications  of  the CP-violating  phases  $\Phi_1$, $\Phi_2$  and
$\Phi_3$ for the three EDMs $d_{\rm Tl}$, $d_n$ and $d_{\rm Hg}$.
The Thallium EDM was found to be nearly independent of $\Phi_3$, since
the electron    EDM   $d^E_e$    turns    out   to    be   the    dominant
contribution.  Moreover, we  have noticed  that one-loop  chargino and
neutralino effects may cancel each another in $d^E_e$.
The  neutron EDM in  the CQM  was also
found  to be nearly  independent of  $\Phi_1$ and  $\Phi_3$, due  to an
accidental cancellation  between the  Weinberg operator $d^G$  and the
down-quark CEDM  $d^C_d$. Unlike in  the CQM, this cancellation  is no
longer present  when the QCD sum  rule approach is used,  leading to a
strong dependence on $\Phi_3$.  
In the PQM,  like in the CQM, the neutron
EDM  does  not  depend strongly  on $\Phi_3$,  since  there  is  a
cancellation between  the chargino and gluino effects  on the dominant
strange-quark EDM $d^E_s$.
The  Mercury EDM  $d_{\rm Hg}$  in the  MCPMFV scenario  was  found to
receive its biggest contribution from the electron EDM $d^E_e$ and the
down-quark  CEDM  $d^C_d$.  
%Moreover,  a  non-vanishing  gluino  phase
%$\Phi_3$  can easily  drive $d_{\rm  Tl}$ to  quite large  values that
%could even exceed the current experimental limit.
A non-vanishing  gluino  phase
$\Phi_3$  can drive $d_{\rm  Hg}$ to  quite large  values that
could easily exceed the current experimental limit.
Finally, the  leading effect on  the deuteron EDM comes  from $d^C_d$,
whilst $d^G$ and $d^E_d$ remain sub-leading. Since the deuteron EDM is
expected  to  be a  factor  $\sim  100$  more sensitive  than  $d_{\rm
Hg}$~\cite{Semertzidis:2003iq,OMS},  it  will  lead  to  much  tighter
constraints on the CP-violating phases~\cite{Lebedev:2004va}.

Our  detailed study  has shown  that the  three measured  EDMs provide
correlated  constraints  on  the   6  CP-violating  phases  in  MCPMFV
scenario,   leaving   open  the   possibility   of  relatively   large
contributions  to other CP-violating  observables. In  particular, the
deuteron  EDM~\cite{Semertzidis:2003iq,OMS} 
will  probe the  unconstrained CP  phases of  the MCPMFV
scenario.

The analytic  expressions for the  EDMs are implemented in  an updated
version  of the  code  {\tt  CPsuperH2.0}. This  new  feature of  {\tt
CPsuperH2.0} will be particularly  valuable in future explorations
for possible new-physics phenomena in the $K$- and $B$-meson systems.

\vspace{-0.2cm}
\subsection*{Acknowledgements}
\vspace{-0.3cm}
\noindent
We thank Yannis Semertzidis for valuable information on the deuteron EDM.
The work of J.S.L. was supported in part by the National Science Council of
Taiwan, R.O.C.  under Grant No. NSC 96-2811-M-008-068.
The work of  A.P. was supported  in part by  the STFC
research grant: PP/D000157/1.

\newpage
%\section*{Appendices}

\def\theequation{\Alph{section}.\arabic{equation}}
\begin{appendix}

\setcounter{equation}{0}
\section{Calculation of $C_S$}\label{sec:cs}

In  this appendix we  calculate the  coefficient $C_S$.   Our starting
point is the interaction Lagrangians:
\begin{eqnarray}
{\cal L}_{C_S}\ =\ C_S \,\bar{e}i\gamma_5e\,\bar{N}N\,,\qquad
{\cal L}_{\rm 4f} &=& C_{qe}\,\bar{q}q\,\bar{e}i\gamma_5 e\;.
\end{eqnarray}
Given the relation $\langle N|{\cal L}_{C_S}|N\rangle = 
\langle N|{\cal L}_{\rm 4f}|N\rangle$, one may identify
\begin{equation}
\left[(C_S)^{4f}\right]_q = C_{qe}\,\frac{\langle N| \bar{q} q |N\rangle}
{\langle N| \bar{N} N |N\rangle}\; ,
\end{equation}
where $q$ could be a light quark, e.g.~$u,d$, or a heavy one such as the
$b$ quark. 

We first derive the light-quark contribution to $(C_S)^{4f}$, 
$\left[(C_S)^{4f}\right]_{q=u,d}$. To this end, we need to know
$\langle N| \bar{u} u |N\rangle$ and $\langle N| \bar{d} d |N\rangle$.
Using the relation~\cite{Demir:2003js}
\begin{equation}
(m_u+m_d)\, \langle N| \bar{u}u + \bar{d}d |N\rangle \simeq
90\,{\rm MeV}\,\langle N| \bar{N} N |N\rangle\,,
\end{equation}
and assuming  that the triplet contribution  vanishes, i.e.
\begin{equation}
\langle N| \bar{u}u - \bar{d}d |N\rangle = 0\,,
\end{equation}
we obtain
\begin{equation}
\frac{\langle N| \bar{u} u |N\rangle}{\langle N| \bar{N} N |N\rangle}\ =\
\frac{\langle N| \bar{d} d |N\rangle}{\langle N| \bar{N} N |N\rangle}\ 
\simeq\ \frac{1}{2}\,\frac{90\,{\rm MeV}}{(m_u+m_d)}\ \simeq\
\frac{29\,{\rm MeV}}{m_d}\ =\
\left(\frac{m_u}{m_d}\right)\,\frac{29\,{\rm MeV}}{m_u}\ ,
\end{equation}
with $m_u/m_d=0.55$. Putting everything together, we find that
\begin{eqnarray}
\left[ (C_S)^{4f}\right]_u 
\!&\simeq &\! C_{ue}\,\left(\frac{m_u}{m_d}\right)\,\frac{29\,{\rm MeV}}{m_u} 
\ \simeq\ C_{ue}\,\frac{16\,{\rm MeV}}{m_u}\ ,\nonumber \\
\left[ (C_S)^{4f} \right]_d \!&\simeq &\! C_{de}\,\frac{29\,{\rm
    MeV}}{m_d}\ .
\end{eqnarray}
The  very  last expression  was  used to  obtain  the  first term  in
Eq.~(\ref{eq:cs4f}).

As for the heavy-quark contribution to $C_S$, there are two approaches
that can be  considered.  We illustrate these by  taking the $b$-quark
as  an  example.   The  first   way  is  to  include  the  heavy-quark
contribution directly to $(C_S)^{4f}$~\cite{Demir:2003js}, i.e.
\begin{equation}
\left[(C_S)^{4f}\right]_b\ =\ C_{be}\frac{66\,{\rm MeV}
  (1-0.25\kappa)}{m_b}\ .
\end{equation}
The second  method uses the QCD  trace anomaly and the  heavy quark is
integrated   out    in   the   gluon-gluon-Higgs    vertex   $(C_S)^g$
[cf.~(\ref{eq:csg})]:
\begin{equation}
\left[(C_S)^g\right]_{b}\ =\ (0.1\,{\rm GeV})\, \frac{m_e}{v^2}
\sum_{i=1}^3
\frac{\left[g^S_{H_igg}\right]_b\,g^P_{H_i\bar{e}e}}{M_{H_i}^2}\
=\ C_{be}\, 
\frac{2\,(0.1\,{\rm GeV})}{3\,m_b}\ ,
\end{equation}
where       we       made        use       of       the       relation
$[g^S_{H_igg}]_b=2/3\,g^S_{H_i\bar{b}b}$   [cf.~(\ref{eq:gsHgg})  and
(\ref{eq:cff})].  Notice that, apart from the $\kappa$-dependent term,
the two approaches are equivalent to each other.

\setcounter{equation}{0}
\section{Calculation of $C_P$ and $C^\prime_P$}\label{sec:cp}

Here we compute the  iso-scalar and iso-triplet coefficients $C_P$ and
$C^\prime_P$  that are relevant  in the  determination of  the Mercury
EDM. To this end, we start considering the interaction Lagrangians,
\begin{eqnarray}
{\cal L}_{C_P}\ =\ C_P \,\bar{e}e\,\bar{N}i\gamma_5 N
+C^\prime_P \,\bar{e}e\,\bar{N}i\gamma_5\tau_3 N
\; ,\qquad 
{\cal L}_{\rm 4f}\ =\ C_{eq}\,\bar{e}e\,\bar{q}i\gamma_5 q\; .
\end{eqnarray}
Imposing  the relation  $\langle N|{\cal  L}_{C_P}|N\rangle  = \langle
N|{\cal L}_{\rm  4f}|N\rangle$, we may project out  the iso-scalar and
iso-triplet contributions as follows:
\begin{equation}
\left[(C_P)^{4f}\right]_q\ =\ C_{eq}\,\frac{\langle N| \bar{q}i\gamma_5
  q |N\rangle} 
{\langle N| \bar{N} i\gamma_5 N |N\rangle}\,, \quad
\left[(C^\prime_P)^{4f}\right]_q\ =\ 
C_{eq}\,\frac{\langle N| \bar{q}i\gamma_5 q |N\rangle}
{\langle N| \bar{N} i\gamma_5\tau_3 N |N\rangle}\ .
\end{equation}

As  in Appendix~A,  we need  to  consider the  light- and  heavy-quark
contributions       separately.        Our      approach       closely
follows~\cite{Anselm:1985cf}.  Thus,  taking   into  account  all  the
relations that follow from isospin invariance, i.e.
\begin{eqnarray}
\langle N|m_u\,\bar{u} i\gamma_5 u \,+\, m_d\,\bar{d}i\gamma_5 d |N\rangle
\!&=&\!
\frac{m_u-m_d}{m_u+m_d}\,m_N\,(-g_A)\,\langle N|\bar{N}i\gamma_5\tau_3
N |N\rangle\,, \nonumber \\
\langle N|\bar{u} i\gamma_5 u |N\rangle \!&=&\!
-\langle N|\bar{d}i\gamma_5 d |N\rangle\; ,
\label{eq:anselm1}
\end{eqnarray}
we obtain
\begin{eqnarray}
m_u\,\langle N|\bar{u} i\gamma_5 u |N\rangle \!&=&\!
m_N\,(-g_A)\,\frac{m_u}{m_u+m_d}\,\langle N|\bar{N}i\gamma_5\tau_3 N
|N\rangle\,, 
\nonumber \\
m_d\,\langle N|\bar{d} i\gamma_5 d |N\rangle \!&=&\!
-m_N\,(-g_A)\,\frac{m_d}{m_u+m_d}\,\langle N|\bar{N}i\gamma_5\tau_3 N
|N\rangle\; , 
\end{eqnarray}
where $(-g_A)=1.25$  is the  axial nucleon form  factor. From  all the
above relations, it is then not difficult to derive that
\begin{eqnarray}
\left[(C^\prime_P)^{4f}\right]_u &=&
\frac{C_{eu}}{m_u}\,m_N\,(-g_A)\,\frac{m_u}{m_u+m_d} \simeq
C_{eu}\,\frac{444~{\rm MeV}}{m_u}\,, \nonumber \\
\left[(C^\prime_P)^{4f}\right]_d &=&
-\frac{C_{ed}}{m_d}\,m_N\,(-g_A)\,\frac{m_d}{m_u+m_d} \simeq
-C_{ed}\,\frac{806~{\rm MeV}}{m_d}\,,
\end{eqnarray}
where we  assume that  $m_N=1~{\rm GeV}$ and  $m_u/m_d=0.55$.  Observe
that the  light quarks  do not contribute  to the  singlet coefficient
$(C_P)^{4f}$.

To  calculate the heavy-quark  contributions to  $C_P$ and  $C'_P$, we
first consider the chiral anomaly relations~\cite{Anselm:1985cf}:
\begin{eqnarray}
\langle N | \partial^\mu J^5_\mu |N\rangle &=&
2\,\langle N|m_u\,\bar{u} i\gamma_5 u \,+\, m_d\,\bar{d}i\gamma_5 d |N\rangle 
\nonumber \\
&&+2\sum_{q=c,s,t,b}\langle N|m_q\,\bar{q} i\gamma_5 q |N\rangle
+6\,\langle N |\frac{\alpha_s}{8\pi}G\tilde{G} |N\rangle\,, \nonumber \\
\langle N | \partial^\mu J^5_\mu |N\rangle &=&
\left(-g_A^{(0)}\right)\, 2\, m_N\, \langle N|\bar{N}i\gamma_5 N|N\rangle\,,
\end{eqnarray}
where $n_l=2$ and  $n_h=4$ are assumed to be the  numbers of the light
and  heavy  quarks, respectively,  and  $g_A^{(0)}=(3/5)  g_A$ in  the
relativistic quark model. After integrating out the heavy quarks 
by employing the relation
\begin{equation}
\left. \langle N|m_q\,\bar{q} i\gamma_5 q |N\rangle \right|_{q=c,s,t,b}
\ =\ -\frac{1}{2}\,\langle N |\frac{\alpha_s}{8\pi}G\tilde{G} |N\rangle\; ,
\label{eq:anselm2}
\end{equation}
we get
\begin{eqnarray}
\langle N|m_u\,\bar{u} i\gamma_5 u \,+\, m_d\,\bar{d}i\gamma_5 d |N\rangle +
\langle N |\frac{\alpha_s}{8\pi}G\tilde{G} |N\rangle\ =\
\left(-g_A^{(0)}\right)\, m_N\, \langle N|\bar{N}i\gamma_5
N|N\rangle\; .
\end{eqnarray}
From~(\ref{eq:anselm1}) and~(\ref{eq:anselm2}), we finally obtain
\begin{eqnarray}
  \label{mqNg5}
\langle N|m_q\,\bar{q} i\gamma_5 q |N\rangle &=&
-\frac{1}{2}\,
m_N\, \left(-g_A^{(0)}\right)\, \langle N|\bar{N}i\gamma_5 N|N\rangle
\nonumber \\ && %\hspace{-0.3cm}
+\frac{1}{2}\,
\frac{m_u-m_d}{m_u+m_d}\,m_N\,(-g_A)\,
\langle N|\bar{N}i\gamma_5\tau_3 N |N\rangle
\end{eqnarray}
for each of the heavy  quarks $q=c,s,t,b$.  The first and second terms
in~(\ref{mqNg5})  give the  heavy-quark contributions  to $(C_P)^{4f}$
and $(C^\prime_P)^{4f}$, respectively.  More explicitly, we have
\begin{eqnarray}
\left[(C_P)^{4f}\right]_{q=c,s,t,b} &=& \frac{C_{eq}}{m_q}
\left[-\frac{1}{2}\, m_N\, \left(-g_A^{(0)}\right)\right]\ \simeq\
-C_{eq}\,\frac{375~{\rm MeV}}{m_q}\,,
\nonumber \\
\left[(C^\prime_P)^{4f}\right]_{q=c,s,t,b} &=& \frac{C_{eq}}{m_q}
\left[\frac{1}{2}\, \frac{m_u-m_d}{m_u+m_d}\,m_N\,(-g_A)\right]\ \simeq\
-C_{eq}\,\frac{181~{\rm MeV}}{m_q}\,,
\end{eqnarray}
where  $m_N=1$  GeV and  $m_u/m_d=0.55$  were  used  in our  numerical
estimates.

\setcounter{equation}{0}
\section{{\tt CPsuperH2.0} Interface}\label{sec:cpsuperh}

\begin{itemize}
%------------------------------------------------------
\item \underline{Input}: For the complex
$A$ parameters of first two generations, part of auxiliary array
{\tt CAUX\_H} is used as
\begin{eqnarray}
      A_e &=& {\tt CAUX\_H(995)}\,, \nonumber \\
      A_u &=& {\tt CAUX\_H(996)}\,, \ \ \ \
      A_c  =  {\tt CAUX\_H(997)}\,, \nonumber \\
      A_d &=& {\tt CAUX\_H(998)}\,, \ \ \ \
      A_s  =  {\tt CAUX\_H(999)}\,.
\end{eqnarray}
\item \underline{Output}: For output, part of auxiliary array
{\tt RAUX\_H} is used.
\begin{itemize}
\item{The electron EDM in units of ${\rm cm}$}:
%      I_DEOE_E     =200 ! d^E_e/e [{\rm cm}]
\begin{equation}
{\tt RAUX\_H(200)} = d^E_e/e = (d^E_e/e)^{\tilde{\chi}^\pm}
+(d^E_e/e)^{\tilde{\chi}^0}+(d^E_e/e)^{\tilde{g}}
+(d^E_e/e)^{H} \,, 
\end{equation}
where the constituent contributions are
\begin{eqnarray}
{\tt RAUX\_H(201)} &=& (d^E_e/e)^{\tilde{\chi}^\pm}\,, \ \ \
{\tt RAUX\_H(202)}  =  (d^E_e/e)^{\tilde{\chi}^0}\, \nonumber \\
{\tt RAUX\_H(203)} &=& (d^E_e/e)^{\tilde{g}}\,, \ \ \ \ \
{\tt RAUX\_H(204)}  =  (d^E_e/e)^{H}\,. \nonumber
\end{eqnarray}
\item{The electric EDM of the $u$ quark in units of ${\rm cm}$}:
%      I_DEOE_U     =210 ! d^E_u/e [{\rm cm}]
\begin{equation}
{\tt RAUX\_H(210)} = d^E_u/e = (d^E_u/e)^{\tilde{\chi}^\pm}
+(d^E_u/e)^{\tilde{\chi}^0}+(d^E_u/e)^{\tilde{g}}
+(d^E_u/e)^{H} \,,
\end{equation}
where the constituent contributions are
\begin{eqnarray}
{\tt RAUX\_H(211)} &=&(d^E_u/e)^{\tilde{\chi}^\pm}\,, \ \ \
{\tt RAUX\_H(212)}  = (d^E_u/e)^{\tilde{\chi}^0}\, \nonumber \\
{\tt RAUX\_H(213)} &=&(d^E_u/e)^{\tilde{g}}\,, \ \ \ \ \
{\tt RAUX\_H(214)}  = (d^E_u/e)^{H}\,.
\end{eqnarray}
\item{The electric EDM of the $d$ quark in units of ${\rm cm}$}:
%      I_DEOE_D     =220 ! d^E_d/e [{\rm cm}]
\begin{equation}
{\tt RAUX\_H(220)} = d^E_d/e = (d^E_d/e)^{\tilde{\chi}^\pm}
+(d^E_d/e)^{\tilde{\chi}^0}+(d^E_d/e)^{\tilde{g}}
+(d^E_d/e)^{H} \,,
\end{equation}
where the constituent contributions are
\begin{eqnarray}
{\tt RAUX\_H(221)} &=&(d^E_d/e)^{\tilde{\chi}^\pm}\,, \ \ \
{\tt RAUX\_H(222)}  = (d^E_d/e)^{\tilde{\chi}^0}\, \nonumber \\
{\tt RAUX\_H(223)} &=&(d^E_d/e)^{\tilde{g}}\,, \ \ \ \ \
{\tt RAUX\_H(224)}  = (d^E_d/e)^{H}\,.
\end{eqnarray}
\item{The electric EDM of the $s$ quark in units of ${\rm cm}$}:
%      I_DEOE_S     =230 ! d^E_s/e [{\rm cm}]
\begin{equation}
{\tt RAUX\_H(230)} = d^E_s/e = (d^E_s/e)^{\tilde{\chi}^\pm}
+(d^E_s/e)^{\tilde{\chi}^0}+(d^E_s/e)^{\tilde{g}}
+(d^E_s/e)^{H} \,,
\end{equation}
where the constituent contributions are
\begin{eqnarray}
{\tt RAUX\_H(231)} &=&(d^E_s/e)^{\tilde{\chi}^\pm}\,, \ \ \
{\tt RAUX\_H(232)}  = (d^E_s/e)^{\tilde{\chi}^0}\, \nonumber \\
{\tt RAUX\_H(233)} &=&(d^E_s/e)^{\tilde{g}}\,, \ \ \ \ \
{\tt RAUX\_H(234)}  = (d^E_s/e)^{H}\,.
\end{eqnarray}
\item{The chromo-electric EDM of the $u$ quark in units of ${\rm cm}$}:
%      I_DC_U       =240 ! d^C_u   [{\rm cm}]
\begin{equation}
{\tt RAUX\_H(240)} = d^C_u = (d^C_u)^{\tilde{\chi}^\pm}
+(d^C_u)^{\tilde{\chi}^0}+(d^C_u)^{\tilde{g}}
+(d^C_u)^{H} \,,
\end{equation}
where the constituent contributions are
\begin{eqnarray}
{\tt RAUX\_H(241)} &=&(d^C_u)^{\tilde{\chi}^\pm}\,, \ \ \
{\tt RAUX\_H(242)}  = (d^C_u)^{\tilde{\chi}^0}\, \nonumber \\
{\tt RAUX\_H(243)} &=&(d^C_u)^{\tilde{g}}\,, \ \ \ \ \
{\tt RAUX\_H(244)}  = (d^C_u)^{H}\,.
\end{eqnarray}
\item{The chromo-electric EDM of the $d$ quark in units of ${\rm cm}$}:
%      I_DC_D       =250 ! d^C_d   [{\rm cm}]
\begin{equation}
{\tt RAUX\_H(250)}  =  d^C_d = (d^C_d)^{\tilde{\chi}^\pm}
+(d^C_d)^{\tilde{\chi}^0}+(d^C_d)^{\tilde{g}}
+(d^C_d)^{H} \,,
\end{equation}
where the constituent contributions are
\begin{eqnarray}
{\tt RAUX\_H(251)} &=&(d^C_d)^{\tilde{\chi}^\pm}\,, \ \ \
{\tt RAUX\_H(252)}  = (d^C_d)^{\tilde{\chi}^0}\, \nonumber \\
{\tt RAUX\_H(253)} &=&(d^C_d)^{\tilde{g}}\,, \ \ \ \ \
{\tt RAUX\_H(254)}  = (d^C_d)^{H}\,.
\end{eqnarray}
\item{The purely gluonic dimension-six Weinberg operator in units of 
${\rm cm}/{\rm GeV}$}:
%      I_DG_WEINBERG=260 ! d^G     [{\rm cm}/GeV]
\begin{equation}
{\tt RAUX\_H(260)}  =  d^G = (d^G)^{H} +(d^G)^{\tilde{g}} \,,
\end{equation}
where the constituent contributions are
\begin{eqnarray}
{\tt RAUX\_H(261)} &=&(d^G)^{H}\,, \ \ \ \ \
{\tt RAUX\_H(262)}  = (d^G)^{\tilde{g}}\,.
\end{eqnarray}
In the distributed version,
Eq.~(\ref{eq:zqh}) is used to evaluate
the function $H(z_1,z_2,z_q)$ for $(d^G)^{\tilde{g}}$. For a full
calculation, especially when $z_q\gsim 0.1$, the user should provide a dedicated routine, see
Fig.~\ref{fig:zqh}.
\item{$C_S$, $C_P$, and, $C_P^\prime$ in units of ${\rm cm}/{\rm GeV}$}:
%      I_CSPP       =270 ! C_S, C_P C'_P [{\rm cm}/GeV]
\begin{eqnarray}
{\tt RAUX\_H(270)} &=&C_S\,, \ \ \ 
{\tt RAUX\_H(271)}  = C_P\,, \ \ \ 
{\tt RAUX\_H(272)}  = C_P^\prime\,.
\end{eqnarray}
\item{The coefficients of four-fermion operators in units of ${\rm cm}/{\rm GeV}^2$}:
%      I_C4FOM      =280 ! C4_ff'/m_f(') [{\rm cm}/GeV^2]: needs 20 slots
\begin{eqnarray}
{\tt RAUX\_H(280)} &=&C_{de}/m_d\,, \ \ \ 
{\tt RAUX\_H(281)}  = C_{se}/m_s\,, \nonumber \\
{\tt RAUX\_H(282)} &=&C_{ed}/m_d\,, \ \ \ 
{\tt RAUX\_H(283)}  = C_{es}/m_s\,, \ \ \
{\tt RAUX\_H(284)}  = C_{eb}/m_b\,, \nonumber \\
{\tt RAUX\_H(285)} &=&C_{ec}/m_c\,, \ \ \ 
{\tt RAUX\_H(286)}  = C_{et}/m_t\,, \nonumber \\
{\tt RAUX\_H(287)} &=&C_{dd}/m_d\,, \ \ \ 
{\tt RAUX\_H(288)}  = C_{sd}/m_s\,, \nonumber \\
{\tt RAUX\_H(289)} &=&C_{bd}/m_b\,, \ \ \ 
{\tt RAUX\_H(290)}  = C_{db}/m_b\,.
\end{eqnarray}
\item{The Thallium EDM in units of $e\,{\rm cm}$}:
%      I_TL         =300 ! d^Tl [e cm]
\begin{equation}
{\tt RAUX\_H(300)}  = d_{\rm Tl} = d_{\rm Tl}(d^E_e) + d_{\rm Tl}(C_S)
\end{equation}
where the constituent contributions are
\begin{eqnarray}
{\tt RAUX\_H(301)} &=&d_{\rm Tl}(d^E_e)\,, \ \ \ 
{\tt RAUX\_H(302)}  = d_{\rm Tl}(C_S)\,. \nonumber 
\end{eqnarray}
\item{The neutron EDM in units of $e\,{\rm cm}$}:
\begin{itemize}
\item{Chiral quark model}:
%      I_N1         =310 ! d^n  [e cm]: Chiral Quark Model
\begin{equation}
{\tt RAUX\_H(310)}  = d_n = d_n(d^E_{u,d}) + d_n(d^C_{u,d}) + d_n(d^G)
\end{equation}
where the constituent contributions are
\begin{eqnarray}
{\tt RAUX\_H(311)} &=&d_n(d^E_{u,d})\,, \ \ \
{\tt RAUX\_H(312)}  = d_n(d^C_{u,d})\,. \ \ \
{\tt RAUX\_H(313)}  = d_n(d^G)\,. \nonumber
\end{eqnarray}
\item{Parton quark model}:
%      I_N2         =320 ! d^n  [e cm]: Parton Quark Model
\begin{equation}
{\tt RAUX\_H(320)}  = d_n = d_n(d^E_u) + d_n(d^E_d) + d_n(d^E_s)
\end{equation}
where the constituent contributions are
\begin{eqnarray}
{\tt RAUX\_H(321)} &=&d_n(d^E_u)\,, \ \ \
{\tt RAUX\_H(322)}  = d_n(d^E_d)\,. \ \ \
{\tt RAUX\_H(323)}  = d_n(d^E_s)\,. \nonumber
\end{eqnarray}
\item{QCD sum rule approach}:
%      I_N3         =330 ! d^n  [e cm]: QCD Sum Rule Technique
\begin{equation}
{\tt RAUX\_H(330)}  = d_n = d_n(d^E_{u,d}) + d_n(d^C_{u,d}) + d_n(d^G) + d_n(C_{bd,db})
\end{equation}
where the constituent contributions are
\begin{eqnarray}
{\tt RAUX\_H(331)} &=&d_n(d^E_{u,d})\,, \ \ \
{\tt RAUX\_H(332)}  = d_n(d^C_{u,d})\,, \nonumber \\
{\tt RAUX\_H(332)} &=&d_n(d^G)\,, \ \ \ \ \,
{\tt RAUX\_H(334)}  = d_n(C_{bd,bd})\,. \nonumber
\end{eqnarray}
\end{itemize}
\item{The Mercury EDM in units of $e\,{\rm cm}$}:
%      I_HG         =340 ! d^Hg [e cm]
\begin{equation}
{\tt RAUX\_H(340)}  = d_{\rm Hg} = d_{\rm Hg}(d^E_e) + d_{\rm Hg}(d^C_{u,d}) +
d_{\rm Hg}(C_{4f}) + d_{\rm Hg}(C_S) + d_{\rm Hg}(C^{(\prime)}_P)
\end{equation}
where the constituent contributions are
\begin{eqnarray}
{\tt RAUX\_H(341)} &=&d_{\rm Hg}(d^E_e)\,, \ \ \
{\tt RAUX\_H(342)}  = d_{\rm Hg}(d^C_{u,d})\,, \nonumber \\
{\tt RAUX\_H(343)} &=&d_{\rm Hg}(C_{4f})\,, \ \ 
{\tt RAUX\_H(344)}  = d_{\rm Hg}(C_S)\,. \ \ \
{\tt RAUX\_H(345)}  = d_{\rm Hg}(C^{(\prime)}_P)\,. \nonumber
\end{eqnarray}
\item{The deuteron EDM in units of $e\,{\rm cm}$}:
%      I_DEUT       =350 ! d^D  [e cm]
\begin{equation}
{\tt RAUX\_H(350)}  = d_D = d_D(d^E_{u,d}) + d_D(d^C_{u,d}) + d_n(C_{4f}) + d_n(d^G)
\end{equation}
where the constituent contributions are
\begin{eqnarray}
{\tt RAUX\_H(351)} &=&d_D(d^E_{u,d})\,, \ \ \
{\tt RAUX\_H(352)}  = d_D(d^C_{u,d})\,, \nonumber \\
{\tt RAUX\_H(353)} &=&d_D(C_{4f})\,, \ \ \ \,
{\tt RAUX\_H(354)}  = d_D(d^G)\,. \nonumber
\end{eqnarray}
\end{itemize}

\item {\tt IFLAG\_H(18)}=1 
is used to print out the Thallium, neutron, Mercury and deuteron EDMs. 
Using the {\tt run} shell-script file distributed,
the sample output obtained with
$\Phi_{A_e}=\Phi_{A_\tau}$,
$\Phi_{A_u}=\Phi_{A_c}=\Phi_{A_t}$, and
$\Phi_{A_d}=\Phi_{A_s}=\Phi_{A_b}$,
and the hierarchy factor 
$\rho_{\tilde{Q}}=\rho_{\tilde{U}}=\rho_{\tilde{D}}=
\rho_{\tilde{L}}=\rho_{\tilde{E}}=1$ is \\ \\
{\tt
~---------------------------------------------------------\\
$~~~~~~~~$Thallium~EDM~in~units~of~[e~cm]:~d\^{}Tl/[e~cm]\\
~---------------------------------------------------------\\
$~~$d\^{}Tl/(e~cm)~[Total]=~-.9371E-24\\
$~~$Each~contribution~to~d\^{}Tl~from\\
$~~~~~~~~~~~~~~$[d\^{}E\_e]=~-.9082E-24\\
$~~~~~~~~~~~~~~$[C\_S~~]=~-.2890E-25\\
~---------------------------------------------------------\\
$~~~~~~~~$Neutron~~EDM~in~units~of~[e~cm]:~d\^{}n/[e~cm]\\
~---------------------------------------------------------\\
$~$(1)~Chiral~Quark~Model\\
$~~$d\^{}n/(e~cm)~~[Total]=~-.4196E-23\\
$~~$Each~contribution~to~d\^{}n~from\\
$~~~~~~~~~~~~~$[d\^{}E\_u~\&~d\^{}E\_d]=~-.3994E-23\\
$~~~~~~~~~~~~~$[d\^{}C\_u~\&~d\^{}C\_d]=~-.2066E-23\\
$~~~~~~~~~~~~~$[~Weinberg-6D~]=~0.1863E-23\\
\\
$~$(2)~Parton~Quark~Model\\
$~~$d\^{}n/(e~cm)~~[Total]=~0.1278E-22\\
$~~$Each~contribution~to~d\^{}n~from\\
$~~~~~~~~~~~~~$[d\^{}E\_u~~~~~~~~]=~-.6868E-25\\
$~~~~~~~~~~~~~$[d\^{}E\_d~~~~~~~~]=~-.2209E-23\\
$~~~~~~~~~~~~~$[d\^{}E\_s~~~~~~~~]=~0.1506E-22\\
\\
$~$(3)~QCD~sum~rule~technique\\
$~~$d\^{}n/(e~cm)~~[Total]=~-.7240E-23\\
$~~$Each~contribution~to~d\^{}n~from\\
$~~~~~~~~~~~~~$[d\^{}E\_u~\&~d\^{}E\_d]=~-.2741E-23\\
$~~~~~~~~~~~~~$[d\^{}C\_u~\&~d\^{}C\_d]=~-.5483E-23\\
$~~~~~~~~~~~~~$[~Weinberg-6D~]=~0.9836E-24\\
$~~~~~~~~~~~~~$[~C\_bd~\&~C\_db~]=~0.1749E-28\\
~---------------------------------------------------------\\
$~~~~~~~~$Mercury~EDM~in~units~of~[e~cm]:~d\^{}Hg/[e~cm]\\
~---------------------------------------------------------\\
$~~$d\^{}Hg/(e~cm)~[Total]=~0.3383E-25\\
$~~$Each~contribution~to~d\^{}Hg~from\\
$~~~~~~~~~~~~~$[d\^{}E\_e~~~~~~~~]=~0.1553E-28\\
$~~~~~~~~~~~~~$[d\^{}C\_u~\&~d\^{}C\_d]=~0.3381E-25\\
$~~~~~~~~~~~~~$[C\_4f~~~~~~~~~]=~-.1094E-29\\
$~~~~~~~~~~~~~$[C\_S~~~~~~~~~~]=~0.2348E-29\\
$~~~~~~~~~~~~~$[C\_P~\&~C\_P\^{}pr~]=~0.2229E-29\\
~---------------------------------------------------------\\
$~~~~~~~~$Deuteron~EDM~in~units~of~[e~cm]:~d\^{}D/[e~cm]\\
~---------------------------------------------------------\\
$~~$d\^{}D/(e~cm)~~[Total]=~-.2598E-22\\
$~~$Each~contribution~to~d\^{}D~from\\
$~~~~~~~~~~~~~$[d\^{}E\_u~\&~d\^{}E\_d]=~-.9236E-24\\
$~~~~~~~~~~~~~$[d\^{}C\_u~\&~d\^{}C\_d]=~-.2604E-22\\
$~~~~~~~~~~~~~$[C\_4f~~~~~~~~~]=~0.7811E-27\\
$~~~~~~~~~~~~~$[~Weinberg-6D~]=~0.9836E-24\\
~---------------------------------------------------------\\
} % {\tt
%------------------------------------------------------
\item {\tt IFLAG\_H(18)}=2
is used to print out the EDMs of the electron and the up, down, and 
strange quarks, the CEDMs of the up, down, and strange quarks, etc.\\
{\tt
$~$---------------------------------------------------------\\
$~$The~Electric~EDMs~of~particles~in~cm:~e,~u,~d,~s:\\
$~$---------------------------------------------------------\\
$~~~~~$d\^{}E\_e/e[Total]:~~0.1553E-26\\
$~~~~~$d\^{}E\_u/e[Total]:~~0.8836E-25\\
$~~~~~$d\^{}E\_d/e[Total]:~~-.1936E-23\\
$~~~~~$d\^{}E\_s/e[Total]:~~-.4355E-22\\
$~~$d\^{}E\_e/e[C,N,Gl,H]:~~0.0000E+00~~-.2833E-26~~0.0000E+00~~0.4386E-26\\
$~~$d\^{}E\_u/e[C,N,Gl,H]:~~0.4467E-28~~0.1755E-26~~0.8717E-25~~-.6115E-27\\
$~~$d\^{}E\_d/e[C,N,Gl,H]:~~0.9825E-25~~-.3481E-26~~-.2042E-23~~0.1170E-25\\
$~~$d\^{}E\_s/e[C,N,Gl,H]:~~0.2211E-23~~-.7832E-25~~-.4595E-22~~0.2631E-24\\
$~$---------------------------------------------------------\\
$~$The~Chromo-Electric~EDMs~of~particles~in~cm:~u,~d:\\
$~$---------------------------------------------------------\\
$~~~~~$d\^{}C\_u~~[Total]:~~-.1208E-24\\
$~~~~~$d\^{}C\_d~~[Total]:~~-.5756E-23\\
$~~$d\^{}C\_u~~[C,N,Gl,H]:~~-.8956E-28~~0.3073E-26~~-.1180E-24~~-.5825E-26\\
$~~$d\^{}C\_d~~[C,N,Gl,H]:~~0.1626E-25~~0.1218E-25~~-.5554E-23~~-.2306E-24\\
$~$---------------------------------------------------------\\
$~$Purely-gluonic~D-6~Weinberg~operator~in~cm/GeV:\\
$~$---------------------------------------------------------\\
$~~$d\^{}G~~~~~~~[Total]:~~0.5786E-23\\
$~~$d\^{}G[Higgs,Gluino]:~~0.2046E-26~~0.5784E-23\\
$~$---------------------------------------------------------\\
$~~$Four-fermion~couplings~needed~for~EDMs~in~cm/GeV\^{}2:\\
$~$---------------------------------------------------------\\
$~~$C4\_de/m\_d:~~~-.2432E-26\\
$~~$C4\_se/m\_s:~~~-.2432E-26\\
$~~$C4\_ed/m\_d:~~~-.2432E-26\\
$~~$C4\_es/m\_s:~~~-.2432E-26\\
$~~$C4\_eb/m\_b:~~~-.1547E-25\\
$~~$C4\_ec/m\_c:~~~-.9727E-28\\
$~~$C4\_et/m\_t:~~~0.3965E-27\\
$~~$C4\_dd/m\_d:~~~-.3565E-25\\
$~~$C4\_sd/m\_s:~~~-.3565E-25\\
$~~$C4\_bd/m\_b:~~~0.1769E-24\\
$~~$C4\_db/m\_b:~~~-.2269E-24\\
$~$---------------------------------------------------------\\
$~~~~$C\_S,~C\_P,~and~C\_P\^{}prime~in~cm/GeV~and~in~1/GeV\^{}2:\\
$~$---------------------------------------------------------\\
$~~$C\_S~~~~~~:~~~0.6710E-27~~0.3400E-13\\
$~~$C\_P~~~~~~:~~~0.6602E-26~~0.3346E-12\\
$~~$C\_P\^{}prime:~~~0.5147E-26~~0.2608E-12\\
$~$---------------------------------------------------------\\
} % {\tt
\end{itemize}

\end{appendix}

%%%%%%%%%%%%%%%%%%%%%%%%%%%%%%%%%%%%%%%%%%%%%%%%%%%%%%%%%%%%%%%%%%%%%%%%%

\newpage 

% Figures...
%
\begin{figure}[htb]
\hspace{ 0.0cm}
\vspace{-0.5cm}
\centerline{\epsfig{figure=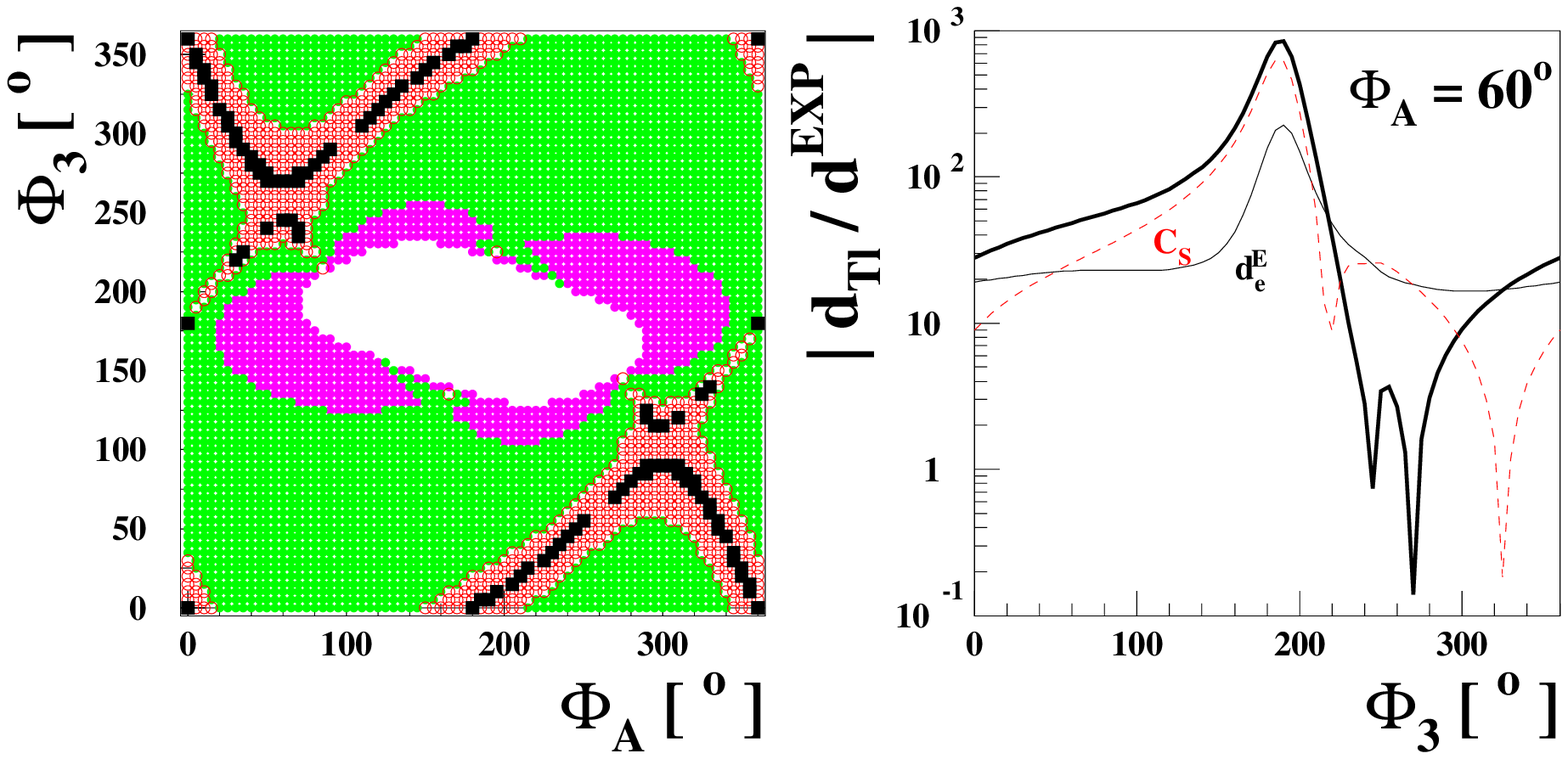,height=16.0cm,width=16.0cm}}
\vspace{-7.5cm}
\caption{\it The absolute value of the Thallium EDM divided by its current experimetal
limit, $d_{\rm Tl}^{\rm EXP}=9\times 10^{-25}\,{\rm e~cm}$, in the $\Phi_3$-$\Phi_A$ plane
(left) and as a functon of $\Phi_3$ taking $\Phi_A=60^\circ$ (right). The trimixing
scenario has been taken. In the left frame, the
plane is divided into 4 regions: $|d_{\rm Tl}/d^{\rm EXP}|<1$ (black),
$1\leq |d_{\rm Tl}/d^{\rm EXP}|<10$ (red),
$10\leq |d_{\rm Tl}/d^{\rm EXP}|<100$ (green), and
$100\leq |d_{\rm Tl}/d^{\rm EXP}|$ (magenta). The unshaded region is not
allowed theoretically. In the right frame, the constituent contributions from
$d^E_e$ and $C_S$ are shown as the thin solid and dashed lines, respectively. 
The thick solid line is for the total EDM.}
\label{fig:tledm.tri}
\end{figure}
\begin{figure}[htb]
\hspace{ 0.0cm}
\vspace{-0.5cm}
\centerline{\epsfig{figure=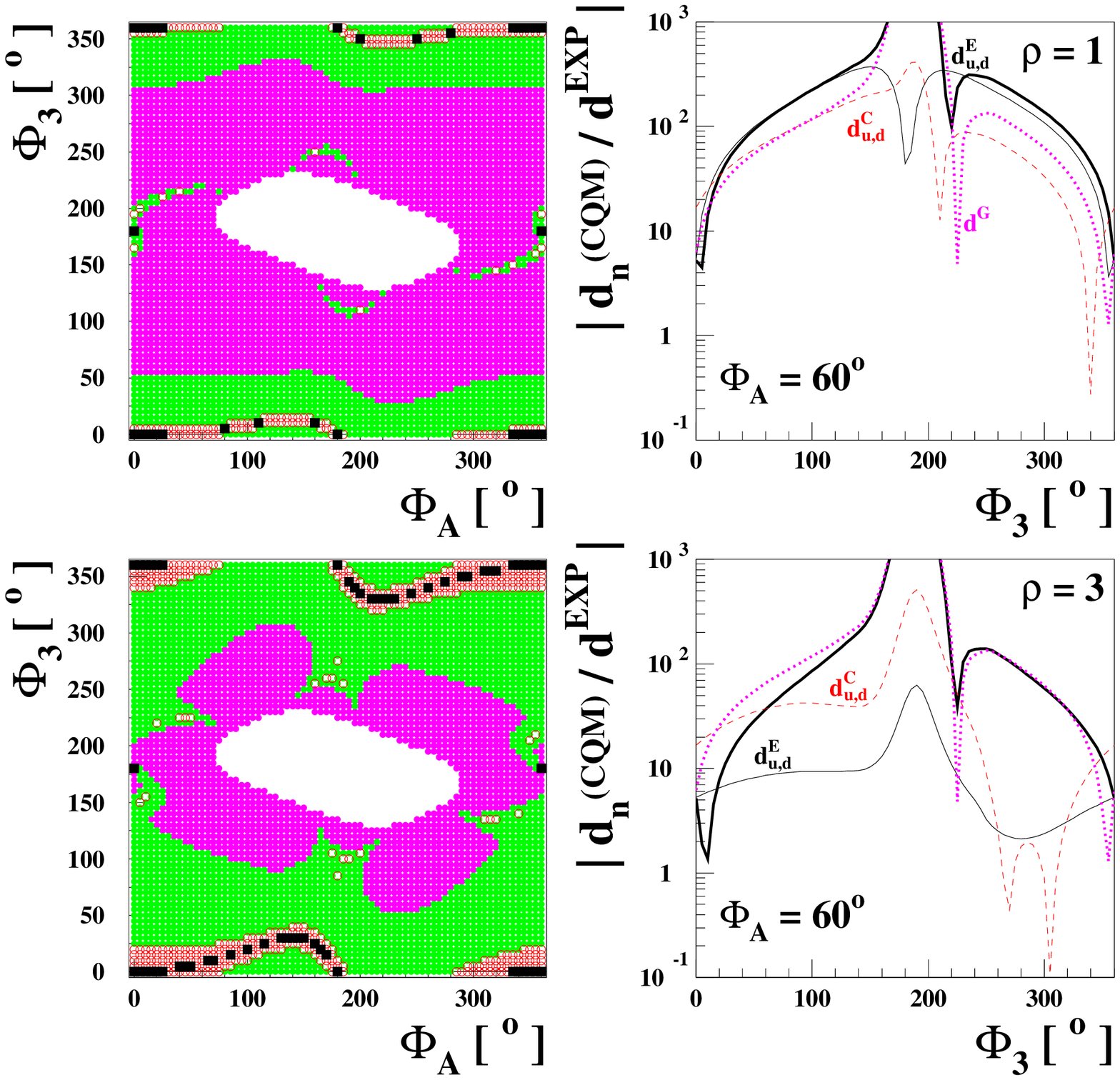,height=16.0cm,width=16.0cm}}
\vspace{-0.5cm}
\caption{\it The absolute value of  the neutron EDM in the CQM divided
by  its  current   experimetal  limit,  $d_{\rm  n}^{\rm  EXP}=3\times
10^{-26}\,{\rm e~cm}$, in the  $\Phi_3$-$\Phi_A$ plane (left) and as a
function of $\Phi_3$  taking $\Phi_A=60^\circ$ (right).  The trimixing
scenario  has been  taken  with the  common  hierachy factor  $\rho=1$
(upper) and $3$  (lower).  In the left frames,  the shaded regions are
the  same as  in Fig.~\ref{fig:tledm.tri}.  In the  right  frames, the
constituent contributions from $d^E_{u,d}$, $d^C_{u,d}$, and $d^G$ are
shown  in the  thin solid,  dashed, dotted  lines,  respectively.  The
thick solid line is for the total EDM.}
\label{fig:n1edm.tri}
\end{figure}
\begin{figure}[htb]
\hspace{ 0.0cm}
\vspace{-0.5cm}
\centerline{\epsfig{figure=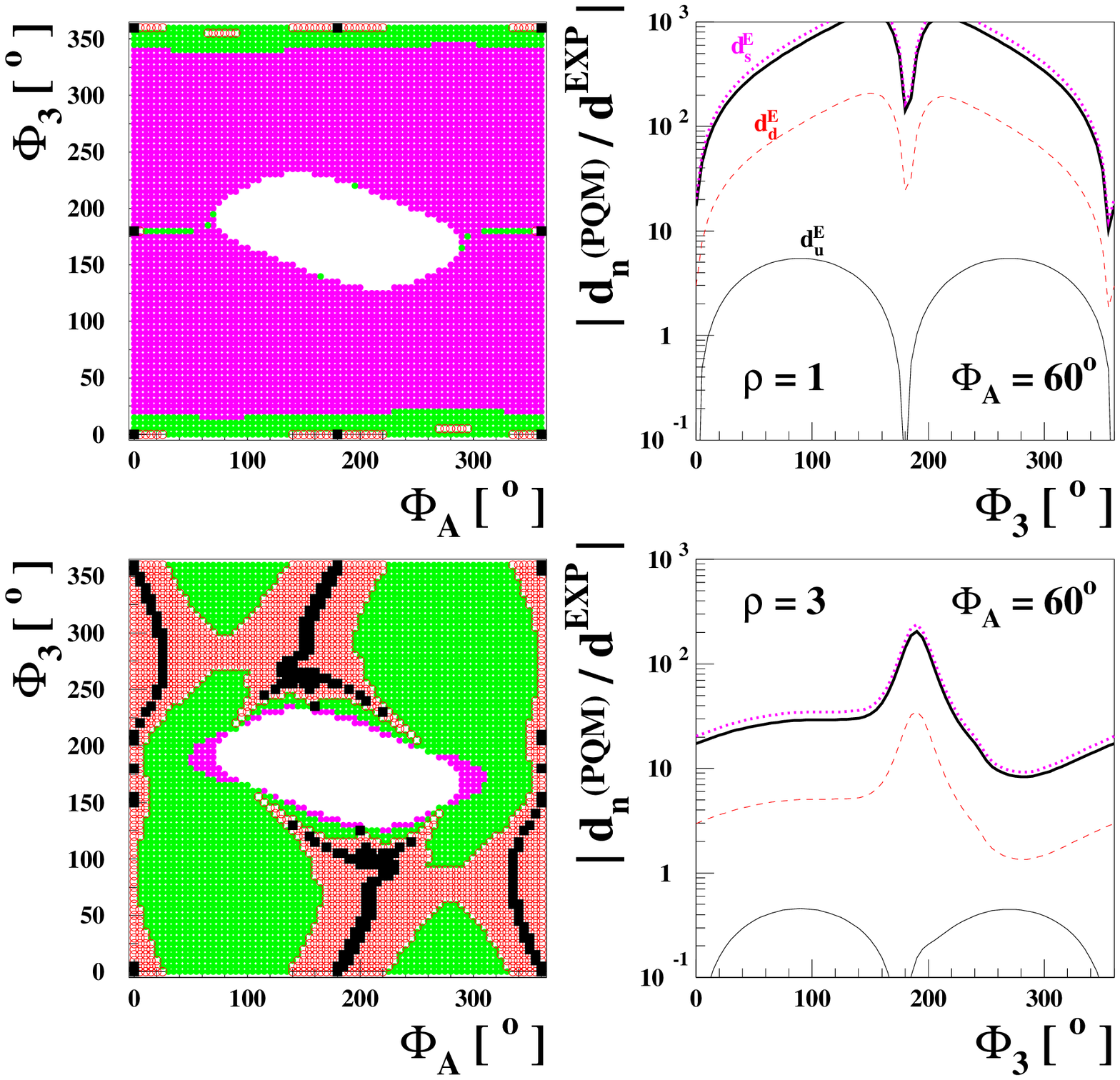,height=16.0cm,width=16.0cm}}
\vspace{-0.5cm}
\caption{\it The  same as  in Fig.~\ref{fig:n1edm.tri}, but  using the
PQM  for  the  calculation.  In  the  right  frames,  the  constituent
contributions from $d^E_u$, $d^E_d$, and $d^E_s$ are shown as the thin
solid, dashed,  dotted lines, respectively.   The thick solid  line is
for the total EDM.}
\label{fig:n2edm.tri}
\end{figure}
\begin{figure}[htb]
\hspace{ 0.0cm}
\vspace{-0.5cm}
\centerline{\epsfig{figure=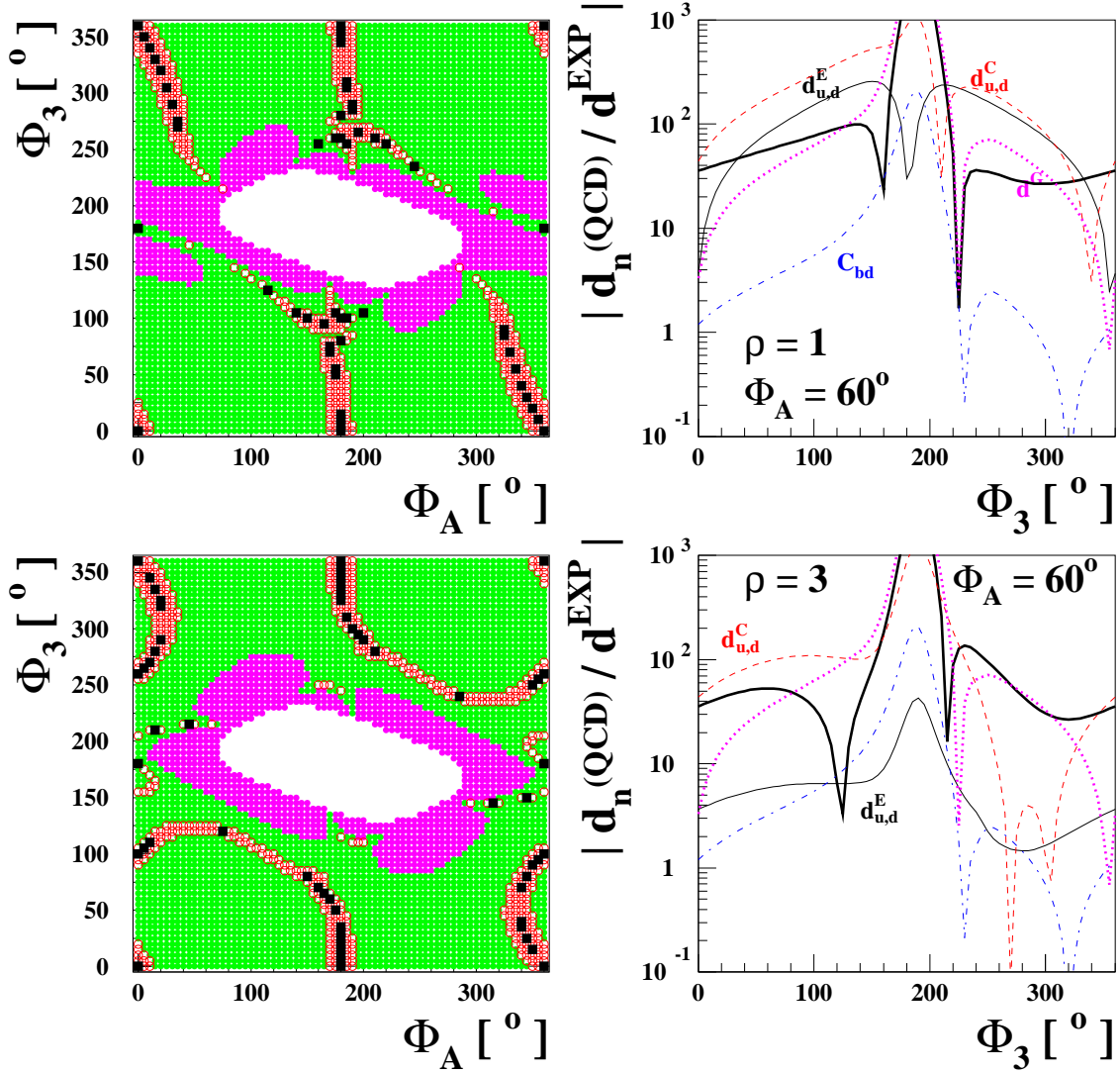,height=16.0cm,width=16.0cm}}
\vspace{-0.5cm}
\caption{\it The same as in Fig.~\ref{fig:n1edm.tri} but using the QCD sum rule approach
for the calculation. In the right frames, the
constituent contributions from $d^E_{u,d}$, $d^C_{u,d}$, $d^G$, and $C_{bd,db}$
are shown as the thin solid, dashed, dotted, dash-dotted lines, respectively.
The thick solid line is for the total EDM.}
\label{fig:n3edm.tri}
\end{figure}
\begin{figure}[htb]
\hspace{ 0.0cm}
\vspace{-0.5cm}
\centerline{\epsfig{figure=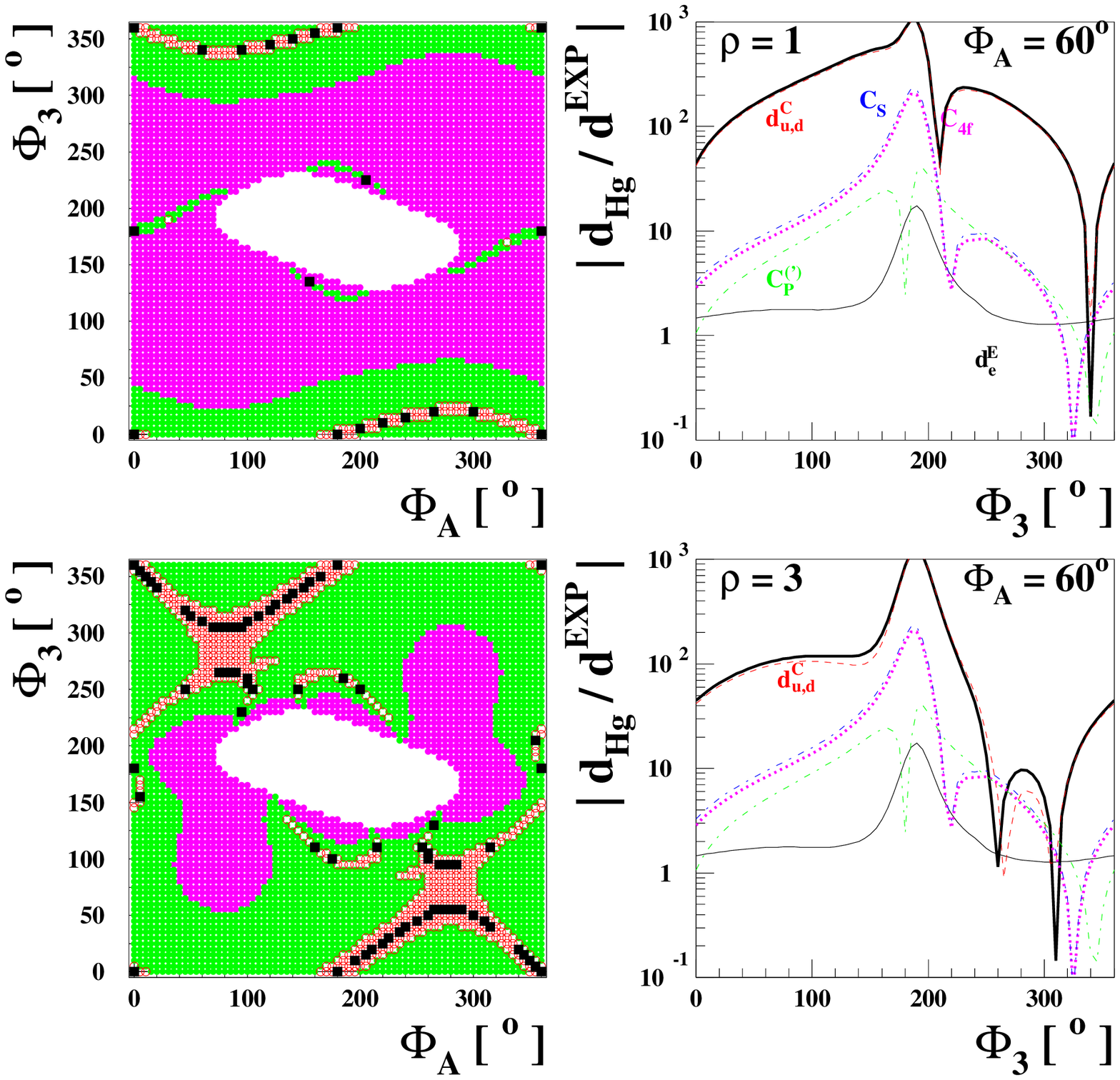,height=16.0cm,width=16.0cm}}
\vspace{-0.5cm}
\caption{\it  The absolute value of the Mercury EDM divided by its current experimetal
limit, $d_{\rm Hg}^{\rm EXP}=2\times 10^{-28}\,{\rm e~cm}$, on the $\Phi_3$-$\Phi_A$ plane
(left) and as a function of $\Phi_3$ taking $\Phi_A=60^\circ$ (right). 
The trimixing scenario has been taken with the common hierachy factor $\rho=1$ (upper) and 
$3$ (lower). In the left frames, the shaded
regions are the same as in Fig.~\ref{fig:tledm.tri}. In the right frames, the
constituent contributions from $d^E_e$, $d^C_{u,d}$, $C_{4f}\equiv C_{dd,sd,bd}$, and $C^{(\prime)}_{S,P}$
are shown as the thin solid, dashed, dotted, and dash-dotted lines, respectively.
The thick solid line is for the total EDM.}
\label{fig:hgedm.tri}
\end{figure}
\begin{figure}[htb]
\hspace{ 0.0cm}
\vspace{-0.5cm}
\centerline{\epsfig{figure=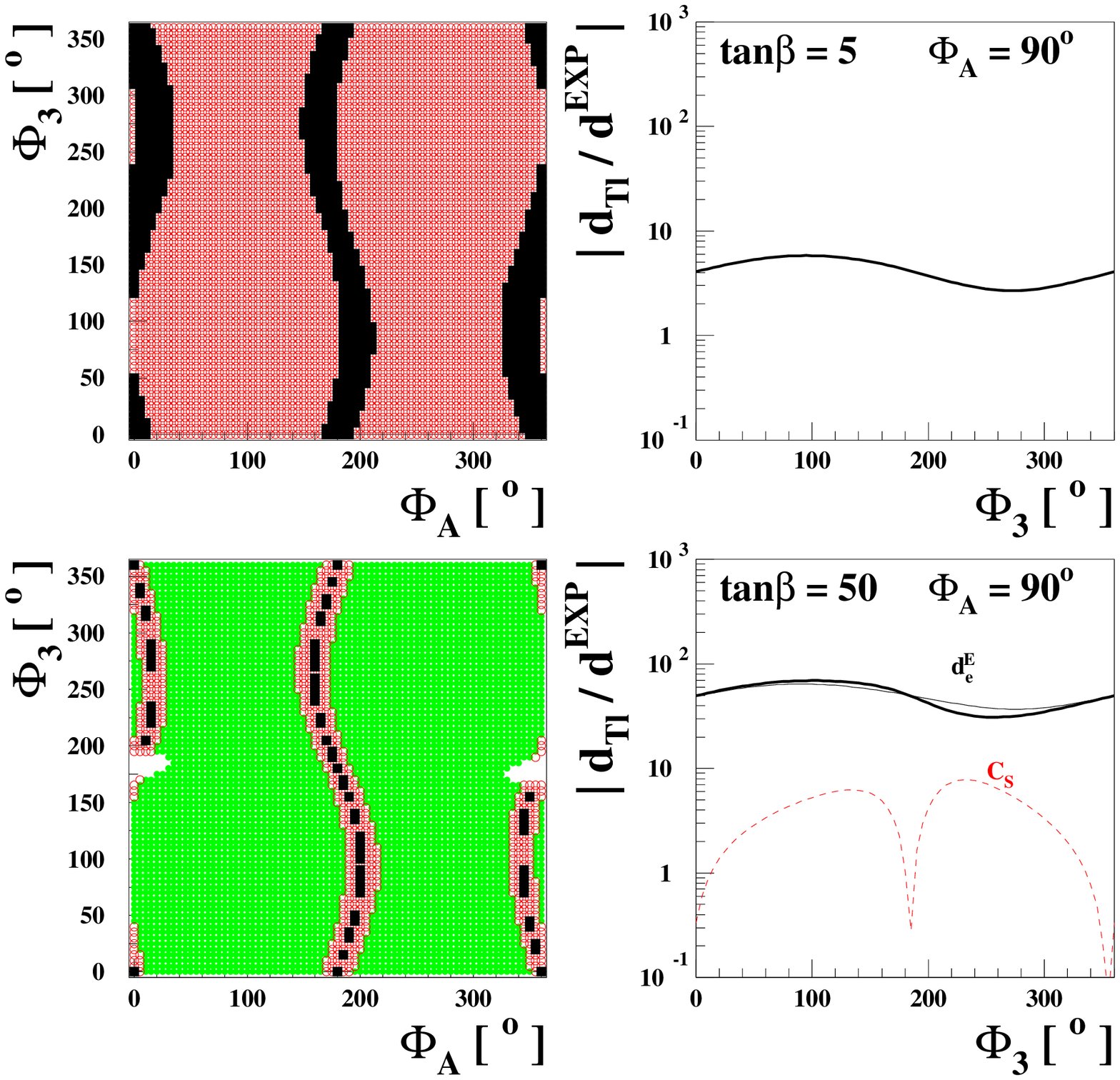,height=16.0cm,width=16.0cm}}
\vspace{-0.5cm}
\caption{\it The Thallium EDM in the CPX scenario. The upper frames are for $\tan\beta=5$
and the lower ones for $\tan\beta=50$ with $\Phi_A=90^\circ$. 
The shaded regions and lines are the same as in
Fig.~\ref{fig:tledm.tri}.}
\label{fig:tledm.cpx}
\end{figure}
\begin{figure}[htb]
\hspace{ 0.0cm}
\vspace{-0.5cm}
\centerline{\epsfig{figure=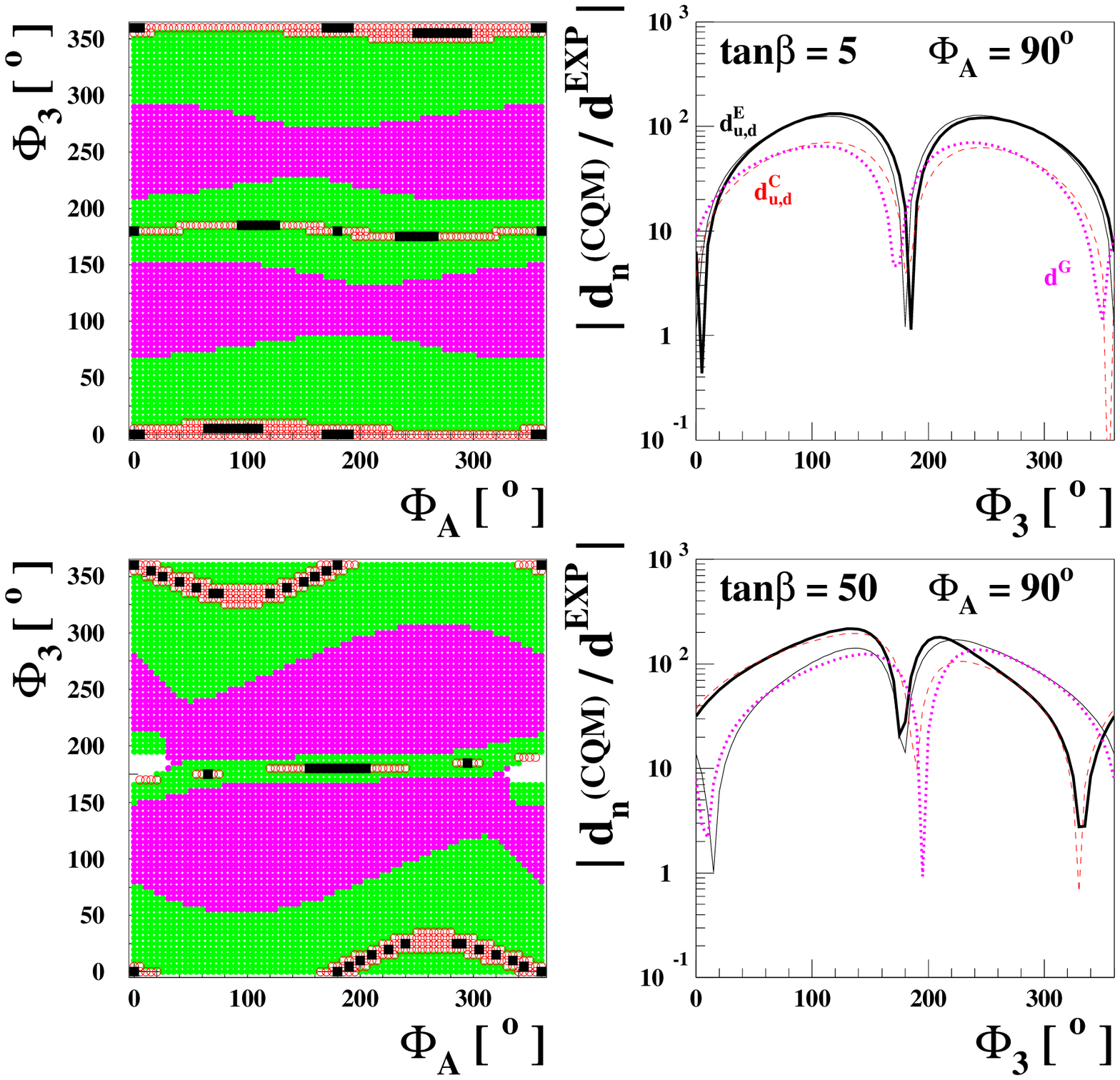,height=16.0cm,width=16.0cm}}
\vspace{-0.5cm}
\caption{\it The  neutron EDM  in the CPX  scenario calculated  in the
CQM. The  upper frames  are for $\tan\beta=5$  and the lower  ones for
$\tan\beta=50$ with  $\Phi_A=90^\circ$.  The shaded  regions and lines
are the same as in Fig.~\ref{fig:n1edm.tri}.}
\label{fig:n1edm.cpx}
\end{figure}
\begin{figure}[htb]
\hspace{ 0.0cm}
\vspace{-0.5cm}
\centerline{\epsfig{figure=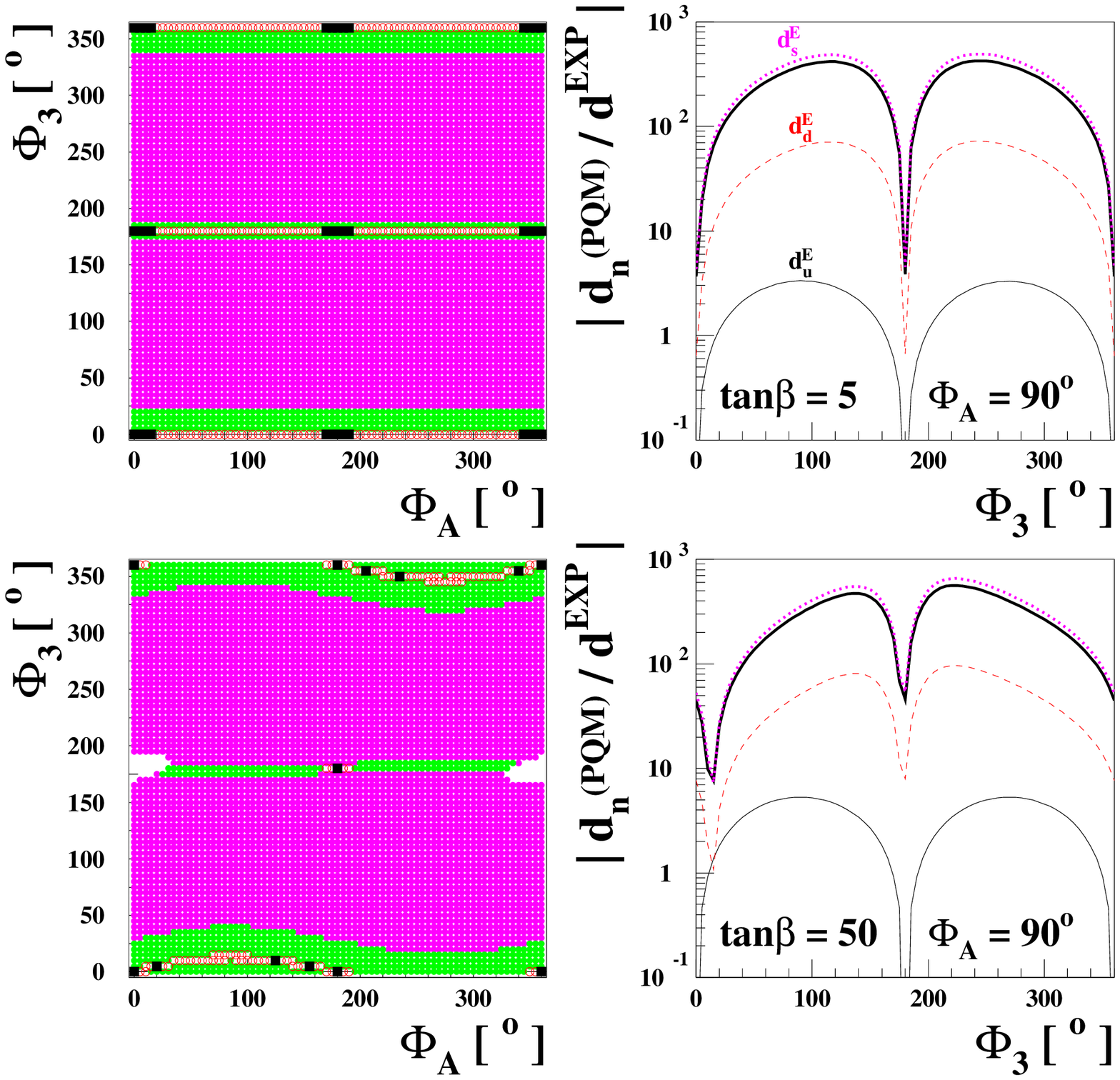,height=16.0cm,width=16.0cm}}
\vspace{-0.5cm}
\caption{\it The  neutron EDM  in the CPX  scenario calculated  in the
PQM. The  upper frames  are for $\tan\beta=5$  and the lower  ones for
$\tan\beta=50$ with  $\Phi_A=90^\circ$.  The shaded  regions and lines
are the same as in Fig.~\ref{fig:n2edm.tri}.}
\label{fig:n2edm.cpx}
\end{figure}
\begin{figure}[htb]
\hspace{ 0.0cm}
\vspace{-0.5cm}
\centerline{\epsfig{figure=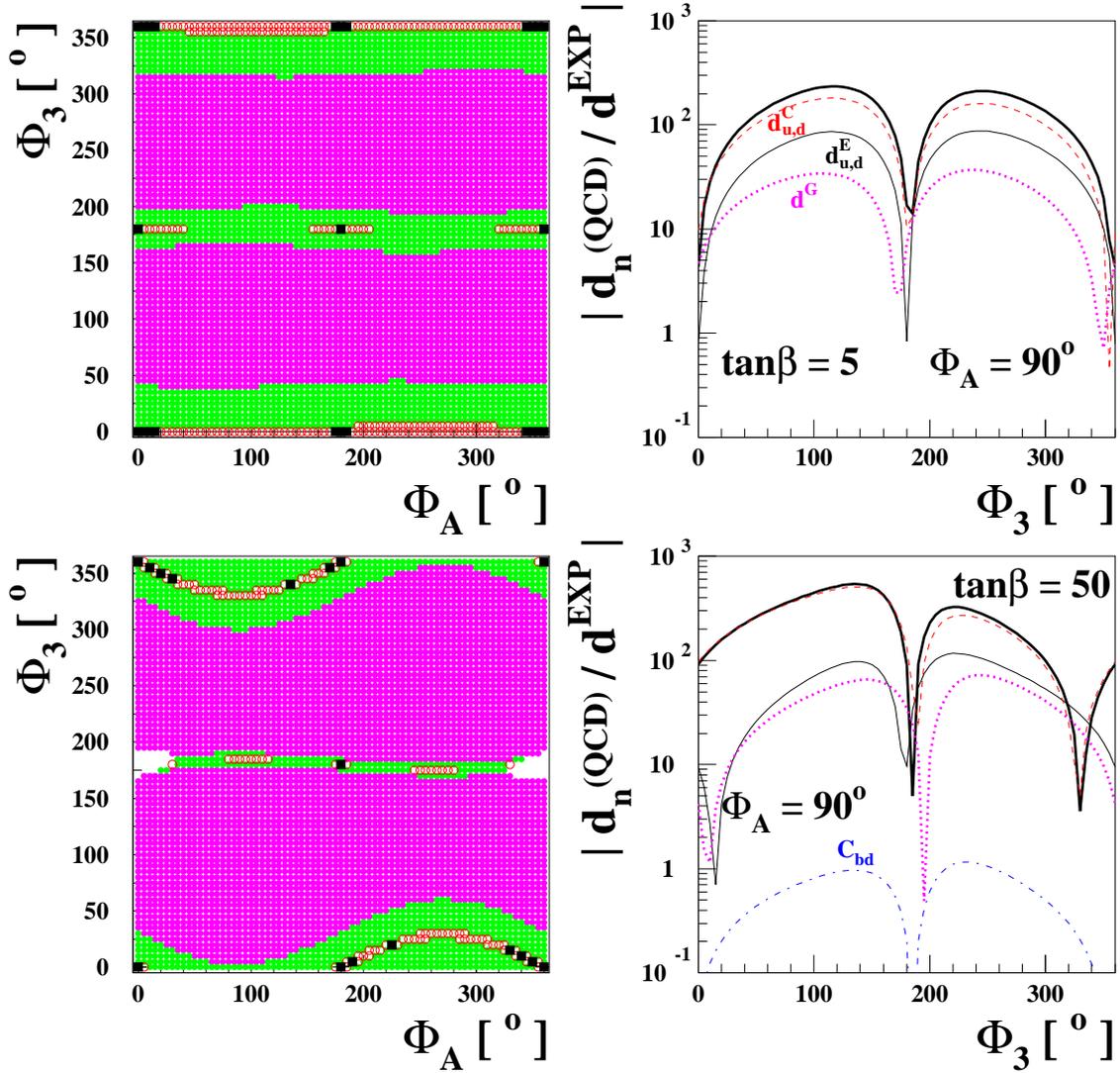,height=16.0cm,width=16.0cm}}
\vspace{-0.5cm}
\caption{\it The neutron EDM in the CPX scenario calculated using the QCD
sum rule approach. The upper frames are for $\tan\beta=5$
and the
lower ones for $\tan\beta=50$ with $\Phi_A=90^\circ$.  The shaded regions and lines are
the same as in Fig.~\ref{fig:n3edm.tri}.}
\label{fig:n3edm.cpx}
\end{figure}
\begin{figure}[htb]
\hspace{ 0.0cm}
\vspace{-0.5cm}
\centerline{\epsfig{figure=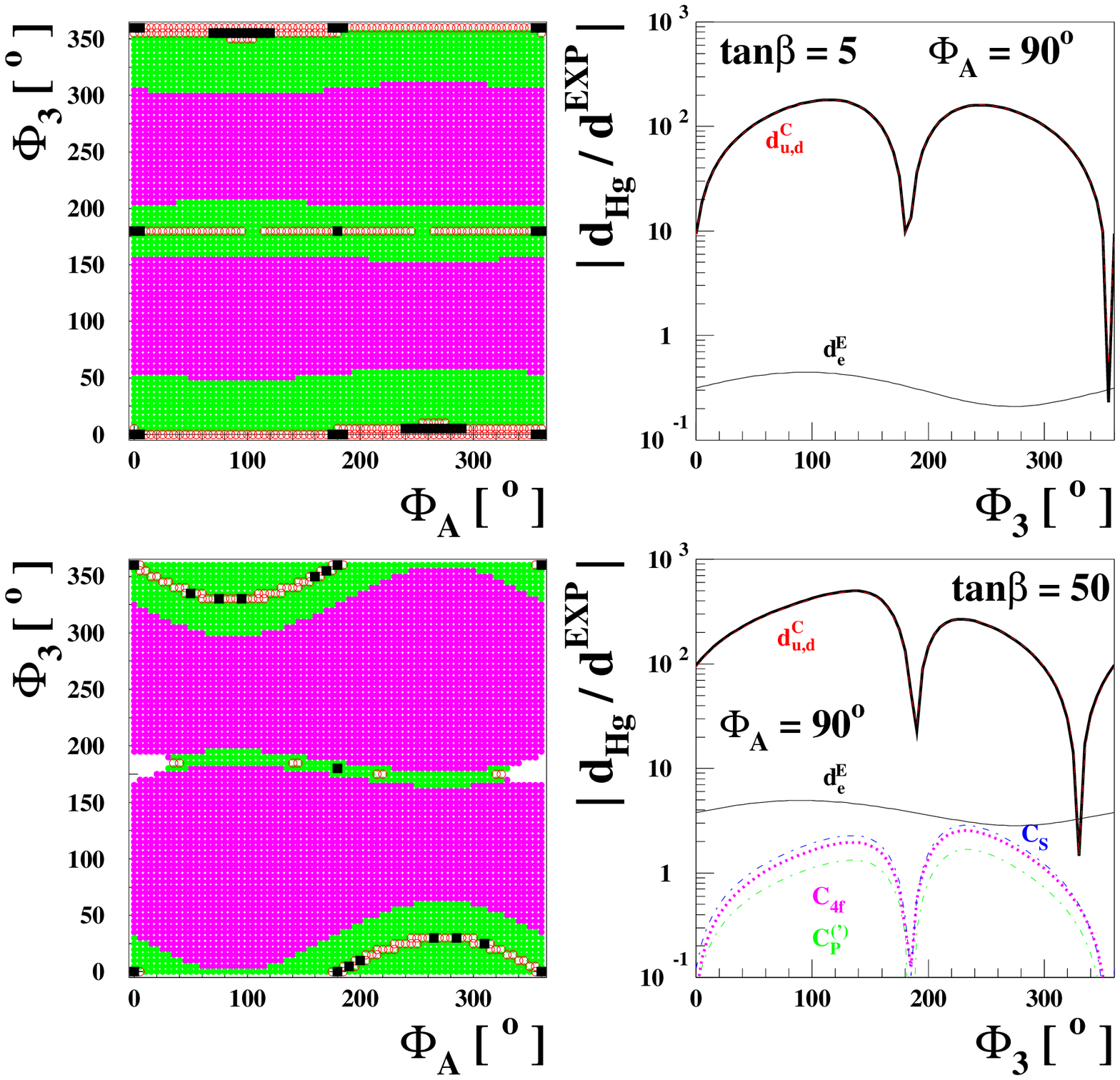,height=16.0cm,width=16.0cm}}
\vspace{-0.5cm}
\caption{\it The Mercury EDM in the CPX scenario.
The upper frames are for $\tan\beta=5$
and the
lower ones for $\tan\beta=50$ with $\Phi_A=90^\circ$.  The shaded regions and lines are
the same as in Fig.~\ref{fig:hgedm.tri}.}
\label{fig:hgedm.cpx}
\end{figure}
\begin{figure}[htb]
\hspace{ 0.0cm}
\vspace{-0.5cm}
\centerline{\epsfig{figure=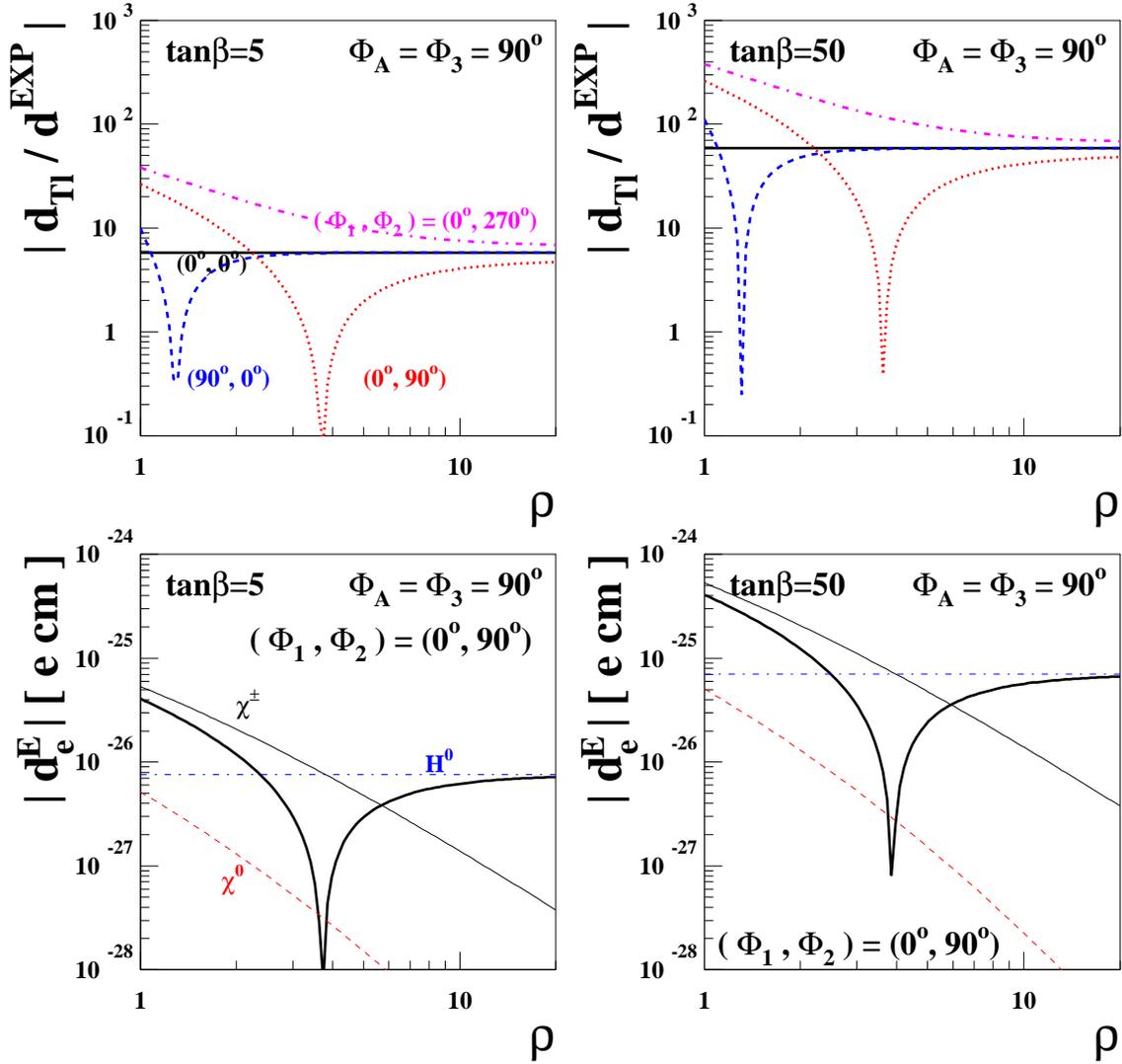,height=16.0cm,width=16.0cm}}
\vspace{-0.5cm}
\caption{\it In the upper frames, we show the Thallium EDM in the CPX scenario
as a function of the common hierarchy factor $\rho$ with several non-trivial
values of $(\Phi_1,\Phi_2)$: $(\Phi_1,\Phi_2)=(0^\circ,0^\circ)$ (solid),
$(90^\circ,0^\circ)$ (dashed),
$(0^\circ,90^\circ)$ (dotted), and
$(0^\circ,270^\circ)$ (dash-dotted).
The left frame is for
$\tan\beta=5$ and the right one for $\tan\beta=50$. The lower frames are for the electron
EDM, which makes the main contribution to the Thallium EDM, in the given scenario,
exemplifying the case with $(\Phi_1,\Phi_2)=(0^\circ,90^\circ)$ from each upper frame. 
Shown separately are
the different contributions to the electron EDM from the chargino- (thin solid), 
neutralino- (thin dashed), and two-loop Higgs- (thin dash-dotted) mediated diagrams.
The thick solid lines are for the total EDM.
We have taken $\Phi_A=\Phi_3=90^\circ$ in all frames.}
\label{fig:tledm.cpx.rho}
\end{figure}
\begin{figure}[htb]
\hspace{ 0.0cm}
\vspace{-0.5cm}
\centerline{\epsfig{figure=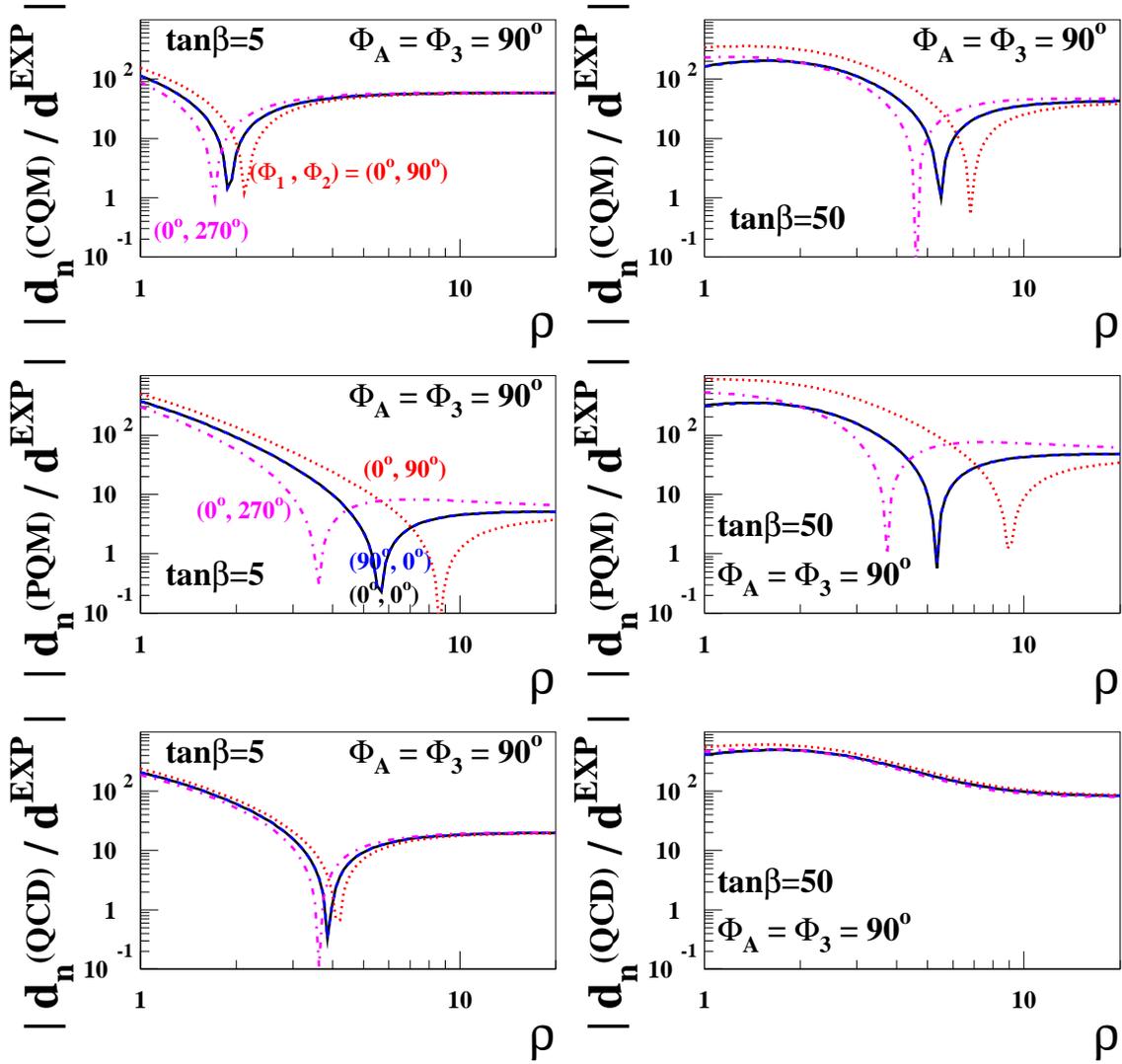,height=16.0cm,width=16.0cm}}
\vspace{-0.5cm}
\caption{\it   The   neutron   EDM    in   the   CPX   scenario   with
$\Phi_A=\Phi_3=90^\circ$ as a function  of the common hierarchy factor
$\rho$ calculated in the CQM  (upper), the PQM (middle), and using the
QCD   sum  rule   approach  (lower)   for  $\tan\beta=5$   (left)  and
$\tan\beta=50$ (right).  The cases  with several non-trivial values of
$(\Phi_1,\Phi_2)$  are considered as  in Fig.~\ref{fig:tledm.cpx.rho}.
The     cases     with     $(\Phi_1,\Phi_2)=(0^\circ,0^\circ)$     and
$(90^\circ,0^\circ)$ are hardly distinguishable from each other.}
\label{fig:n123edm.cpx.rho}
\end{figure}
\begin{figure}[htb]
\hspace{ 0.0cm}
\vspace{-0.5cm}
\centerline{\epsfig{figure=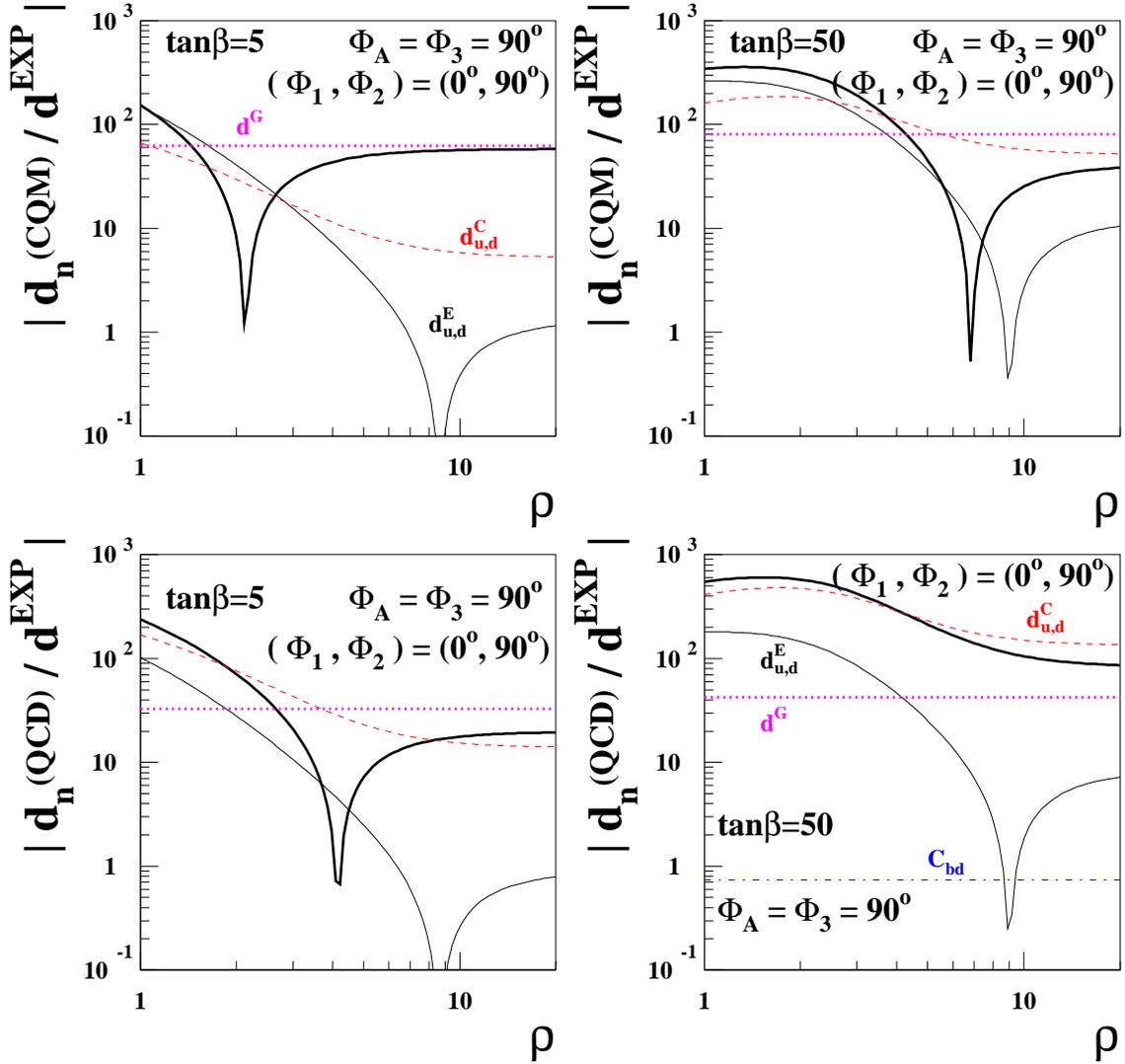,height=16.0cm,width=16.0cm}}
\vspace{-0.5cm}
\caption{\it  Comparison of  the  neutron EDM  calculated  in the  CQM
(upper) and using  the QCD sum rule approach  (lower). Among the lines
in       Fig.~\ref{fig:n123edm.cpx.rho},      the       case      with
$(\Phi_1,\Phi_2)=(0^\circ,90^\circ)$   is  shown  together   with  the
constituent contributions: $d^E_{u,d}$  (thin solid), $d^C_{u,d}$ (thin
dashed),  $d^G$  (thin   horizontal  dotted),  and  $C_{bd,db}$  (thin
horizontal  dash-dotted).  The  thick solid  lines are  for  the total
EDM.}
\label{fig:n13edm.cpx.rho}
\end{figure}
\begin{figure}[htb]
\hspace{ 0.0cm}
\vspace{-0.5cm}
\centerline{\epsfig{figure=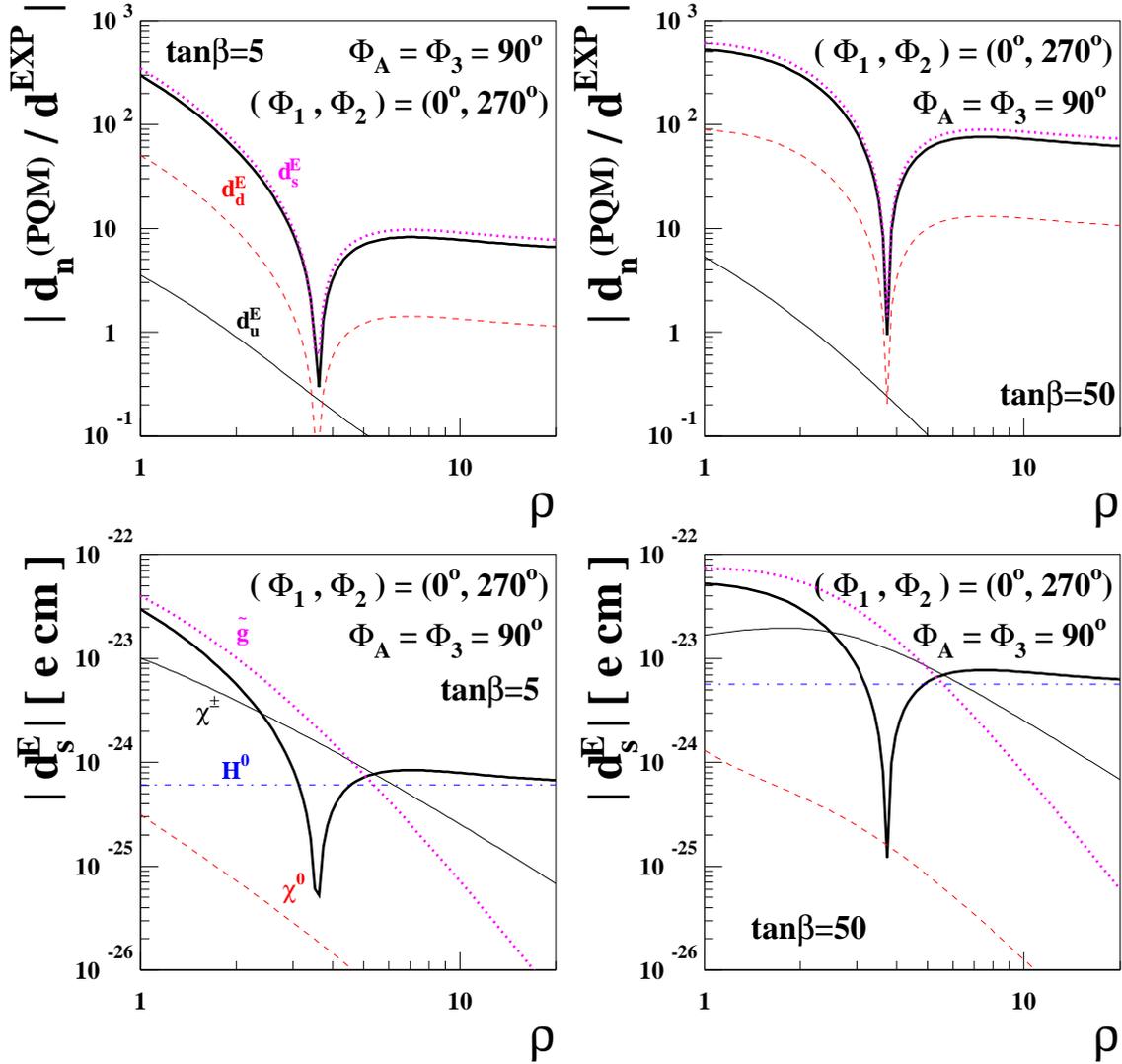,height=16.0cm,width=16.0cm}}
\vspace{-0.5cm}
\caption{\it  The neutron  EDM calculated  in the  PQM.  In  the upper
frames,    among    the    lines    in   the    middle    frames    of
Fig.~\ref{fig:n123edm.cpx.rho},          the         case         with
$(\Phi_1,\Phi_2)=(0^\circ,270^\circ)$  is   shown  together  with  the
constituent contributions: $d^E_u$ (thin solid), $d^E_d$ (thin dashed),
and the  main contribution  from $d^E_s$ (thin  dotted). In  the lower
frames,  the strange-quark  EDM is  shown as  functions of  the common
hierachy  factor $\rho$.  The  thin solid,  dashed, dotted  lines, and
horizontal  dash-dotted  lines  are  for the  contributions  from  the
chargino-, neutralino-, gluino-  and two-loop Higgs-mediated diagrams.
The thick solid lines are for the total EDM.}
\label{fig:n2edm.cpx.rho}
\end{figure}
\begin{figure}[htb]
\hspace{ 0.0cm}
\vspace{-0.5cm}
\centerline{\epsfig{figure=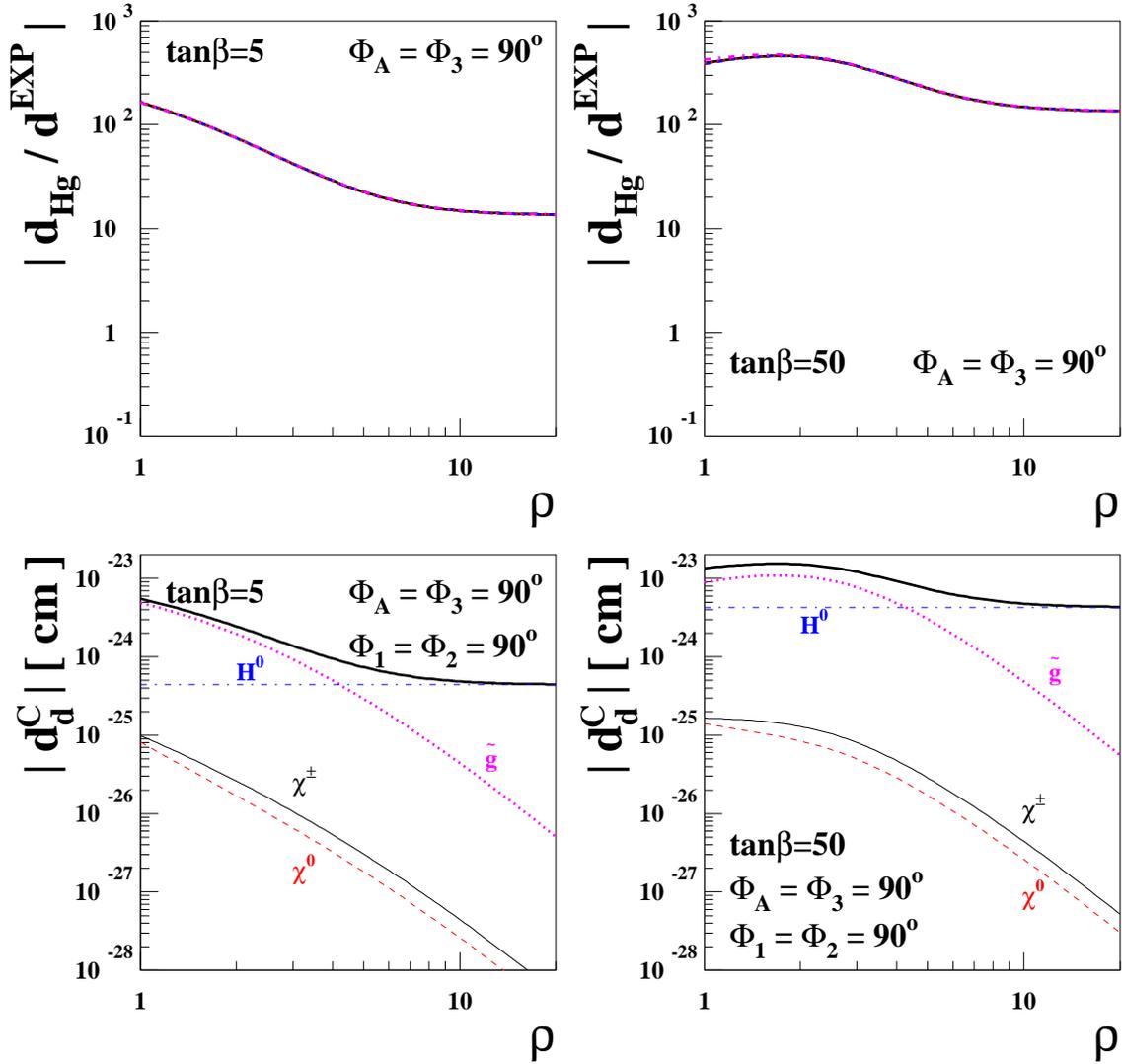,height=16.0cm,width=16.0cm}}
\caption{\it In the upper frames, we show
the Mercury EDM in the CPX scenario with $\Phi_A=\Phi_3=90^\circ$
as a function of the common hierarchy factor $\rho$ for $\tan\beta=5$ (left) and
$\tan\beta=50$ (right).  
It is hardly affected by $(\Phi_1,\Phi_2)$, 
because of the dominance of the contribution from $d^C_{u,d}$, 
see Fig.~\ref{fig:hgedm.cpx}. In the lower frames, we show the dominant CEDM of the down
quark, $d^C_{d}$, as a function of $\rho$.
The lines are the same as in the lower frames of Fig.~\ref{fig:n2edm.cpx.rho}.}
\label{fig:hgedm.cpx.rho}
\end{figure}
\clearpage
\begin{figure}[htb]
\hspace{ 0.0cm}
\vspace{-0.5cm}
\centerline{\epsfig{figure=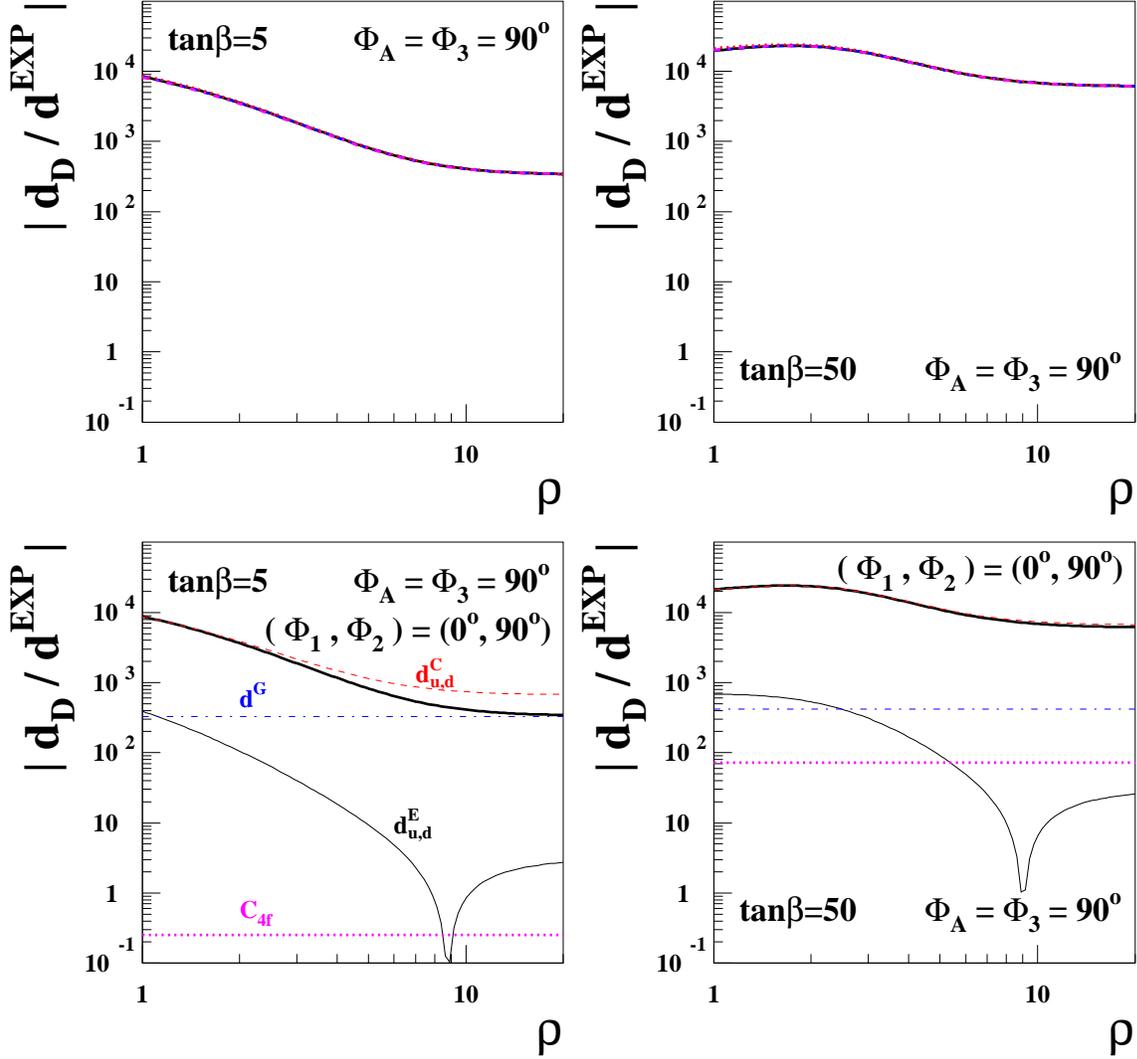,height=16.0cm,width=16.0cm}}
\caption{\it The deuteron EDM in the CPX scenario with $\Phi_A=\Phi_3=90^\circ$
as a function of the common hierarchy factor $\rho$ for $\tan\beta=5$ (left) and
$\tan\beta=50$ (right).  We have taken $d^{\rm EXP}_D = 3\times 10^{-27} {\rm e\,cm}$.
It is hardly affected by $(\Phi_1,\Phi_2)$, see the upper frames.
In the lower frames,
the case with $(\Phi_1,\Phi_2)=(0^\circ,90^\circ)$ is
shown together with the constituent contributions: $d^E_{u,d}$ (thin solid), $d^C_{u,d}$ (thin
dashed), $C_{4f}\equiv C_{dd,sd,bd}$ (thin lower horizontal dotted), and 
$d^G$ (thin upper horizontal dash-dotted). The thick lines are for the total EDM.}
\label{fig:ddedm.cpx.rho}
\end{figure}
%%
%\clearpage
%
\begin{figure}[htb]
\hspace{ 0.0cm}
\vspace{-0.5cm}
\centerline{\epsfig{figure=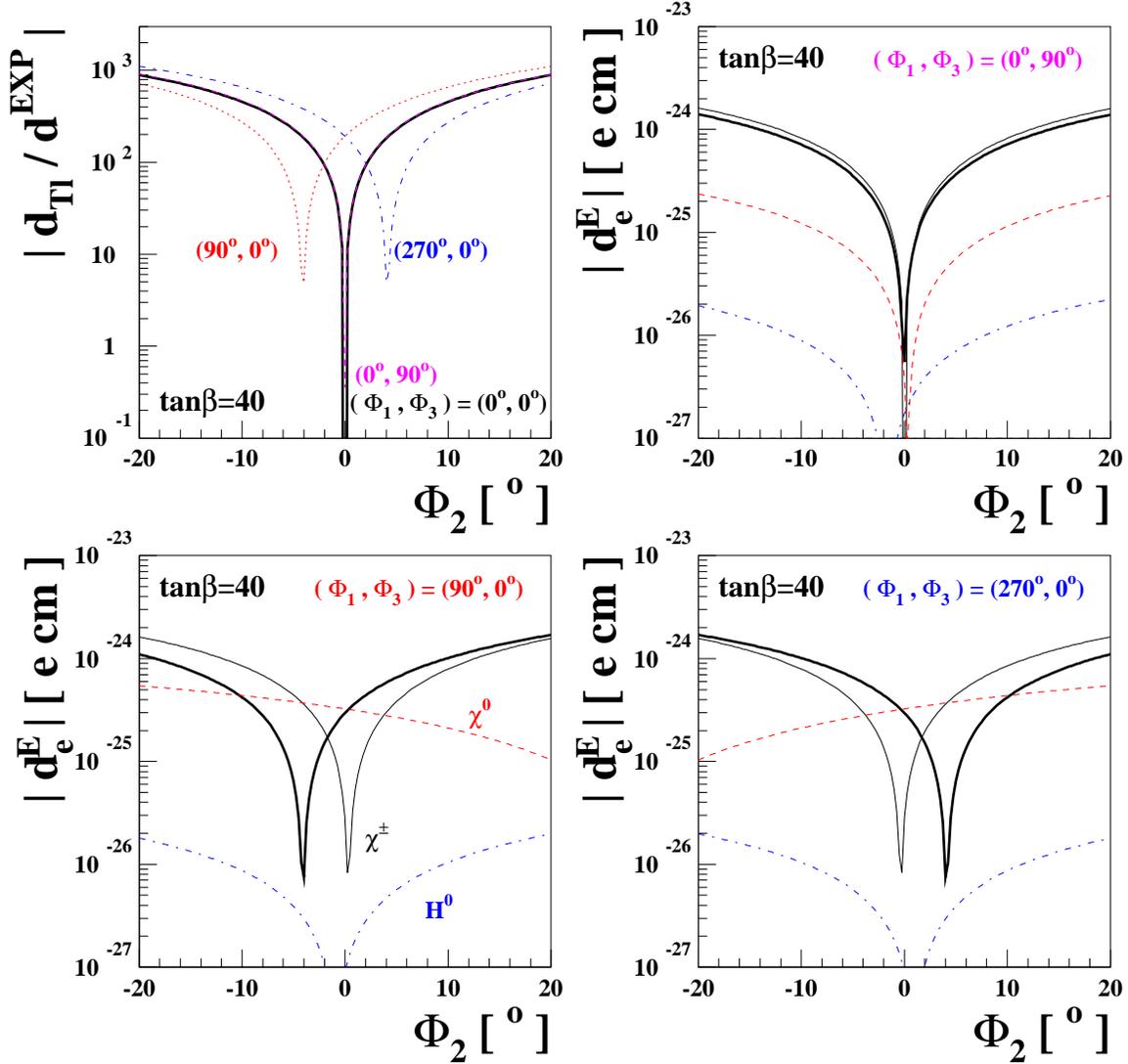,height=16.0cm,width=16.0cm}}
\caption{\it In the upper-left frame, we show
the Thallim EDM in the MCPMFV scenario with $\tan\beta=40$
as a function of $\Phi_2$ for several values of the CP-violating phases $(\Phi_1,\Phi_3)$:
$(0^\circ,0^\circ)$
(solid), $(0^\circ,90^\circ)$ (dashed), $(90^\circ,0^\circ)$ (dotted), and
$(270^\circ,0^\circ)$ (dash-dotted). 
The cases with $(0^\circ,0^\circ)$ and
$(0^\circ,90^\circ)$ are hardly distinguishable from each other due to the dominance of the
electron EDM, $d^E_e$.
In the upper-right, lower-left, and
lower-right frames, we show the electron EDM as a function of $\Phi_2$ when
$(\Phi_1,\Phi_3)=(0^\circ,90^\circ)$,
$(90^\circ,0^\circ)$, and
$(270^\circ,90^\circ)$, respectively. 
The lines are the same as in Fig.~\ref{fig:tledm.cpx.rho}.
}
\label{fig:tledm.sps1a.p2}
\end{figure}
\begin{figure}[htb]
\hspace{ 0.0cm}
\vspace{-0.5cm}
\centerline{\epsfig{figure=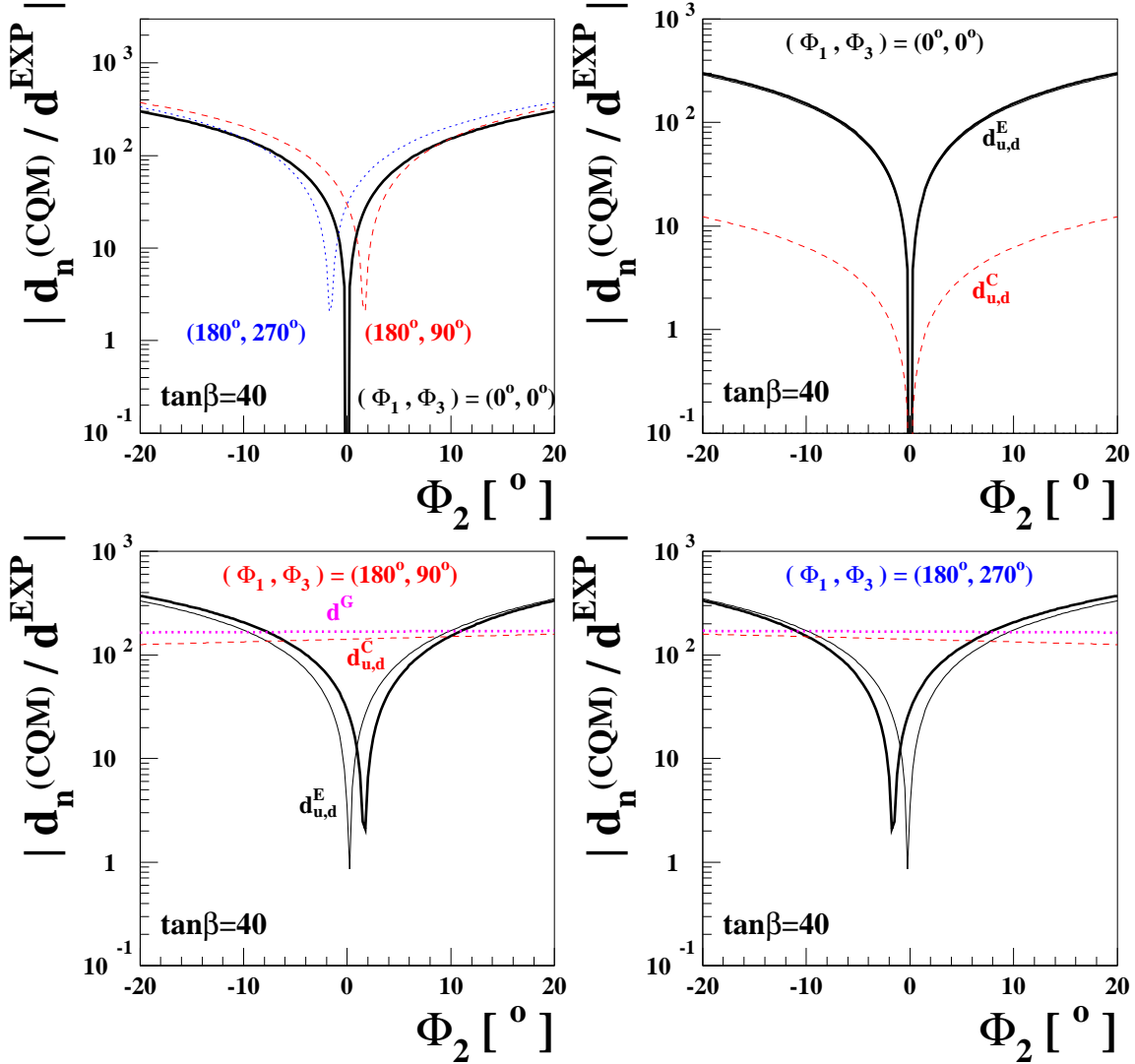,height=16.0cm,width=16.0cm}}
\caption{\it In the upper-left frame, we show
the neutron EDM calculated in the chial quark model
in the MCPMFV scenario with $\tan\beta=40$
as a function of $\Phi_2$ for several values of the CP-violating phases $(\Phi_1,\Phi_3)$:
$(0^\circ,0^\circ)$ (solid), 
$(180^\circ,90^\circ)$ (dashed), and $(180^\circ,270^\circ)$ (dotted).
In the upper-right, lower-left, and
lower-right frames, we show the neutron EDM as a function of $\Phi_2$ when
$(\Phi_1,\Phi_3)=(0^\circ,0^\circ)$,
$(180^\circ,90^\circ)$, and
$(180^\circ,270^\circ)$, respectively, together with its constituent contributions
from $d^E_{u,d}$ (thin solid), $d^C_{u,d}$ (thin dashed), and $d^G$ (thin dotted). 
The thick solid lines are for the total EDM.  }
\label{fig:n1edm.sps1a.p2}
\end{figure}
\begin{figure}[htb]
\hspace{ 0.0cm}
\vspace{-0.5cm}
\centerline{\epsfig{figure=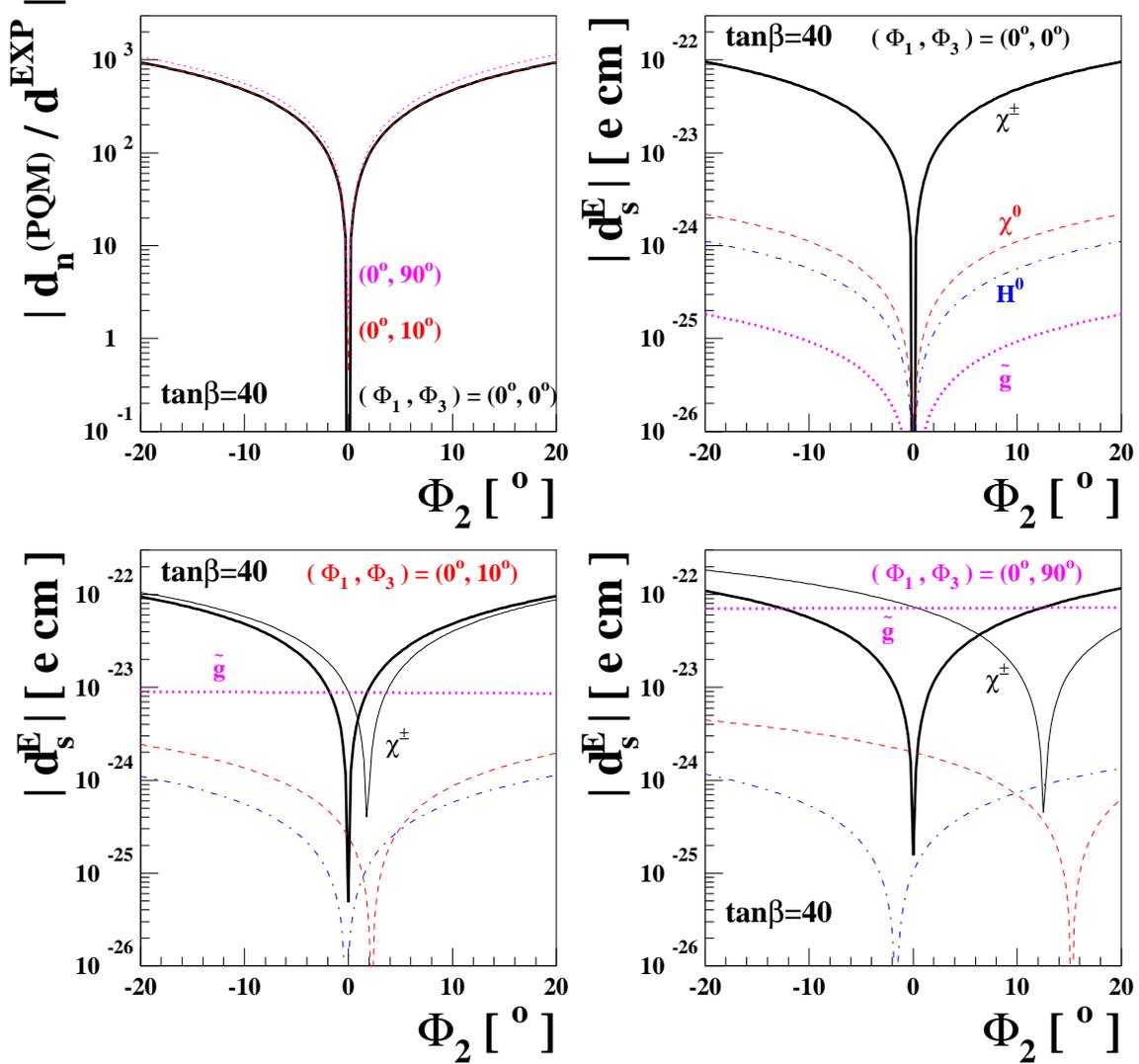,height=16.0cm,width=16.0cm}}
\caption{\it  In  the  upper-left  frame,  we  show  the  neutron  EDM
calculated in the PQM in  the MCPMFV scenario with $\tan\beta=40$ as a
function  of $\Phi_2$ for  several values  of the  CP-violating phases
$(\Phi_1,\Phi_3)$:  $(0^\circ,0^\circ)$  (solid), $(0^\circ,10^\circ)$
(dashed),  and  $(0^\circ,90^\circ)$   (dotted).  All  cases  are  not
distinguishable.  In this model,
the neutron  EDM is dominated  by the strange-quark EDM,  $d^E_s$.  In
the  upper-right,  lower-left, and  lower-right  frames,  we show  the
strange-quark    EDM    as     a    function    of    $\Phi_2$    when
$(\Phi_1,\Phi_3)=(0^\circ,0^\circ)$,     $(0^\circ,10^\circ)$,     and
$(0^\circ,90^\circ)$,  respectively.   The  thin  lines  are  for  the
chargino-  (solid),   neutralino-  (dashed),  gluino-   (dotted),  and
Higgs-mediated (dash-dotted)  diagrams, respectively. The  thick solid
lines are for the total EDM.}
\label{fig:n2edm.sps1a.p2}
\end{figure}
\begin{figure}[htb]
\hspace{ 0.0cm}
\vspace{-0.5cm}
\centerline{\epsfig{figure=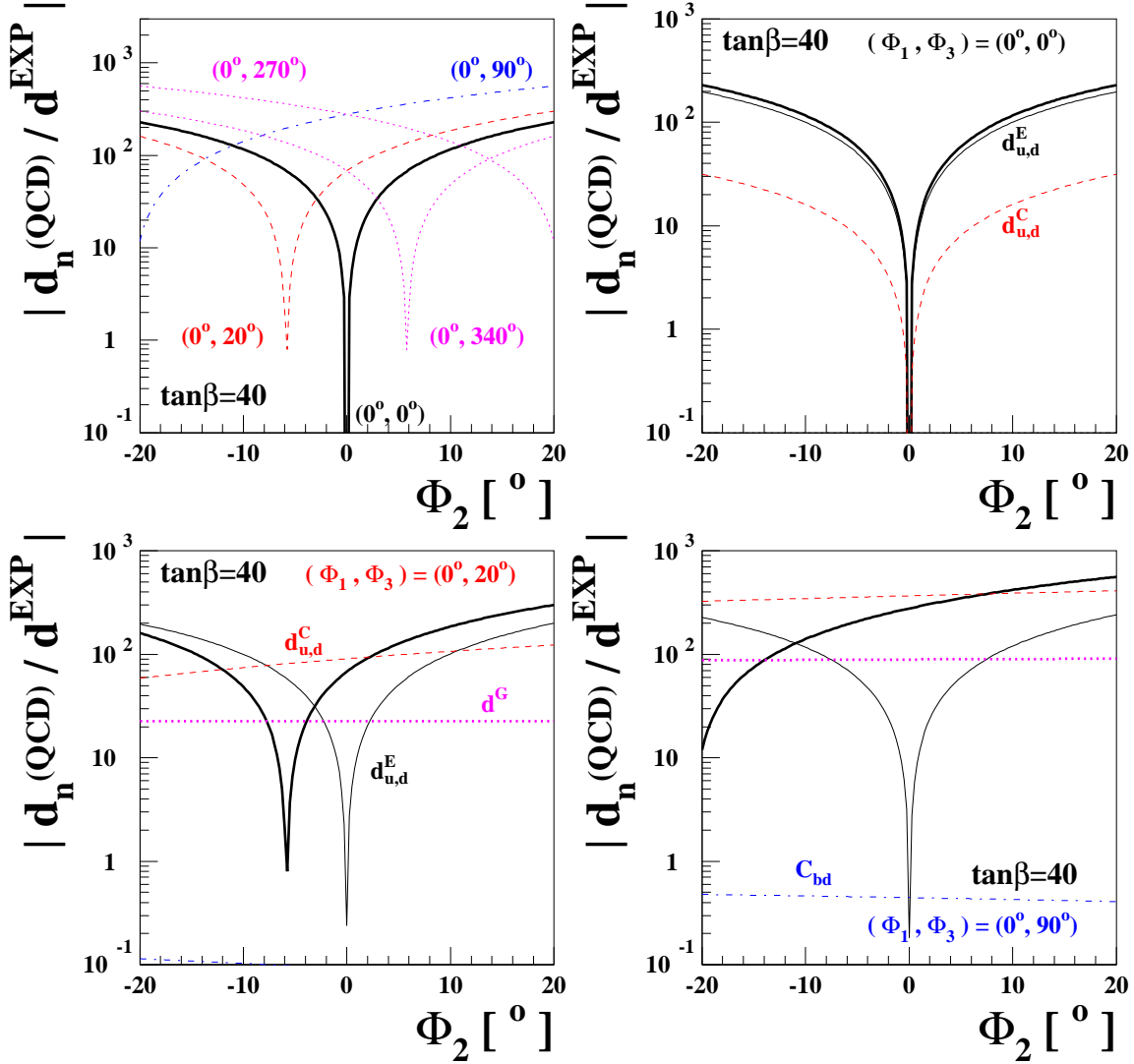,height=16.0cm,width=16.0cm}}
\caption{\it In the upper-left frame, we show
the neutron EDM calculated using the QCD sum rule approach
in the MCPMFV scenario with $\tan\beta=40$
as a function of $\Phi_2$ for several values of the CP-violating phases $(\Phi_1,\Phi_3)$:
$(0^\circ,0^\circ)$ (solid), 
$(0^\circ,20^\circ)$ (dashed), 
$(0^\circ,340^\circ)$ (dotted), 
$(0^\circ,270^\circ)$ (dotted), 
and $(0^\circ,90^\circ)$ (dash-dotted). 
In the upper-right, lower-left, and
lower-right frames, we show the neutron EDM as a function of $\Phi_2$ when
$(\Phi_1,\Phi_3)=(0^\circ,0^\circ)$,
$(0^\circ,20^\circ)$, and
$(0^\circ,90^\circ)$, respectively. The lines are the same as in the lower frames of
Fig.~\ref{fig:n13edm.cpx.rho}.
}
\label{fig:n3edm.sps1a.p2}
\end{figure}
\begin{figure}[htb]
\hspace{ 0.0cm}
\vspace{-0.5cm}
\centerline{\epsfig{figure=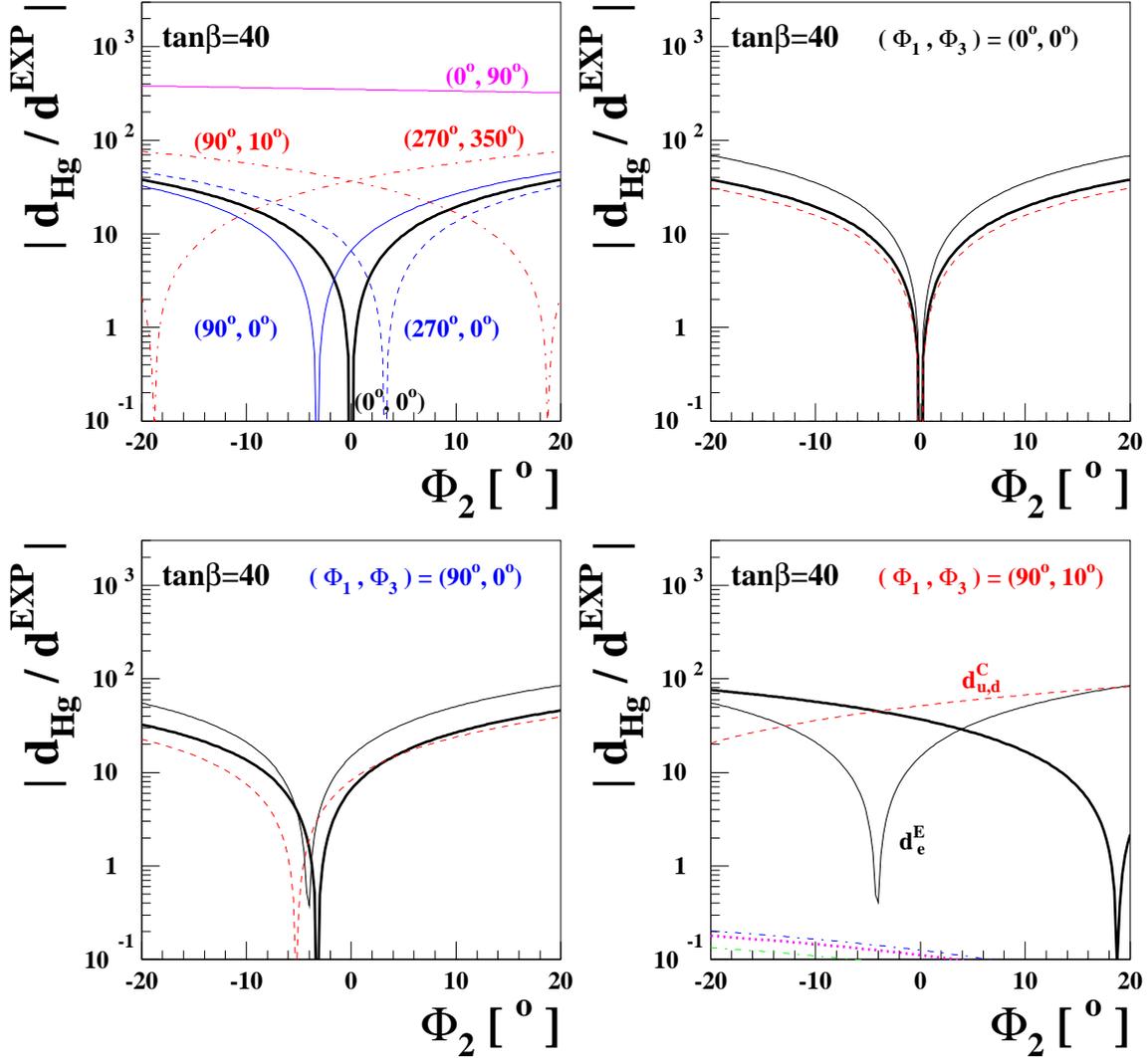,height=16.0cm,width=16.0cm}}
\caption{\it In the upper-left frame, we show
the Mercury EDM calculated using the QCD sum rule approach
in the MCPMFV scenario with $\tan\beta=40$
as a function of $\Phi_2$ for several values of the CP-violating phases $(\Phi_1,\Phi_3)$:
$(0^\circ,0^\circ)$ (solid),
$(90^\circ,0^\circ)$ (solid),
$(270^\circ,340^\circ)$ (dashed),
$(90^\circ,10^\circ)$ (dash-dotted),
$(270^\circ,350^\circ)$ (dash-dotted),
and $(0^\circ,90^\circ)$ (solid).
In the upper-right, lower-left, and
lower-right frames, we show the Mercury EDM as a function of $\Phi_2$ when
$(\Phi_1,\Phi_3)=(0^\circ,0^\circ)$,
$(90^\circ,0^\circ)$, and
$(90^\circ,10^\circ)$, respectively.
The lines are the same as in the lower-right frame of Fig.~\ref{fig:hgedm.cpx}.}
\label{fig:hgedm.sps1a.p2}
\end{figure}
\begin{figure}[htb]
\hspace{ 0.0cm}
\vspace{-0.5cm}
\centerline{\epsfig{figure=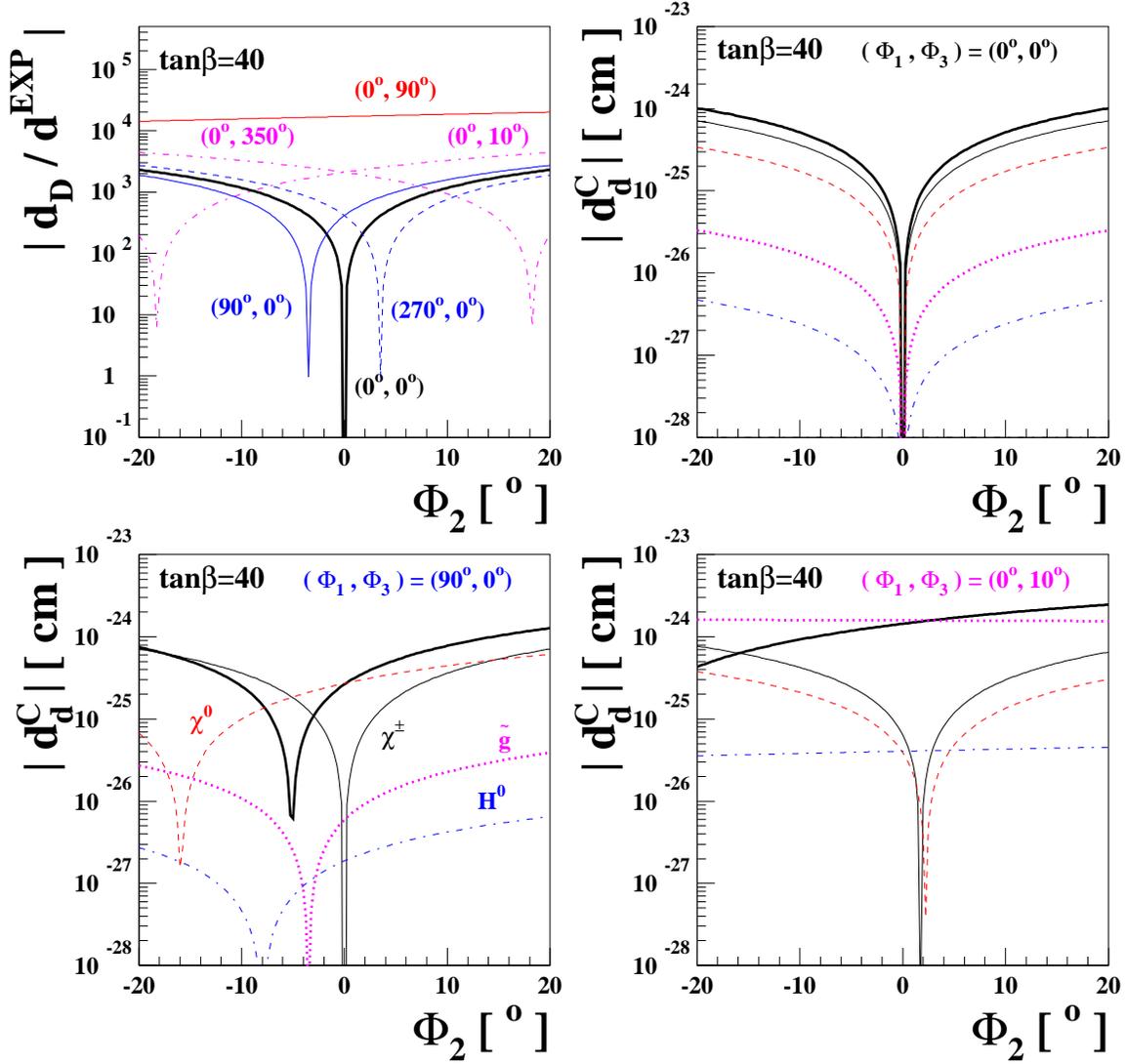,height=16.0cm,width=16.0cm}}
\caption{\it In the upper-left frame, we
show the deuteron EDM calculated using the QCD sum rule techniques
in the MCPMFV scenario with $\tan\beta=40$
as a function of $\Phi_2$ for several values of the CP-violating phases $(\Phi_1,\Phi_3)$:
$(0^\circ,0^\circ)$ (solid),
$(90^\circ,0^\circ)$ (solid),
$(270^\circ,0^\circ)$ (dashed),
$(0^\circ,10^\circ)$ (dash-dotted),
$(0^\circ,350^\circ)$ (dash-dotted),
and $(0^\circ,90^\circ)$ (solid).
$d^{\rm EXP}_D=3\times 10^{-27}$ is taken. 
The dominant contribution comes from the CEDM of the down quark, $d^C_d$.
In the upper-right, lower-left, and
lower-right frames, we show the CEDM of the down quark as functions of $\Phi_2$ when
$(\Phi_1,\Phi_3)=(0^\circ,0^\circ)$,
$(90^\circ,0^\circ)$, and
$(90^\circ,10^\circ)$, respectively. The lines are the same as in 
Fig.~\ref{fig:n2edm.sps1a.p2}.
}
\label{fig:ddedm.sps1a.p2}
\end{figure}

\end{document}

%% file: def_colors
\definecolor{GreenYellow}   {cmyk}{0.15,0,0.69,0}
\definecolor{Yellow}        {cmyk}{0,0,1,0}
\definecolor{Goldenrod}     {cmyk}{0,0.10,0.84,0}
\definecolor{Dandelion}     {cmyk}{0,0.29,0.84,0}
\definecolor{Apricot}       {cmyk}{0,0.32,0.52,0}
\definecolor{Peach}         {cmyk}{0,0.50,0.70,0}
\definecolor{Melon}         {cmyk}{0,0.46,0.50,0}
\definecolor{YellowOrange}  {cmyk}{0,0.42,1,0}
\definecolor{Orange}        {cmyk}{0,0.61,0.87,0}
\definecolor{BurntOrange}   {cmyk}{0,0.51,1,0}
\definecolor{Bittersweet}   {cmyk}{0,0.75,1,0.24}
\definecolor{RedOrange}     {cmyk}{0,0.77,0.87,0}
\definecolor{Mahogany}      {cmyk}{0,0.85,0.87,0.35}
\definecolor{Maroon}        {cmyk}{0,0.87,0.68,0.32}
\definecolor{BrickRed}      {cmyk}{0,0.89,0.94,0.28}
\definecolor{Red}           {cmyk}{0,1,1,0}
\definecolor{OrangeRed}     {cmyk}{0,1,0.50,0}
\definecolor{RubineRed}     {cmyk}{0,1,0.13,0}
\definecolor{WildStrawberry}{cmyk}{0,0.96,0.39,0}
\definecolor{Salmon}        {cmyk}{0,0.53,0.38,0}
\definecolor{CarnationPink} {cmyk}{0,0.63,0,0}
\definecolor{Magenta}       {cmyk}{0,1,0,0}
\definecolor{VioletRed}     {cmyk}{0,0.81,0,0}
\definecolor{Rhodamine}     {cmyk}{0,0.82,0,0}
\definecolor{Mulberry}      {cmyk}{0.34,0.90,0,0.02}
\definecolor{RedViolet}     {cmyk}{0.07,0.90,0,0.34}
\definecolor{Fuchsia}       {cmyk}{0.47,0.91,0,0.08}
\definecolor{Lavender}      {cmyk}{0,0.48,0,0}
\definecolor{Thistle}       {cmyk}{0.12,0.59,0,0}
\definecolor{Orchid}        {cmyk}{0.32,0.64,0,0}
\definecolor{DarkOrchid}    {cmyk}{0.40,0.80,0.20,0}
\definecolor{Purple}        {cmyk}{0.45,0.86,0,0}
\definecolor{Plum}          {cmyk}{0.50,1,0,0}
\definecolor{Violet}        {cmyk}{0.79,0.88,0,0}
\definecolor{RoyalPurple}   {cmyk}{0.75,0.90,0,0}
\definecolor{BlueViolet}    {cmyk}{0.86,0.91,0,0.04}
\definecolor{Periwinkle}    {cmyk}{0.57,0.55,0,0}
\definecolor{CadetBlue}     {cmyk}{0.62,0.57,0.23,0}
\definecolor{CornflowerBlue}{cmyk}{0.65,0.13,0,0}
\definecolor{MidnightBlue}  {cmyk}{0.98,0.13,0,0.43}
\definecolor{NavyBlue}      {cmyk}{0.94,0.54,0,0}
\definecolor{RoyalBlue}     {cmyk}{1,0.50,0,0}
\definecolor{Blue}          {cmyk}{1,1,0,0}
\definecolor{Cerulean}      {cmyk}{0.94,0.11,0,0}
\definecolor{Cyan}          {cmyk}{1,0,0,0}
\definecolor{ProcessBlue}   {cmyk}{0.96,0,0,0}
\definecolor{SkyBlue}       {cmyk}{0.62,0,0.12,0}
\definecolor{Turquoise}     {cmyk}{0.85,0,0.20,0}
\definecolor{TealBlue}      {cmyk}{0.86,0,0.34,0.02}
\definecolor{Aquamarine}    {cmyk}{0.82,0,0.30,0}
\definecolor{BlueGreen}     {cmyk}{0.85,0,0.33,0}
\definecolor{Emerald}       {cmyk}{1,0,0.50,0}
\definecolor{JungleGreen}   {cmyk}{0.99,0,0.52,0}
\definecolor{SeaGreen}      {cmyk}{0.69,0,0.50,0}
\definecolor{Green}         {cmyk}{1,0,1,0}
\definecolor{ForestGreen}   {cmyk}{0.91,0,0.88,0.12}
\definecolor{PineGreen}     {cmyk}{0.92,0,0.59,0.25}
\definecolor{LimeGreen}     {cmyk}{0.50,0,1,0}
\definecolor{YellowGreen}   {cmyk}{0.44,0,0.74,0}
\definecolor{SpringGreen}   {cmyk}{0.26,0,0.76,0}
\definecolor{OliveGreen}    {cmyk}{0.64,0,0.95,0.40}
\definecolor{RawSienna}     {cmyk}{0,0.72,1,0.45}
\definecolor{Sepia}         {cmyk}{0,0.83,1,0.70}
\definecolor{Brown}         {cmyk}{0,0.81,1,0.60}
\definecolor{Tan}           {cmyk}{0.14,0.42,0.56,0}
\definecolor{Gray}          {cmyk}{0,0,0,0.50}
\definecolor{Black}         {cmyk}{0,0,0,1}
\definecolor{White}         {cmyk}{0,0,0,0}